\documentclass[12pt,preprint]{aastex}
\usepackage{epsfig}

\usepackage{url}
\usepackage{color}
\usepackage{lscape}

\makeatletter
\def\url@leostyle{%
  \@ifundefined{selectfont}{\def\UrlFont{\sf}}{\def\UrlFont{\small\ttfamily}}}
\makeatother
\newcommand\gcn{GCN Circ.}

\date{\today\\
Accepted for publication in the Astrophysical Journal Supplement Series.}
\shorttitle{BAT Transient Monitor}
\shortauthors{Krimm et al.}

\begin{document}
\title{The {\em Swift}/BAT Hard X-ray Transient Monitor}
\author{ H. A. Krimm\altaffilmark{1,2}, S. T. Holland\altaffilmark{3,1},  R. H. D. Corbet\altaffilmark{1,4}, A. B. Pearlman\altaffilmark{5,1,4}, P. Romano\altaffilmark{6},  \\ J. A. Kennea\altaffilmark{7},  J. S. Bloom\altaffilmark{8}, S. D. Barthelmy\altaffilmark{9}, W. H. Baumgartner\altaffilmark{1,4}, J. R. Cummings\altaffilmark{1,4}, N. Gehrels\altaffilmark{9}, A. Y. Lien\altaffilmark{10}, C. B. Markwardt\altaffilmark{9},  D. M. Palmer\altaffilmark{11},  T. Sakamoto\altaffilmark{12}, \\ M. Stamatikos\altaffilmark{13}, T. N. Ukwatta\altaffilmark{14,9}
}

\altaffiltext{1}{Center for Research and Exploration in Space Science and Technology (CRESST) and NASA Goddard Space Flight Center, Greenbelt,  MD 20771, USA}
\altaffiltext{2}{Universities Space Research Association, 10211
Wincopin Circle, Suite 500, Columbia, MD 21044, USA}
\altaffiltext{3}{Space Telescope Science Institute, 3700 San Martin Dr., Baltimore, MD 21218, USA}
\altaffiltext{4}{Department of Physics and Center for Space Sciences and Technology, University of Maryland Baltimore County, Baltimore, MD 21250, USA}
\altaffiltext{5}{Department of Applied Physics, California Institute of Technology, 1200 E. California Blvd., Pasadena, CA 91125 USA}
\altaffiltext{6}{INAF, Istituto di Astrofisica Spaziale e Fisica Cosmica, Via U. La Malfa 153, I-90146 Palermo, Italy}
\altaffiltext{7}{Department of Astronomy and Astrophysics, Pennsylvania State University, University Park, PA 16802 USA}
\altaffiltext{8}{Department of Astronomy, University of California, Berkeley, Berkeley CA, 94720-3411, USA}
\altaffiltext{9}{NASA Goddard Space Flight Center, Greenbelt, MD 20771, USA}
\altaffiltext{10}{NASA Postdoctoral Program Fellow, Goddard Space Flight Center, Greenbelt, MD 20771, USA}
\altaffiltext{11}{Los Alamos National Laboratory, B244, Los Alamos, NM, 87545, USA}
\altaffiltext{12}{Department of Physics and Mathematics, College of Science and Engineering, Aoyama Gakuin University, 5-10-1 Fuchinobe, Chuo-ku, Sagamihara-shi, Kanagawa 252-5258, Japan}
\altaffiltext{13}{Department of Physics and Center for Cosmology and Astro-Particle Physics, Ohio State University, Columbus, OH 43210, USA}
\altaffiltext{14}{Department of Physics and Astronomy, Michigan State University, East Lansing, MI 48824, USA}


\begin{abstract}

\noindent The {\em Swift}/Burst Alert Telescope (BAT) hard X-ray transient monitor provides near real-time coverage of the X-ray sky in the energy range 15$-$50~keV.  The BAT observes 88\% of the sky each day with a detection sensitivity of 5.3~mCrab for a full-day observation and a time resolution as fine as 64~seconds.  The three main purposes of the monitor are (1) the discovery of new transient X-ray sources, (2) the detection of outbursts or other changes in the flux of known X-ray sources, and (3) the generation of light curves of more than 900 sources spanning over eight years.  The primary interface for the BAT transient monitor is a public web page.   Between 2005 February 12 and 2013 April 30, 245 sources have been detected in the monitor, 146 of them persistent and 99 detected only in outburst.  Among these sources, 17 were previously unknown and were discovered in the transient monitor.   In this paper, we discuss the methodology and the data processing and filtering for the BAT transient monitor and review its sensitivity and exposure.  We provide a summary of the source detections and classify them according to the variability of their light curves.  Finally, we review all new BAT monitor discoveries; for the new sources that are  previously unpublished, we present basic data analysis and interpretations.  

\noindent \end{abstract}


\section{Introduction}

\noindent In the history of X-ray astronomy, many of the most important results have come from X-ray surveys and monitors, both of which 
usually cover the entire sky or large portions of the sky.  Broadly speaking, the difference between a survey and monitor is the time frame.
Typically, a survey is either integrated over a long time period ($\gtrsim$ 1 yr) or else built up from small numbers of observations of each part of the sky to produce a catalog of sources and their fluxes.  A monitor, on the other hand, operates on shorter time scales ($\sim$ 1 day) with multiple revisits to the same part of the sky to track short time scale variations in known sources and to make the initial discovery of new sources. There have been many wide-field X-ray and $\gamma$-ray surveys, including those by {\em ROSAT} \citep{voges99} in soft X rays, the {\em XMM-Newton} Slew Survey \citep{warw12} in medium energy X rays, {\em INTEGRAL} IBIS/ISGRI \citep{bird10,kriv12} in hard X rays to soft $\gamma$\ rays, the {\em Fermi} Large Area Telescope \citep[LAT;][]{atwo09,nola12} and AGILE \citep{pitt09} in $\gamma$ rays, and the Milagro observatory \citep{atki04,abdo07} in TeV $\gamma$\ rays.

Given the large and rapid variations of most X-ray sources and the strong interest in the field to study outbursts of Galactic X-ray binaries, cataclysmic variables, blazars, etc. as soon after onset as possible, a rapid monitor is a very powerful tool for quickly alerting the astronomical community to changes in flux for known sources.  A large fraction of hard X-ray sources, particularly X-ray binaries, spend long periods of quiescence punctuated by short periods of intense activity.   This long latency means that there are many X-ray sources that have not been active during the era of sensitive wide-field X-ray telescopes; the only reliable way to discover these sources in outburst is with a rapid response monitor.  Finally, the archival light curves produced in an X-ray monitor provide a record of activity for multiple sources on short time scales, tracking outbursts, state changes, and periodic variations, which can be correlated with other observations or used to derive a long-term history of a source. 
 
From 1996 to 2011, the most important X-ray monitor was the All-Sky Monitor (ASM) on the {\em Rossi X-ray Timing Explorer} \citep[{\em RXTE};][]{levi96}, which operated in the 1.5-12 keV band, covering most of the sky every 90 minutes and producing light curves for nearly 200 X-ray sources.  In recent years, this energy band  (2-20 keV) has been covered by the {\em Monitor of All-sky X-ray Image}/Gas Slit Camera \citep[{\em MAXI}/GSC;][]{hiro11,sugi11}.  {\em INTEGRAL}/IBIS is also an effective monitor above 15~keV, though the {\em INTEGRAL} observing plan in concentrated near the Galactic center. {\em Fermi}/LAT has a field of view of about 20\% of the sky and scans continuously, covering the whole sky every three hours to monitor $\gamma$-ray sources between 20 MeV and 300 GeV.  Other instruments use the earth-occultation technique \citep{harm02} to perform as effective X-ray monitors.  These include the Burst and Transient Source Experiment \citep[BATSE;][]{harm02} onboard the {\em Compton Gamma-Ray Observatory (CGRO)} from 1991 to 2000 and the currently operating {\em Fermi} Gamma-ray Burst Monitor \citep[GBM;][]{case11,wils12}.   
Table~\ref{tab-comparison} shows a comparison of the {\em Swift}/BAT monitor with most of these monitoring instruments operating in the hard X-ray band.  We do not include {\em INTEGRAL} in the table because it has a much more concentrated observing plan than the other truly all-sky monitors.

When the {\em Swift} mission \citep{gehr04} was first conceived, it was understood that the Burst Alert Telescope \citep{bart05} would be very useful for serendipitous hard X-ray survey science in addition to its primary role in gamma-ray burst (GRB) science. The  combination of the broad sky coverage integral to the {\em Swift} observing plan and the large field of view of the BAT makes the BAT an ideal survey instrument.  These survey capabilities have been exploited on two different time scales.  The methodology of the BAT hard X-ray survey \citep{mark05, tuel10, baum13, segr10, cusu10} is to combine data covering many years of observations to achieve a deep limiting sensitivity  with the goal of detecting as many nearby active galactic nuclei (AGN) as possible and deriving time-averaged spectra of these extragalactic sources.  Light curves on short time scales are produced in the BAT survey, but these are intended as an archival record of source flux variations, rather than as a means of tracking source behavior in real time.  The main goal of the BAT transient monitor, on the other hand, is to detect and disseminate variations in the flux from bright hard X-ray objects in near real-time, rather than integrated over long time scales.  

 With its broad spatial and spectral coverage and its rapid response, the BAT hard X-ray monitor has become one of the most important monitors in its energy range, 15-50~keV.   The BAT monitor began operations in 2006 October \citep{atel904} and since that time has provided continuous coverage during all times when the {\em Swift} satellite was operational.  The publicly available monitor web page,   
 http://swift.gsfc.nasa.gov/docs/swift/results/transients/, provides light curves for 972 astrophysical sources on two time scales: single {\em Swift} pointings from 64~s to $\sim 1000$\ s (see Section~\ref{methodology} for details) and one-day averages.   All sources in the monitor catalog are tracked, whether or not they are currently detected and light curves starting from 2005 February have been constructed from archival data.  
 All light curves can be downloaded from the web page in either Flexible Image Transport System (FITS) or ASCII format.  The data products are described in more detail in Section~\ref{products}.

In this paper, we describe first how the BAT monitor data are produced and analyzed (Section~\ref{methodology}), covering both the generation of light curves for known sources  (Section~\ref{methodology-basic}) and production of the mosaic images that are used for new source discoveries (Section~\ref{methodology-mosaics}).  In Section~\ref{sensitivity} we discuss the overall sensitivity and exposure of the monitor.  Section~\ref{results} covers the results derived from the BAT monitor from 2005 February 12 through 2013 April 30.  In Section~\ref{results-known}, we discuss previously known sources and in Section~\ref{results-new} we present the observations and interpretations of each of the seventeen new sources discovered with the BAT transient monitor.  Section~\ref{discussion} is a brief summary of overall activities.

\section{BAT Monitor Processing}\label{methodology}

\noindent The primary mission of the {\em Swift} satellite is the rapid detection and study of gamma-ray bursts.  Since GRBs are isotropically distributed, the design of the Burst Alert Telescope, the GRB triggering instrument for {\em Swift},  is based on the need for a large field of view combined with good sensitivity.  Also critical to the {\em Swift} mission is that the BAT angular resolution is sufficient to localize bursts onboard to within the field of view of the two {\em Swift} narrow-field instruments (NFIs), the X-Ray Telescope \citep[XRT; $23.6^{\prime} \times 23.6^{\prime}$;][]{burro05} and UltraViolet/Optical Telescope \citep[UVOT; $17^{\prime} \times 17^{\prime}$;][]{romi05}.  The optimal instrument design for such goals is a coded-mask imager.  The BAT telescope is composed of a mask constructed of 52,000 $5 \times 5 \times 1$\ mm lead tiles distributed in a half-filled random pattern and a detector array of 32,768 $4 \times 4 \times 2$\ mm CdZnTe detectors positioned 1~m below the mask.   A point source is imaged (using a fast Fourier transform) when at least part of the mask lies between the source and the detector.  This configuration results in a BAT field of view (FOV) with the greatest sensitivity in the center and diminishing sensitivity toward the edges as the coding fraction (percentage of the detector array shadowed by the mask) falls.  The field down to 5 (10)\% coding is 2.29 (1.94)~sr .  The field out to 0\% coding is 2.85~sr. The angular resolution also varies with location in the FOV, with a point-spread function ranging from $22^{\prime}$\ in the center to $\sim 14^{\prime}$\ at $50^{\circ}$\ off-axis.  As shown in Section~\ref{methodology-mosaics}, detected source positions can be found to much better than this, normally $\lesssim 4^{\prime}$.

The {\em Swift} observing plan is driven by GRB research and the rapid slewing capabilities of the spacecraft, and secondarily, by the goal of observing as far from the Sun as possible, to facilitate GRB follow-up observations.  When convolved with the observing constraints of {\em Swift}'s near-earth orbit, the avoidance of the Sun, Moon and Earth limb, and the large BAT FOV, the result is that BAT will observe, on average, 80\%$-$94\% (10th to 90th percentile) of the sky each day.  This large coverage makes BAT an ideal instrument for a wide-field X-ray monitor.

The GRB triggers are generated automatically on-board the spacecraft, as discussed in \citet{saka08}.  Most GRBs trigger on time scales of $<64$~s, using a rate trigger.  However, {\em Swift}/BAT has another mode, called the image trigger, which is sensitive to bursts on time scales of 64~s to a full pointing ($\lesssim 20$\ min).  For image triggers the onboard processor first constructs ``scaled maps'' in the 15$-$50~keV band, with the count rate in each detector scaled relative to a full scale value of 255 ($2^8-1$). Next, the scaled maps are convolved with the lead mask pattern using a fast Fourier transform to produce tangent-plane images of the BAT FOV. Point sources found in the images are compared to a catalog of known sources.  An image trigger is generated for either a statistically significant new source or a known source found at a flux level above an outburst threshold specific to that source.  So as to be sensitive to bursts of different durations, scaled maps are produced on multiple time scales, starting at 64~s, and increasing in duration by factors of two up to the full duration of a {\em Swift} pointing.   Since {\em Swift} is in low-earth orbit, the maximum pointing duration is $\approx 1200\ s$, although this varies considerably depending on the observing plan.  In this paper the term ``pointing'' refers to a single continuous observation pointed at the same sky location.
64-s  and full-pointing scaled maps are transmitted to the ground, along with a sampling of other time scales. All of the active non-slewing observing time of {\em Swift} is covered by one or more scaled map.  Although they are produced on-board for a different purpose, GRB detection, the scaled maps are the basic data product for the BAT transient monitor.  

\subsection{Basic data processing}\label{methodology-basic}

\noindent In order for the BAT monitor processing to proceed as rapidly as possible, we use data produced in a customized pipeline, which runs only on BAT data and more quickly than the {\em Swift} Data Center (SDC) pipeline.    Though the custom pipeline is reliable, its use does restrict the data products available for the BAT monitor.  To fit all {\em Swift} data p7roducts into download passes and to downlink data in order of priority, some large data products, such as the multi-energy detector plane histograms (DPHs) used in the BAT hard X-ray survey \citep{tuel10} are broken into pieces, which are reassembled at the SDC.  The custom pipeline lacks the tools to reassemble data products, so only small products such as scaled maps and attitude files can be reliably used.  There is no reduction in sensitivity or coverage for scaled maps when compared with DPHs; the only limitation is that the BAT monitor is restricted to a single energy band (15 - 50 keV).  Although {\em Swift}/BAT commenced operations in 2004 December, the BAT monitor archive begins on 2005 February 12, since there was a change at that time to the data formats of the star camera housekeeping files required for monitor processing, meaning that older data are incompatible with the processing script.

\subsubsection{Pre-processing}\label{preprocess}

\noindent Our custom pipeline, run at the Goddard Space Flight Center, produces Flexible Image Transport System (FITS) files for each type of {\em Swift}/BAT data transmitted to the ground.  These products are organized by data downlink pass.   When a new data pass is produced and available, the transient monitor pre-processing script organizes and, depending on data type, either concatenates or indexes by day the relevant data for use in the BAT monitor.  These data consist of (1) spacecraft attitude and orbital element files, (2) {\em Swift} star camera housekeeping, (3) maps of enabled BAT detectors, and (4) scaled maps.   Along with the attitude and star camera files, the script produces what we call ``bad time intervals," which mark times during which there are gaps in the attitude data or an invalid star camera solution. 
The scaled maps are flagged to indicate short maps ($< 64\ s$), long maps ($\geq 64\ s$) and full pointing maps.  Full pointing maps are defined as the longest duration maps covering a particular time interval.  For some pointings, in particular those interrupted by GRB observations, there may not be a full-pointing map.  In this case, the flags indicate which long maps are to be grouped together to cover a pointing.   

\subsubsection{Filtering and corrections}\label{filtering}

\noindent There are several levels of data filtering, most of which follow closely those employed in the BAT survey \citep{tuel10}.  First the ``bad times'' (Section \ref{preprocess}) are rejected.  The script also makes sure that the time of each scaled map is covered in the spacecraft attitude file.  Next we use the {\ttfamily aspect}\footnote{This tool and the others mentioned in this paper are distributed as part of the FTOOLs package: http://heasarc.gsfc.nasa.gov/docs/software/ftools/ftools\_menu.html} 
tool to find the median attitude for each map.  This is necessary because the attitude at the beginning or end of the exposure is sometimes less well settled than the attitude in the middle.  Any detectors disabled by the BAT flight software are masked so that they are not included in any solutions. In addition, a ``global pattern mask'' is used to mask detectors that have significantly higher than average variance compared to Poisson statistics.  This is effective in filtering noise due to differential illumination of the sides of detectors by bright off-axis sources.
Finally, the tool {\ttfamily bathotpix} is used to mask detectors that are hot (noisy) in a particular map and to reject maps where there are more than 15 hot  detectors, beyond those masked or disabled for other reasons.
A by-product of the imaging is a map that delineates, as a function of sky position, the partial coding fraction, or percentage of the BAT detector array that is illuminated through the mask.  Source flux is only calculated when the partial coding fraction is at least 10\%.

Even though the {\em Swift} observing plan prevents pointing the narrow-field instruments near the Sun, Moon or Earth limb, the BAT field of view is so large that any of these objects can be within the field and thus occult the sky behind them.  The position of the earth limb is tracked with {\ttfamily batoccultmap} as it moves through the field during a pointing and two corrections are made (the angular sizes of the Sun and moon are too small to significantly affect source detection and are not corrected for).  First, the partial coding map is multiplied by the occultation map to reduce the coding in occulted regions and secondly the sky image is divided by the occultation map to correct for losses due to occultation.

Purely geometric projection corrections are handled automatically in the BAT tools, as are distortions due to the very small warp in the lead mask.  However, additional corrections must be made for the passive materials above the detector array and for the collimation effect of the 5-cm thick composite honeycomb panel supporting the BAT mask \citep{bart05}.  Both of these effects are energy-dependent and corrections that been derived empirically for the BAT hard X-ray survey are applied in the transient monitor. 
Another important correction is to remove what we call the ``fixed pattern noise'' from the detector array.  This pattern, described fully in Section 3.3 of \citet{tuel10}, is based on trends in the long-term running average cleaned rate for each individual detector in the array.

Due to a flight software issue, there are rare cases when the highest scaled count value calculated for a map is greater than the 8-bit maximum map value (255).  In these cases, the value written ``wraps" to a lower number (modulo 255).   A filter is applied to remove such cases from all light curves and averages.  This is found to happen only when there are two or more very bright sources in the field of view and affects less than 0.1 \% of the data, mostly in late 2009 when 1A~0535+262, which is near the Crab in the sky, was in an exceptionally bright outburst.  

Finally, it is a known property of coded mask imaging that systematic errors arising from the presence of bright sources in the field of view are spatially correlated. If two or more spacecraft pointings have exactly the same orientation on the sky (to within a few arcminutes) then fluctuations due to systematics (either positive or negative) will tend to accumulate in a particular location in the BAT field of view and hence at a particular equatorial or Galactic sky coordinate.  
For example, a $2\sigma$\ positive fluctuation in multiple single pointings at the same sky location would accumulate and grow to a $\gtrsim 7\sigma$\ positive point in the daily averages.  For such points, the apparent significance would be much higher than it should be because the systematic error bars are underestimated.  

To mitigate this effect, starting on  2005 September  17, the {\em Swift} mission operations team instituted a procedure known as ``roll angle dithering.''  In successive pointings at the same target (same field center), the spacecraft roll is changed to a value within $\pm 1^{\circ}$\ of the original value. The maximum size of the change is chosen to be small enough so that it does not adversely affect operations of  the {\em Swift} NFIs (XRT and UVOT), but it does ensure that systfematic errors do not accumulate in BAT images. A rotation by $\pm 1^{\circ}$\ means that any mask element at least 1 mm/tan($1^{\circ}) = 57$\ mm from the center will be shifted by at least one element width, leaving only $\approx 1.4$\% of the array ``undithered''.
The roll dithering procedure is carried out for most targets. However, there are certain situations in which it is not done. Since the dithering must be commanded, there is no dithering for automatic targets (ATs), which are GRBs or other transients that trigger on-board that lead to an automatic observation.  Similarly there is no dithering for those targets of opportunity (ToOs) observations that are uploaded outside the normal observing plan. There are also other times when a decision is made not to do the dithering, either because the precise orientation of a source in the UVOT or XRT field is required (e.g. for UV grism observations), or for NFI calibration purposes. Finally, for part of 2005 and early 2006, the dithering commands were generated by hand, and sometimes this step was forgotten in calculating the daily observing schedule.

To identify ``no-dither'' times, the pre-processing script produces a draft as-flown science timeline of {\em Swift} observations by concatenating the published pre-planned science timeline with the actual spacecraft attitude.  During this process, a flag is set for each pointing indicating whether or not the spacecraft roll angle was changed between successive observations at the same nominal sky coordinates. No-dither pointings remain in the light curves and are included in the daily averages, but a flag (2 for automated targets, 1 for other no-dither cases, 0 for roll-dithering active) is set  in  the final source light curves for such pointings.  We follow this course because it is a random process whether or not any given point in the sky will show this effect; hence most sources are unaffected.  The flag is a warning to investigate carefully any unusual light curve peak during a no-dither pointing.

\subsubsection{Source imaging, cleaning and masking}\label{cleaning}

\noindent The core processing tool is {\ttfamily batfftimage}, which uses a fast Fourier transform to deconvolve the illumination pattern of BAT detectors with the known random pattern of closed and open coded mask elements to produce an image in sky coordinates.  The native coordinates of the image are tangent plane coordinates, which are registered to equatorial coordinates using the FITS World Coordinate System convention.   As discussed in \citet{tuel10}, tangent plane coordinates provide a distortion-free system for a coded-mask imager over the entire field of view.  The image contains the reconstructed distribution of point sources plus background within the BAT field of view.  
This coded mask deconvolution technique also produces, across the sky image, systematic  noise due to the diffuse sky background and also bright point sources.  The {\ttfamily batclean} tool was developed to ``clean'' BAT detector plane images of these sources of noise.  The cleaning is carried out in two steps.  First, we fit a 14-element background model to the data.  This includes a constant term, terms proportional to each of the two orthogonal directions of the array, their squares and cross-products and corresponding terms for detectors on different sides of array ``sandwiches'' \citep[see][]{bart05}.  After the background fit is subtracted, the sky image is searched for bright sources ($> 9\sigma$).  For each bright source, {\ttfamily batclean} forward projects (ray traces) along the source direction to determine the model illumination pattern expected from the source.  Each source model is then added to the background model and fitted to the detector plane.  At this stage, the map is also ``balanced'' to remove systematics due to variations between individual detectors from geometry and detector quality.  This process is explained in Section 3.2 of \citet{tuel10}.

Very bright sources can also illuminate the array through the shield enclosing the space between the coded mask and detector array.  Diffuse illumination through the shield itself is not a problem, but shadowing by the mask support structures on the edge of the mask can add significant noise.  This problem is handled by ray tracing to map the shadows and then masking the shadowed detectors.   The final sky image is created using the balanced and masked detector plane image with the fit from {\ttfamily batclean} subtracted.  

 \subsubsection{Source detection}\label{detection}

\noindent The final step is to search the sky image for sources.The tool {\ttfamily batcelldetect} is used to fit a point-spread function to all sources in a user-supplied catalog and to high significance points not in the catalog, but found using a sliding cell method. In the sliding cell method, a small window or cell\footnote{All of the sliding cell parameters are the default values for  {\ttfamily batcelldetect.} The cell is a circular annulus with outer radius 30 pixels and inner radius 6 pixels, the step size is 0.05 pixels, and the detection threshold is 5$\sigma$.}  in which flux and background are calculated is systematically moved across the image to reveal sources above a preset threshold.   Through this process a count rate and background variance (statistical error) can be determined for all catalog sources, whether or not they are formally detected in the image.  Since the final image is missing the now-cleaned bright sources, fluxes for cleaned sources are determined from the intermediate stage image (with only diffuse background cleaned).  Catalog entries were chosen to contain known hard X-ray sources, mostly Galactic binaries and blazars, along with other classes of sources that have a possibility of being detected by BAT.  The distribution of source classes and detection statistics are covered in Section \ref{results}.

A catalog file is produced for each processed sky image and at the end of the processing, all new catalog files are concatenated and then split by source so that the light curve for each individual source can be updated.  As the source light curves are produced, data are combined for multiple time-contiguous intervals within a single spacecraft pointing.  Also a weighted average rate is calculated for each source for each universal time (UT) calendar day.  Statistical errors for the day are combined in quadrature.  The daily average light curves also contain entries providing the total exposure for the day,  the exposure weighted by the partial coding fraction, and the exposure time for which roll-angle dithering was done.  In both the pointing (orbit) level and daily average level some data are produced that are considered to be of low quality.  Such data are flagged to indicate either a large ($<-10\sigma$) negative fluctuation or a statistical error more than four times the mean statistical error for the source or both.  Flagged points are excluded from the light curve plots and daily averages.

 \subsubsection{Systematic Errors}\label{statistics}

\noindent Although every effort is made to reduce systematic errors in the transient monitor analysis, using data cuts and corrections, the overall errors remain larger than what is expected from purely Gaussian statistics. The systematics are accounted for in two systematic error terms described here.

\begin{figure}
 \plotone{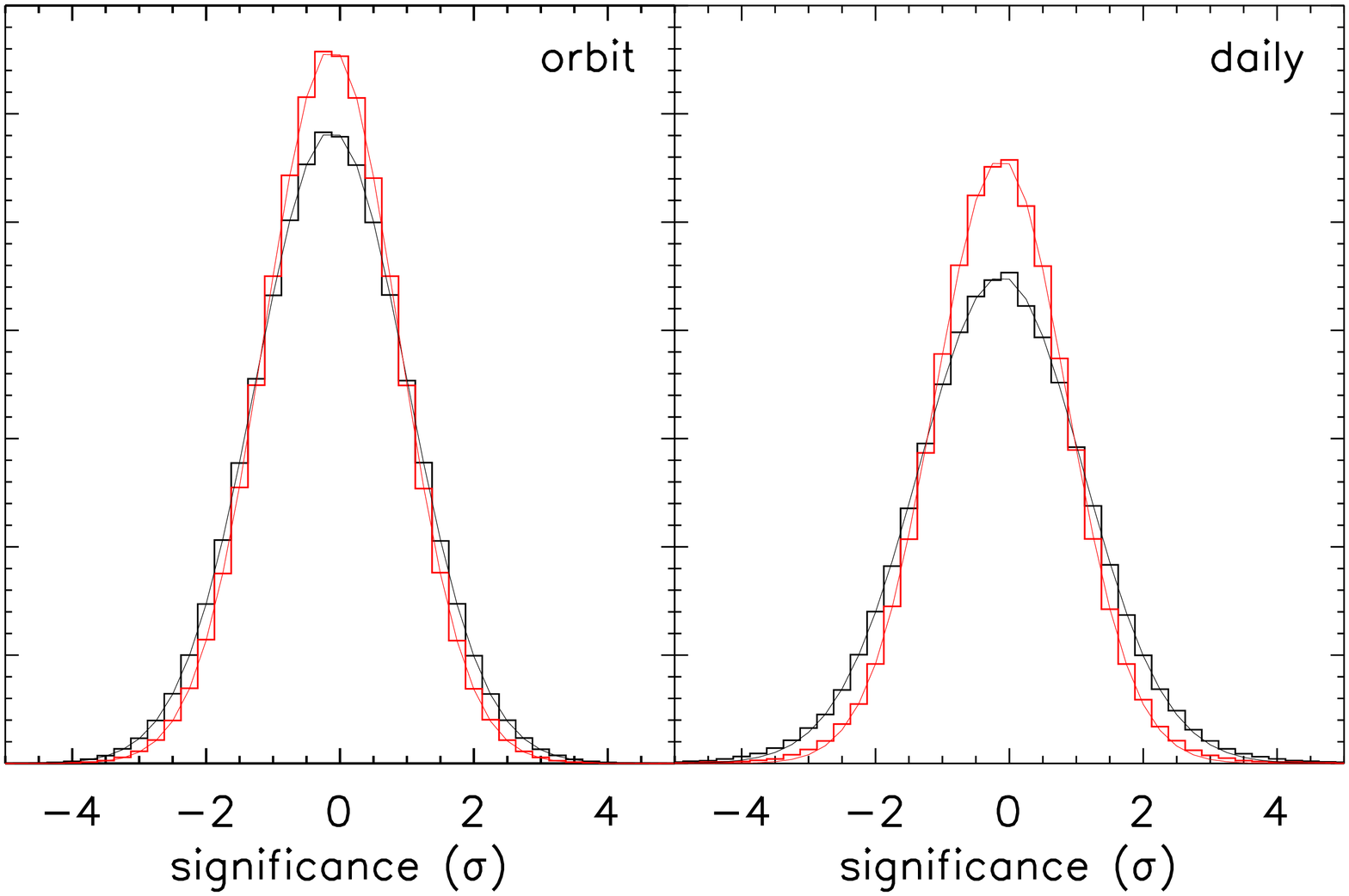}  
 \caption{Study of the significance distribution for blank sky points.  The black histograms and Gaussian fit curves are for the unadjusted statistical errors.  In the red histograms and Gaussian fit curves, the errors have been increased by a factor of 1.126 for the orbit light curves (left) and by a factor of 1.222 for the daily light curves (right) to force the distributions to be Gaussian with a width of unity.  Note that, since these figures include data for the entire duration of the monitor, the correction factors are weighted averages of the 2005 and post-2005 values given in the text. All BAT monitor systematic errors are increased by either the orbit or daily factor, as appropriate.}\label{signif_fig}
\end{figure}

In order to understand residual systematics in the distributions of counts from catalog sources, the BAT transient monitor catalog includes 106 ``blank'' points in the sky, randomly distributed across the sky and chosen to be at least 10 arc minutes from any reported X-ray source. Since there are no sources in these locations, the distribution of significances of counts from these locations should follow a Gaussian distribution with zero mean and width of unity. As seen in Figure~\ref{signif_fig}, there are no systematic biases toward either high or low significance, however, the width of the significance histograms (black in Figure~\ref{signif_fig}) are larger than one, which indicates that the statistical errors underestimate the true distribution of errors. The statistical errors must therefore be increased by a systematic factor that makes the width of the distribution unity. This correction is applied as a multiplicative factor that increases all statistical error values in the transient monitor.  
For the period after roll dithering was instituted (from 2006 onward; see Section~\ref{filtering}), the mean correction was found to be 12.2\% for the orbital data and 20.5\% for daily averages. (A larger correction of 16.7\% for orbital and 44.5\% for daily is applied to the 2005 data.) No significant variations in the correction factors are found after 2005. The daily average correction is always larger than the orbital correction, because systematic errors increase as the 
 integration time increases.

\begin{figure}
 \plotone{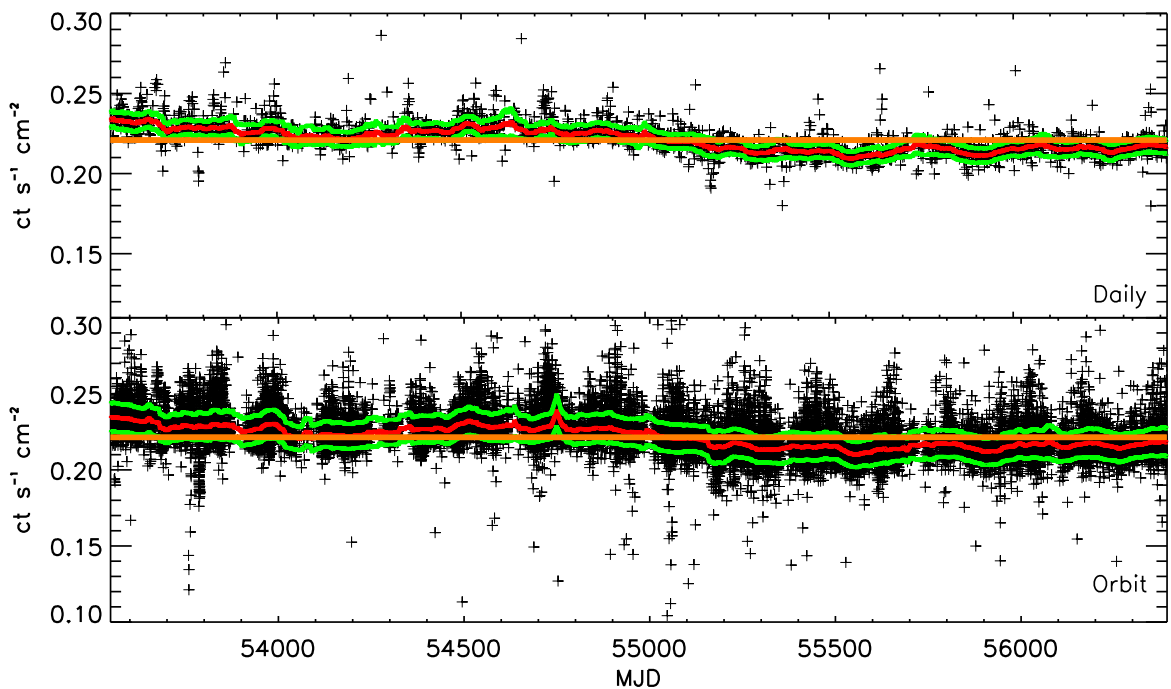}  
 \caption{Light curve of the Crab in the BAT transient monitor.   The top plot shows the daily averages and the bottom plot the orbit-by-orbit variations.  In each plot, the red curve is the trend based on 60-day sliding windows and the green curves show one standard deviation based on the scatter in the data points.  The orange line indicates the overall average rate and shows significant deviations in the Crab rate as was found in \citet{wils11}.  Note that the Crab flux has been below the long-term average (over the entire transient monitor light curve) of $0.221\ \rm{ct\  cm^{-2}\ s^{-1}}$ since approximately August 2009 (MJD 55046). }\label{crab_fig}
\end{figure}

A second systematic error is derived from an empirical analysis of the Crab light curve for which it was found that there was more scatter in the data points than could be explained by statistical variations alone. Since the transient monitor data are not corrected using the BAT response matrix, these errors are expected to affect the measured flux by $\sim10$\%.  We studied this effect by determining the deviation of the data from the long-term trend of the Crab light curve, which was calculated using a 60-day sliding window (see Figure~\ref{crab_fig}).  It was found that the residual scatter in the orbit light curve had a standard deviation of 3.01\% of the Crab trend rate, and in the daily light curve 1.82\% of the trend rate. This value is applied to all light curves, but only makes an important contribution to bright sources.

It is important to note that there are strong spatial correlations in the BAT observations of a given source that can place the Crab or another bright source in the same location in the field of view for many days at a time. Swift is not a scanning survey instrument. Its observing program is driven by the random location of gamma-ray bursts on the sky, and gamma-ray burst afterglows are typically observed for many days. Thus in any given $\sim$week-long interval, the bright sources are likely to be at the same locations in the BAT field of view and the same systematics will apply.  Even though bright sources are cleaned (see Section~\ref{cleaning}), the cleaning is not perfect and residual effects of uncleaned sources can lead to  the short time-scale coherent structure seen in the light curves.

 \subsubsection{Data Products}\label{products}

\noindent For each source, all data products are produced and updated for each processing run and immediately made available on the monitor web site.  Each catalog source has a separate page on the monitor web site containing images and also links to download  data products.  The products that can be downloaded are two light curves:  orbit (pointing)-level and daily average, both of which extend back to the start of the monitor\footnote{The exception is for sources recently added to the monitor catalog, for which the light curve plots and tables initially extend only back to the time of the addition of the source to the catalog, but are completed to the start of the monitor with roughly yearly reprocessing of the monitor data.}, and are available in FITS and ASCII formats.  Each source that is also detected in the BAT hard X-ray survey has a link directly from the monitor source page to the corresponding survey source page. Data storage capacity limits us from serving the tangent plane images from which the light curves are generated or the raw scaled maps. From the light curves we generate and display three plots for each source: an orbit-level light curve plot covering the past thirty days, and daily average light curve plots covering the entire mission and the past year.  Although BAT does not have an explicit Sun-constraint like the {\em Swift} NFI's, Sun-avoidance does affect the BAT monitor light curves.  When a source position is near that of the Sun, its sky coverage is significantly reduced, causing gaps in the monitor light curve and an increase in the size of the error bars.  This effect is most prominent in mid-December when the many sources near the Galactic center show these effects.   

The main monitor web page includes a table listing all catalog sources,  and there are also several other subsidiary pages including tables of currently detected sources, historically detected sources, black hole transients and flare stars.

\subsection{Daily mosaics}\label{methodology-mosaics}

\noindent In addition to deriving light curves of known sources, the BAT transient monitor is also useful for discovering previously undiscovered sources.  Details of the transient monitor discoveries are found in Section~\ref{results-new}.  Here we discuss the methodology of the search.

All sky images derived from full-pointing maps are combined into a series of mosaic maps.  The images used for this have been cleaned of both background and bright sources, so the mosaic maps do not include sources bright enough to have been cleaned (SNR$> 9\sigma$\ in a single image). Since the purpose of the mosaics is to search for previously unknown sources, the absence of bright sources does not affect results.
Except for the time scales, the procedure for producing the mosaics is the same as that outlined in Section 3.5 of \citet{tuel10}.  The sky is divided into six facets in Galactic coordinates and the maps are accumulated on five different time scales: 1-day, 2-day, 4-day, 8-day and 16-day.  Along with flux maps, two auxiliary mosaic maps are created on each time scale.  The first is a coded exposure mosaic map that gives, for each point in the sky, the temporal exposure scaled by the fractional coding.  This is derived by combining individual coding images (Section~2.1.2) multiplied by the exposure of the image.  The second auxiliary  mosaic map is of the average variance for each part of the sky.

Since the daily mosaics usually provide the first position determination for newly discovered sources, it is important to understand the position accuracy as a function of source brightness.  To investigate this, we ran the source detection program on all levels of daily mosaics, but with the option in {\ttfamily batcelldetect} of allowing the source fit position to vary ({\ttfamily posfit=YES}), choosing  {\ttfamily posfitwindow = 7.2}~(arc minutes).  This way we could compare the derived positions of known sources to the  best catalog positions.  We had to make several cuts.  First of all, as discussed above, the images used to make the daily mosaics do not include cleaned sources.  Since sources near the cleaning threshold are present in some images and not in others contributing to the same mosaic, we must exclude all sources that have been cleaned at any time in the monitor process.  This removes 77 sources, but still leaves many detected sources.  Secondly, we must exclude confused sources since neither positions nor fluxes are accurate in such cases.  We removed sources that are listed as confused in the BAT survey catalog \citep{tuel10}.  This deletes 17 sources that had not already been excluded.  

\begin{figure}
 \plotone{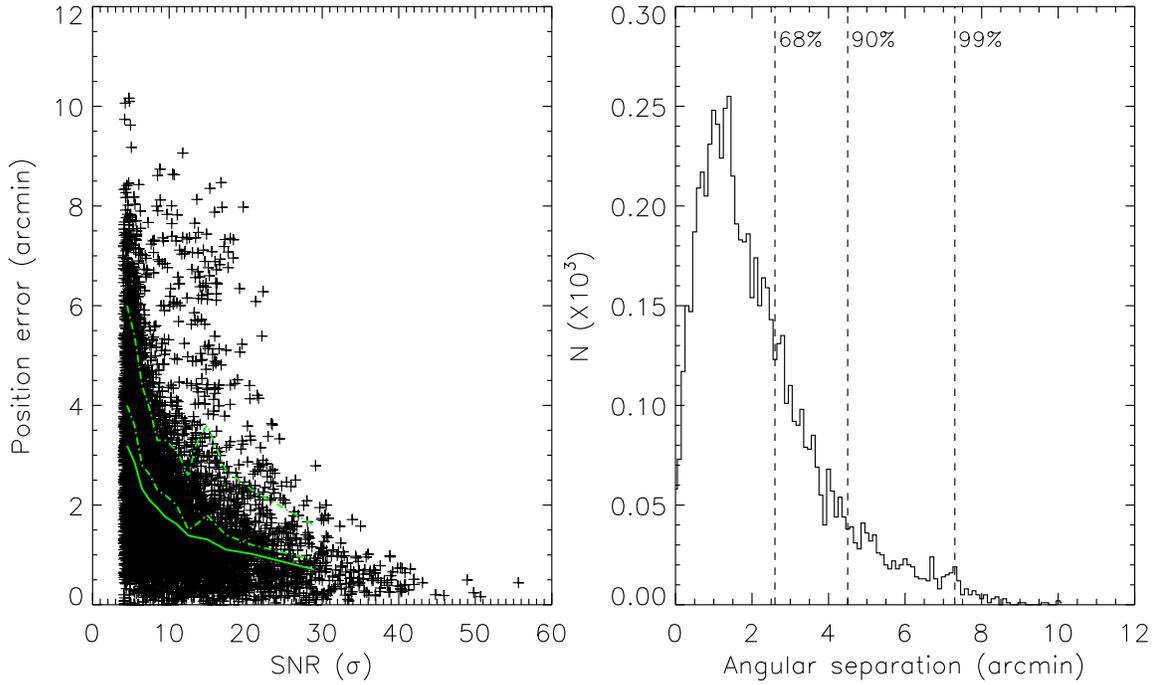}  
 \caption{Left: Scatter plot of position error versus source detection significance.  We have parameterized the distribution and show in the green curves (from bottom to top): the best fit to the distribution (solid), the 68\% confidence limit (C.L.), and the 90\% C.L. Right:  Histogram of angular separations between the BAT position and the best catalog position (averaged over all values of SNR). Source positions have an accuracy of better than what is indicated by the vertical lines in, from left to right, 68\%, 90\% or 95\% of cases.}\label{daily1_fig}
\end{figure}

We can then compare, for each detection, the signal to noise ratio (SNR) from the fixed-position (archival) fits to the position from the varying-position fits (undetected sources will not contribute since they fall below the SNR threshold).  The results are shown in Figure~\ref{daily1_fig}, left panel.  We use SNR from the fixed-position fits since when the position is allowed to vary within a wide radius, the program will not always fit a peak near the source, but will sometimes produce a best fit position in the direction toward another bright source and, in so-doing, over-estimate the source flux through contamination from the bright source.   The overall distribution of position errors is shown in the right panel of Figure~\ref{daily1_fig}, showing that in 99\% of cases, the position is fit to within 7.25~arc minutes, smaller than the field of view of the {\em Swift} XRT.  
In the left-hand panel, one can see that although the plot is filled in, there are almost no cases where a significantly detected source (SNR $> 6\sigma$) yields a position worse than 8~arc minutes.  This means that a null detection in the {\em Swift} XRT of a new source is quite unlikely to be due to poor positioning in the BAT transient monitor.

\section{Sensitivity and exposure}\label{sensitivity}

\noindent The average exposure of the BAT transient monitor is calculated from the one-day mosaics.  A given patch of sky is considered to be exposed for a particular day, if during that day, it was at least 10\% coded for at least one observation.  With this definition, we can calculate, for each day, what fraction, or percentage, of the sky is exposed.   Examining the distribution of daily exposure fractions we find that 
the mean daily exposure percentage is 87\% (the range from 10th to 90th percentile is 78\% to 95\%).\footnote{The same calculation changing the definition of ``exposed'' to 20\% yields 79\% mean daily exposure (10th to 90th percentile range of 70\% to 87\%).}

\begin{figure}
 \plotone{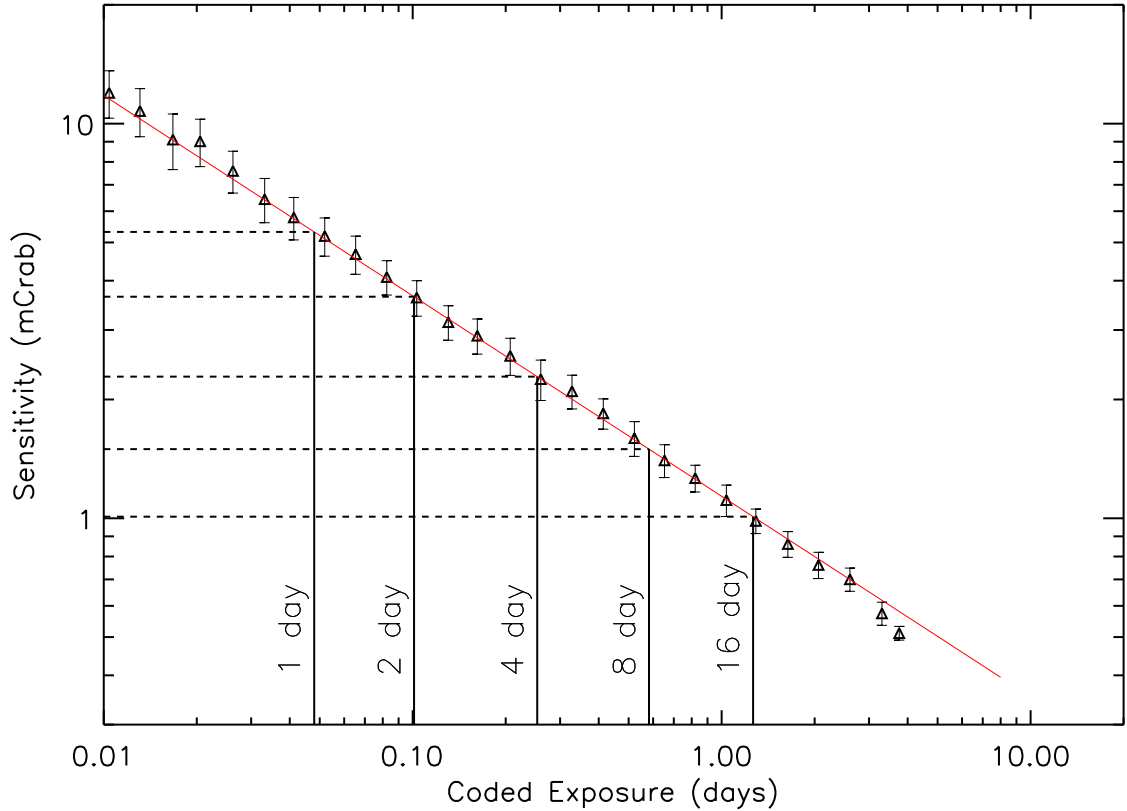}  
 \caption{Sensitivity ($1\sigma$) in mCrab units is plotted versus coded exposure for the daily mosaics. For BAT images, the coded exposure is the product of the actual temporal exposure and the partial coding fraction.  Therefore, even though the mosaics are built by accumulating all images over a given 1, 2, 4, 8 or 16-day period, the actual coded exposure for any given point in the sky is much less than full time period of the accumulation.  The vertical lines indicate the median coded exposure for each accumulation period and the horizontal lines show the equivalent median sensitivity. The red line is a fit to the data.}\label{mcrab_expo_fig}
\end{figure}

\begin{figure}
 \plotone{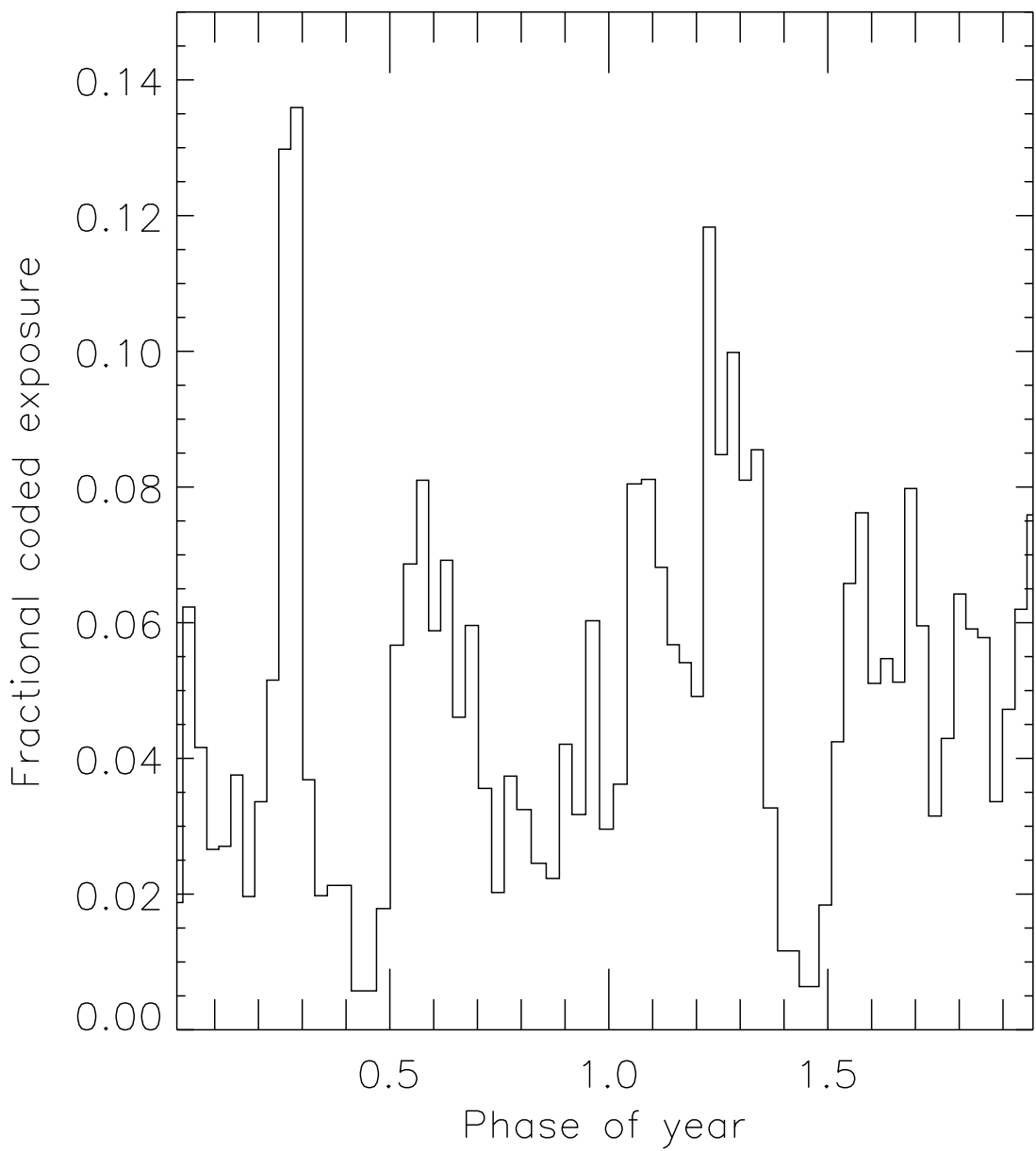}  
 \caption{A representative plot showing, for a source near the ecliptic plane, the daily average coded exposure divided by 86400, the number of seconds in a day.  The source represented is Seyfert 1.5 source 4U 0517+17, with data binned on 10-day intervals for calendar years 2009 and 2010.  The coded exposure for this particular source is very low near phase 0.4, when the source is located closest to the Sun, although it does not drop to zero, since a source very near the Sun can still be in the BAT field of view.  The greatest exposure is around phase 0.25, but there is large variation due to the variable {\em Swift} observing program and spacecraft orientation considerations.  The mean fractional coded exposure is 0.05. }\label{codex}
\end{figure}

The sensitivity of the daily mosaics depends on the exposure at the position of the source.    The average sensitivity as a function of exposure is shown in Figure~\ref{mcrab_expo_fig}, which is derived by comparing the coded exposure mosaic maps to the variance mosaic maps (see Section~\ref{methodology-mosaics}).  In Figure~\ref{mcrab_expo_fig}, the horizontal axis represents coded exposure, which is the product of the partial coding fraction and the temporal exposure.  It is clear from the vertical lines on this figure, which represent the median coded exposure for each mosaic time, that the coded exposure is well below the total accumulation time.  This is understood by considering how the BAT exposure is accumulated.  First of all, since BAT only covers $\approx 15$\% of the sky (to 10\% coding) at any time, we expect a typical sky point to be exposed for only $24\ {\rm hr}\ \times 0.18 \approx 4$\ hr per day.  In addition, most of the BAT field is only partially coded, further reducing the median coded exposure to $\sim 1$\ hr per day, ($\approx 5$\%).  The coverage is more uniform on longer time scales, so for the 16-day mosaics, the median coded exposure is $\approx 8$\% (1.3 days).    The coded exposure is also quite variable throughout the year as seen in Figure~\ref{codex}. Here we see the low coded exposure when the source is near the Sun and other large variations related to the observing program and to Sun-angle considerations.

The vertical axis of Figure~\ref{mcrab_expo_fig} represents variance, or $1\sigma$\ sensitivity, in units of mCrab.  Although it is now known \citep{wils11} that the Crab is not strictly constant, it is still useful to use the average Crab rate in our band as a yardstick.  For the BAT 15-50 keV band, 1 mCrab is $0.00022\ {\rm ct\ cm^{-2}\ s^{-1}}$, which using a power-law spectral index $\Gamma = 2.15$\ \citep{tuel10} corresponds to $1.26 \times 10^{-11}\ {\rm erg\ cm^{-2}\ s^{-1}}$. The comparison of source flux to Crab flux is strictly true only for sources with the same spectral index, but the systematic error for sources with different indices is small.
The relationship between sensitivity and coded exposure is linear with a slope of -0.5, as expected.  The horizontal dashed lines in the figure indicate the approximate sensitivity of each time scale of mosaic.  
The mean variance for one-day mosaics is 5.3~mCrab, for two-day mosaics 3.6~mCrab, for four-day mosaics 2.3~mCrab, for eight-day mosaics 1.5~mCrab, and 16-day mosaics 1.0~mCrab.

\begin{figure}
 \plotone{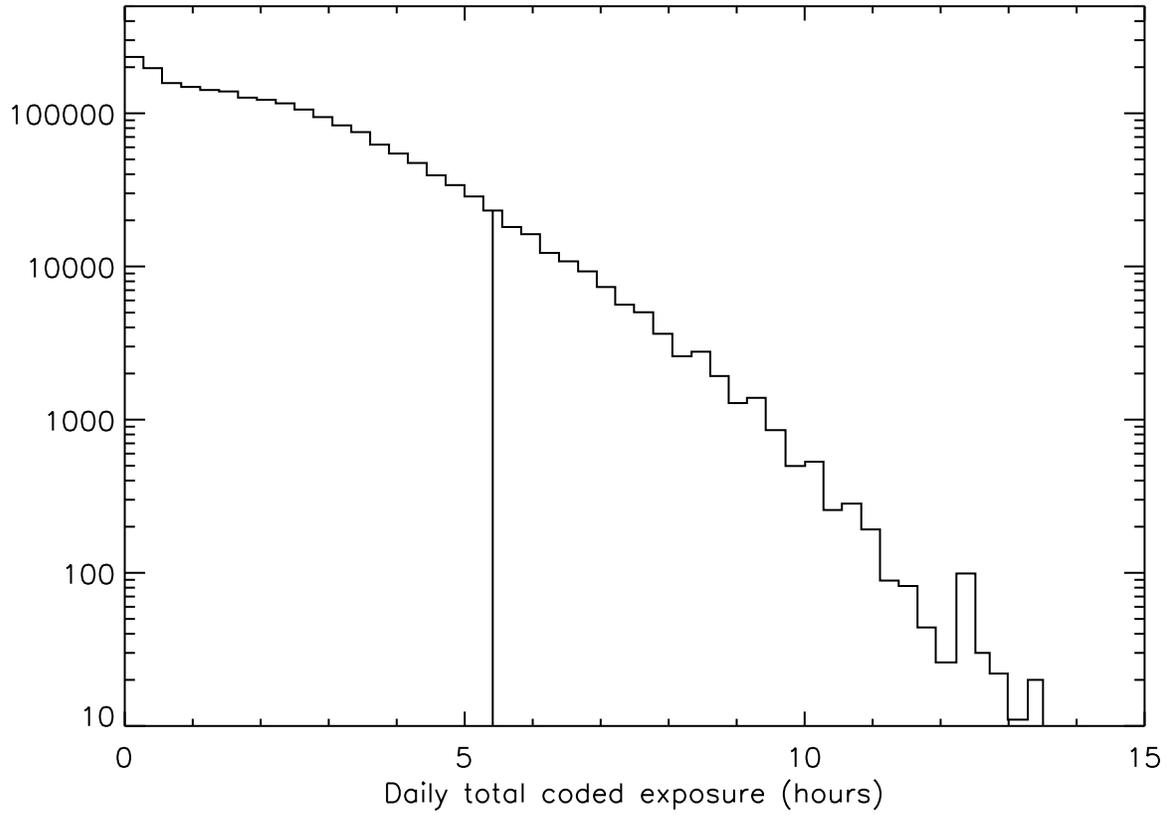}  
 \caption{A histogram of the daily total coded exposure for sources detected in the BAT transient monitor.  95\% of the exposures are less than 5.4 hours per day (indicated by the vertical line.)}\label{expo_fig}
\end{figure}

Exposure  for catalog sources is shown in Figure~\ref{expo_fig}, which shows a histogram of the daily total coded exposures for each daily observation of each source in the catalog.  While daily exposures can extend as long as $> 12$\ hr in rare cases, 95\% of coded exposures are less than 5.4~hr per day and 50\% are less than 1.7~hr.

\section{Results}\label{results}

\noindent Over the 6.5 years that the BAT transient monitor has operated, it has been a rich source of discovery of new Galactic and extragalactic sources and has provided an ongoing and archival resource of light curves for several hundred hard X-ray sources.  The light curves of known hard X-ray sources and sources expected to produce hard X rays in outburst are monitored automatically and daily rates and orbital rates are determined whether or not the source is actually detected, so that upper limits can be derived.  A description of the criteria for considering a source detected in the monitor and a summary of the sources detected is given in Section~\ref{results-known}.  As described in Section~\ref{methodology-mosaics}, the BAT transient monitor also allows for the discovery of previously unknown sources. This has also proven to be quite fruitful, with 17 sources discovered over the period from 2007 June through  2013 March.  Each of these new sources is discussed in detail in Section~\ref{results-new}.  

\subsection{Previously known sources}\label{results-known}

\noindent As of 2013 April 30, the input catalog to the BAT transient monitor (apart from blank sky points and provisional sources) contains 975 sources, covering most known hard X-ray sources, well-localized $\gamma$-ray sources, and a strong sampling of flare stars and active galaxies visible in the northern sky as monitored in the MOJAVE program \citep{list09}. 
Out of this list of sources, 245 have been detected in the transient monitor in the daily averages.   In order to systematically determine when a source is detected in the monitor, we examine two quantities for each catalog source based on the daily average count rates:  $M$, the mean count rate; and $P_7$, the peak count rate for days when the source was found at $\geq 7\sigma$\ significance.  The distributions of these quantities were studied and compared to samples of source light curves to determine detection criteria.  A source is considered detected if it meets either of the following criteria: $M \geq 3.0$\ mCrab (0.3\% of the mean rate of the Crab) or $P_7 \geq 30$\ mCrab. The numbers of sources meeting each of the criteria separately and collectively are shown in Table~\ref{tab-criteria}.  From this table we see 223 sources are found by these criteria.  A review of the Astronomer's Telegrams finds that there are 22 additional sources that are not in the list of 223, but which had significant outbursts during the transient monitor era.  These were found either by integrating monitor results over periods of longer than a day, by an onboard BAT trigger (usually for a short-duration event), or from an outburst report on the source from another instrument such as {\em RXTE}/PCA, {\em Fermi}/LAT or MAXI.  Seven of these sources are new {\em Swift}/BAT discoveries, which are discussed in Section~\ref{results-new}.  The 20 sources are indicated in Table~\ref{tab-catalog} by a footnote reference in the ``Class'' column.

All 245 detected sources are listed in Table~\ref{tab-catalog}.    The information listed for each source in the table is (1) the name, as listed in the BAT monitor catalog and web page (in most cases, but not all, this is the most common name in the literature), (2) J2000 equatorial coordinates, (3) source type (see the caption to Table~\ref{tab-class} for acronym definitions), (4) mean flux  $M$ in mCrab, (5) peak count rate $P_7$ in mCrab (sources with values of zero have no days when the source is detected at $> 7\sigma$), (6) scaled variability index $V$ (see below), (7)~normalized excess variance $F_{\rm var}$ (see below) and (8) error in $F_{\rm var}$. We classify each source by type based on classifications in SIMBAD\footnote{http://simbad.u-strasbg.fr/simbad/} and literature searches.  The total numbers in each broad classification are summarized in the first two columns of Table~\ref{tab-class}.

In order to study the variability of the detected sources, we calculate two parameters that quantify the variability.  The first is the scaled variability index based on a simple $\chi^2$\ criterion \citep[cf.][]{abdo09,abdo10}:

\begin{equation}
V = \left( \sum\frac{(F_i - F_{\rm avg})^2}{(\sigma_i^2 + \sigma_{i,\rm syst}^2)} \right)/ (N-1),\label{eq-1}
\end{equation}

\noindent where the $F_i$'s are the individual measurements of a source flux, $F_{\rm avg}$\ is the (weighted) average flux for the source, and  $\sigma_i$\ and $\sigma_{i,\rm syst}$\  are, respectively, the statistical and systematic errors on each flux measurement.  The sum is over all $N$\ observations meeting the criteria for inclusion in the published light curve.   Due to the presence of the systematic error in the denominator, the value of $V$\ has a floor of $V \gtrsim 0.75$ (seen most clearly in Figure~\ref{fig-meanvar}).  When divided by the number of degrees of freedom, $N - 1$, the variability index is a reasonably good measure of intrinsic variability for persistent sources.  However, to fully classify both persistent sources and those with outbursts, we need to include a second  measure, called the normalized excess variance, which is the variance with statistical and systematic fluctuations subtracted out.  This calculated as defined in \cite{abdo09} and \cite{vaug03}, with an error evaluated as in \cite{vaug03}:

\begin{equation}
F_{\rm var} = \sqrt{\frac{\sum{(F_i - F_{\rm avg})^2}}{(N-1)F_{\rm avg}^2} - \frac{(\sigma_i^2 + \sigma_{i,\rm syst}^2)}{N F_{\rm avg}^2}} \label{eq-2}
\end{equation}

\begin{figure}
 \plotone{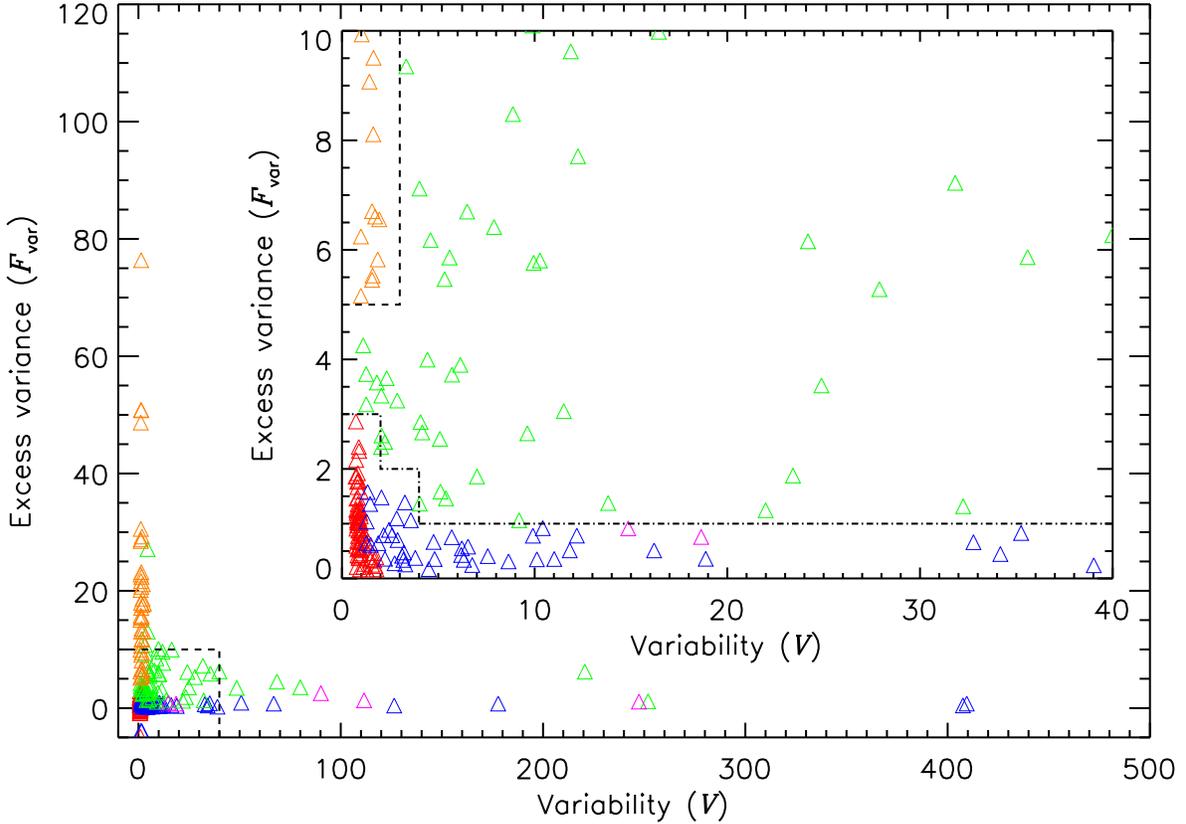}  
 \caption{Excess variance plotted with respect to variability (see main text, Section~\ref{results-known}, for definitions) for sources detected in the BAT transient monitor.  The colors indicate source variability identification based on this plot.  Orange points represent flaring sources ($\lesssim 1$\ day outbursts), green points outburst sources ($> 1$\ day outbursts), blue persistent variable sources, red steady sources and magneta periodic sources.  In the main plot, the dashed lines delineate the extent of the inset.  In the inset, the dashed lines indicate the divisions between flaring and outburst sources, while the dot-dashed lines divide the outburst sources from the persistent sources, as discussed in the text.  The division between steady and variable sources also depends on the source mean flux (Figure~\ref{fig-meanvar}).}\label{fig-var}
\end{figure}

By plotting $F_{\rm var}$\ vs. $V$, we can break the BAT sources down by variability and persistence as shown in Figure~\ref{fig-var}.   
The two panels of Figure~\ref{fig-var} show where in the $F_{\rm var} - V$\ space detected sources fall.  In the full plot we see an ``L''-shaped distribution with most of the sources (190/242) clustered in the region shown in the inset.  The wings of the ``L'' show that there is a bias in these parameters with source strength.  The 16 sources with large variability ($V > 40$) along the bottom of the main plot (to the right of the dashed line; note also that the vertical axis of the plot is extended below zero for clarity) are mostly very bright sources ($M > 55$\ mCrab) including the highly variable source Cygnus X-1, whose plot points ($V = 1369, F_V = 0.46$) would lie to the right of the full figure.   Among the highly variable sources, the five marked in green are sources with very large outbursts; in fact 1A 0535+262 ($V = 221$) has the brightest peak of any source at 5300 mCrab in its 2009 outburst.   The other four are a bit less bright ($M < 40$\ mCrab) and consist of 
GX 339$-$4 ($V = 49$) and GX 304$-$1 ($V = 70$), both discussed below in Section~\ref{outburst}, GS 0834$-$430 ($V = 90$), which had a single moderately large ($\approx 270$\ mCrab) outburst in 2012, and the recently discovered source Swift J1745.1-2624 (Section~\ref{sw1745}) with $V = 252$.
The other $V > 60$ sources are persistent.   There are 36 sources along the left side of the plot with large excess variance ($F_{\rm var} > 10$).  All but one are weak sources ($M < 1.2$\ mCrab).  The exception is GRO J1655$-$40, which even though normally undetectable, produced so much flux in its 2005 outburst that its average remains $M > 4$\ mCrab.

In the zoomed-in inset of Figure~\ref{fig-var} there is less correlation with mean flux and we can use this figure to separate sources into 
four source classes based on their variability characteristics.   We do this by comparing visual inspection of individual source  light curves with their position in  $F_{\rm var} - V$\ space.    The four main morphological categories we use are (1) Steady: persistent sources with low variability, (2) Variable: persistent sources with high variability, (3) Outburst: transient sources with a low quiescent level punctuated by episodes of high flux lasting for from several days to many months, and (4) Flaring:  transient sources with brief ($\lesssim 1$\ d) high flux episodes.  A few variable sources can also be sub-categorized as (5) Periodic, a  classification (see Section~\ref{periodic}) based on the relative intensity in the power spectrum of high frequency peaks compared to the average.  

Excess variance is sensitive to short episodes with an increase in count rate above a normally low background, and as such is a very good discriminator between persistent sources (steady, variable and periodic; red, blue and magenta plot points, respectively, in Figure~\ref{fig-var}) at low $F_{\rm var}$, below the dot-dash lines in the inset, and transient sources (flaring and outburst; orange and green plot points, respectively) at high $F_{\rm var}$, above the dot-dash lines.   
We then use variability to further distinguish among transient sources:  those with low $V$\ and very high $F_{\rm var}$, to the left of the dashed lines in Figure~\ref{fig-var} are classified as flaring and those to the right are classified as outburst.
Table~\ref{tab-vfvar} gives the specific divisions in terms of $V, F_{var}$\ and $M$\ between the four categories: steady, variable/periodic, outburst and flaring.  Variable and periodic sources are divided using a further criterion discussed in Section~\ref{periodic}.

\subsubsection{Steady sources}\label{steady}

\begin{figure}
 \plotone{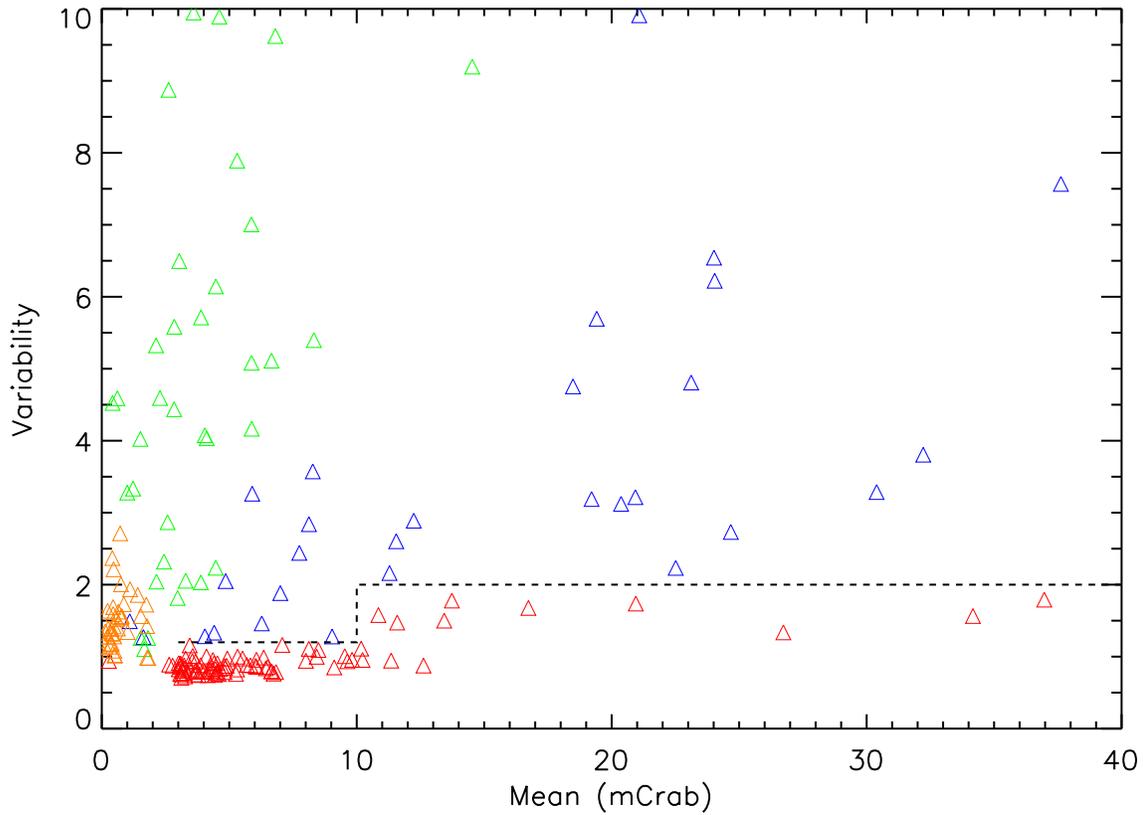}  
 \caption{Variability plotted with respect to mean count rate in mCrab for sources detected in the BAT transient monitor.  The colors indicate source variability identification based on Figure~\ref{fig-var}.  Orange points represent flaring sources ($\lesssim 1$\ day outbursts), green points outburst sources ($> 1$\ day outbursts), blue persistent variable sources, and red steady sources.  The dashed lines indicate the divisions between the {\em steady} and {\em variable} identifications.}\label{fig-meanvar}
\end{figure}

\noindent In the low part of the $F_{\rm var} - V$\ plane (below the dot-dash line in Figure~\ref{fig-var}) we find the persistent sources, which we divide into three categories, steady, variable and periodic.   After examining individual source light curves, we recognize that the division between steady and variable/periodic is dependent on the source brightness because it is more difficult for this method to identify variability in faint sources.   This is reflected in Figure~\ref{fig-meanvar}, where the dashed lines indicate the steady/variable dividing lines.  For bright sources ($M > 10$\ mCrab) we find that setting the threshold at  $V > 2$ is robust, with only  the blazar 3C~273 ($M = 13.5$\ mCrab; $V = 1.5$) showing variations in the light curve visible to the eye, but falling into the steady category. However, for $M < 10$\ mCrab, visual inspection shows four sources with $1.2 < V < 2$\ and significant variation in the light curves, while only two sources in this range of $V$\ do not show such variability. No sources with $V < 1.2$\ at any brightness level show variability in their light curves.    Therefore we set the dividing line between variable and steady sources so that a source with $M < 10$\ mCrab is considered variable for $V > 1.2$ and a source with $M >10$\ mCrab is variable for $V > 2$. 

The steady sources, by this classification, are found in the lower left of Figure~\ref{fig-var} (red points). Steady sources tend to be weaker than the variable sources (median $M = 4.8$\ mCrab) although (excepting the Crab nebula at $M = 1000$\ mCrab) they range as high as $M = 37.0$\ mCrab.    In Figure~\ref{fig-meanvar} we see plotted the relationship between variability and mean rate.  First, outburst sources and, to an even greater extent, flaring sources, have low mean rates.  This is because the mean rate for such sources is the average over long periods below the detection threshold and only short periods of detectability.   The variable and steady sources by contrast have, with only a few exceptions (see below), mean values above the 3 mCrab threshold.  Figure~\ref{fig-meanvar} shows that there is a band of steady sources mostly with $0.75 < V < 2$\ covering a broad range of mean count rates.   These sources can be bright, but have a low intrinsic variability.   The most exceptional example is the Crab nebula, which, despite its recently discovered hard X-ray variability \citep{wils11}, has slow enough variation to be classed as a steady source ($V = 1.46$). For other examples we look at the lower right corner of Figure~\ref{fig-meanvar}.  The two steady sources are the LMXB/NS systems 4U 1822$-$371, with $M = 34.1$\ mCrab and $V = 1.57$\ and GX 9+1 with $M = 37.0$\ mCrab and $V = 1.579$,  while nearby on the plot, HMXB/NS X~Per and LMXB/NS 4U~1735$-$44 have respectively, $M = 30.4$\ mCrab and 32.2 mCrab and $V = 3.28$\ and 3.81 (variable).  Examination of the light curves shows that 4U 1822$-$371 and GX 9+1 have nearly flat light curves, while X~Per shows a broad hump in 2009-2010 and 4U~1735$-$44 has a long term dipping and rising trend in brightness.  So the variability parameter is a good measure for these bright sources.  

 It is still of course quite possible that some of the relatively weak sources are variable, but that the statistical noise in their light curves is large enough to bury a variability signal on the time scale of a day in the BAT monitor.   Indeed many of these sources show variability in other energy bands (e.g. the {\em RXTE} ASM) or on longer time scales \citep{tuel10}.  However, for consistency we restrict the variability study to the one-day time scale and realize that some weak variable sources are included in the steady class.  

Among the weak sources whose variability is not found by our metric is the famous blazar 3C~454.3, which had flares detected in the BAT in 2005 \citep{giom06}, 2009 \citep{pacc10} and 2010 and the HMXB/NS system IGR J16393$-$4643, for which phase analysis of BAT and {\em RXTE} PCA light curves reveal a likely orbital period of 4.24 days \citep{atel2570}, improving the results of \citet{thom06}.  Although this period can be extracted from the BAT monitor data, it does not reveal itself in the $V$\ value of this weak ($M = 6.0$\ mCrab) source, which has $V = 0.87$.  
The three steady sources with $M < 3$\ mCrab are IGR J17062$-$6143 ($M = 2.8$\ mCrab), which was added to the list since \citet{atel3785} clearly shows that it is detected in the BAT monitor, IGR J17062$-$6143 ($M = 2.6$\ mCrab), a weak source that triggered BAT onboard \citep{atel4219}, 
and Swift J1112.2$-$8238 ($M = 0.3$\ mCrab), a known transient (Section~\ref{sw1112}) but with slow and weak enough variation to fall into the steady classification.

The remainder of the steady sources include 39 of the 42 AGNs detected in the BAT monitor, all of them weak ($M < 13$\ mCrab). There are 22 LMXB/NS systems including the moderately bright sources GX 9$+$1 ($M = 37.0$\ mCrab), an atoll source \citep[e.g.][]{iari05}; the accretion disk corona system 4U 1822$-$371 \citep[$M = 34.1$\ mCrab;][]{jonk01}; and the Galactic bulge source GX 9$+$9 \citep[$M = 20.9$\ mCrab;][]{hert88,harr09}.  Among HMXBs we have the candidate SFXT IGR J16418$-$4532 \citep{sgue06} with a 3.75-day orbital period discovered in {\em Swift}/BAT and {\em RXTE}/PCA light curves \citep{atel779}, which had a flare in the BAT in 2011 \citep{roma12a}.
None of the black hole or black hole candidates detected in the BAT monitor are classed as steady sources

\subsubsection{Variable sources}\label{variable}

\noindent The variable source class contains 55 sources.   These are sources that are normally detected in the BAT (median $M = 24.0$\ mCrab), but show variation 
without the high levels of excess variance seen in outburst sources. The highest variability found is for the extremely bright and variable Cygnus~X-1, with a calculated value of $V = 1357$, a factor of more than three larger than the next most variable source, Vela~X-1 at $V = 410$.  The other sources with $V > 100$\ are also among the brightest BAT sources, GRS~1915$+$105, 4U~1700$-$377, and Cygnus~X-3.   Sco~X-1, with the highest mean flux of any BAT source, $M = 1225$\ mCrab, has $V = 39$.  Some of the variable sources actually have very long (multi-year) outbursts.  These include the {\em Swift} discovered transient, Swift ~J1753.5$-$0127 \citep[e.g.][]{mill06}, a BHC discovered in 2005 \citep{atel546} that has been in outburst ever since, and the accretion-powered X-ray pulsar 4U~1626$-$67, which underwent a torque reversal and significant (and so-far sustained) increase in flux in early 2008 \citep{atel1426,came10}.

Also accepted to the variable class are the Seyfert 2 galaxy NGC~4388, which is known to have variations in hard X rays on the 3-6 month time scale \citep{fedo11}; the HMXB/$\mu$Quasar SS~433, which is observed in BAT to follow the 164-day superorbital period \citep{ogil01}; and the LMXB/NS and thermonuclear burster KS 1741$-$293 (AX J1744.8$-$2921), which has triggered BAT onboard \citep{atel3632} and which shows long-term variability in the monitor light curve. 

\subsubsection{Periodic sources}\label{periodic}

\noindent The periodic sources are a subset of the variable sources, which fall in the same part of the $V - F_{\rm var}$\ plane).  Periodic sources are found by performing a systematic search of the power density spectra (PDS) of source light curves.  A source with periodicity in its light curve will show a peak in the PDS at the frequency of this periodicity.  The stronger the periodic signal, the sharper the peak in the PDS.  In this work we do not attempt to measure the period frequency for any source. We simply compare  the ratio $R$\ of the PDS maximum to the mean as a crude but effective way to separate strongly periodic sources in the BAT monitor from other variable sources.
It was found that setting the threshold to $R > 14$\ identifies five known periodic sources, while excluding sources without reported periodicities.  The five periodic sources are (1) Hercules X-1 with a well-known 35-day eclipse period \citep{bahc72}, (2) GX 301$-$2 with a 40.8-day orbital period \citep{whit78}, (3) LMC X-4 for which BAT clearly sees the 30.48-day period attributed to the precession of the accretion disk \citep{lang81,heem89}, (4) SMC X-1 where we detect the $\approx 55$-day superorbital period \citep{trow07}, and (5) EXO 2030+375 with a 45.9-day orbital period \citep{parm89}. Setting the threshold lower would start to include such quasi-periodic sources as GX 354$-$0, EXO~1657$-$419 and GX 1+4, which began a series of regular outbursts in late 2008 and has a 303.8-day orbital period \citep{pere99}, but which does not show periodic modulation on any scale in the BAT \citep{corb08}. Looking at other sources with confirmed superorbital periods \citep{sood07} with ratios near threshold, we find 4U~1636$-$536, with a $\approx 46$-day superorbital period found in the BAT \citep{farr09} has $R = 10.0$;  GRS~1915$+$105 with a 33.5-day orbital and 590-day superorbital period \citep{rau03} has $R = 8.8$;  GX 339$-$4, an outburst source that shows variation that is not commensurate with the $\approx 240$-day period \citep{ogil01}; and Vela X$-$1, with a 93.3-day superorbital period \citep{khru83} has $R = 9.1$ and a very complicated light curve, which does not show this period.  Detailed phase analysis to definitively confirm detection of these periods in the BAT data are beyond the scope of this work.

\subsubsection{Outburst sources}\label{outburst}

\begin{figure}
 \plotone{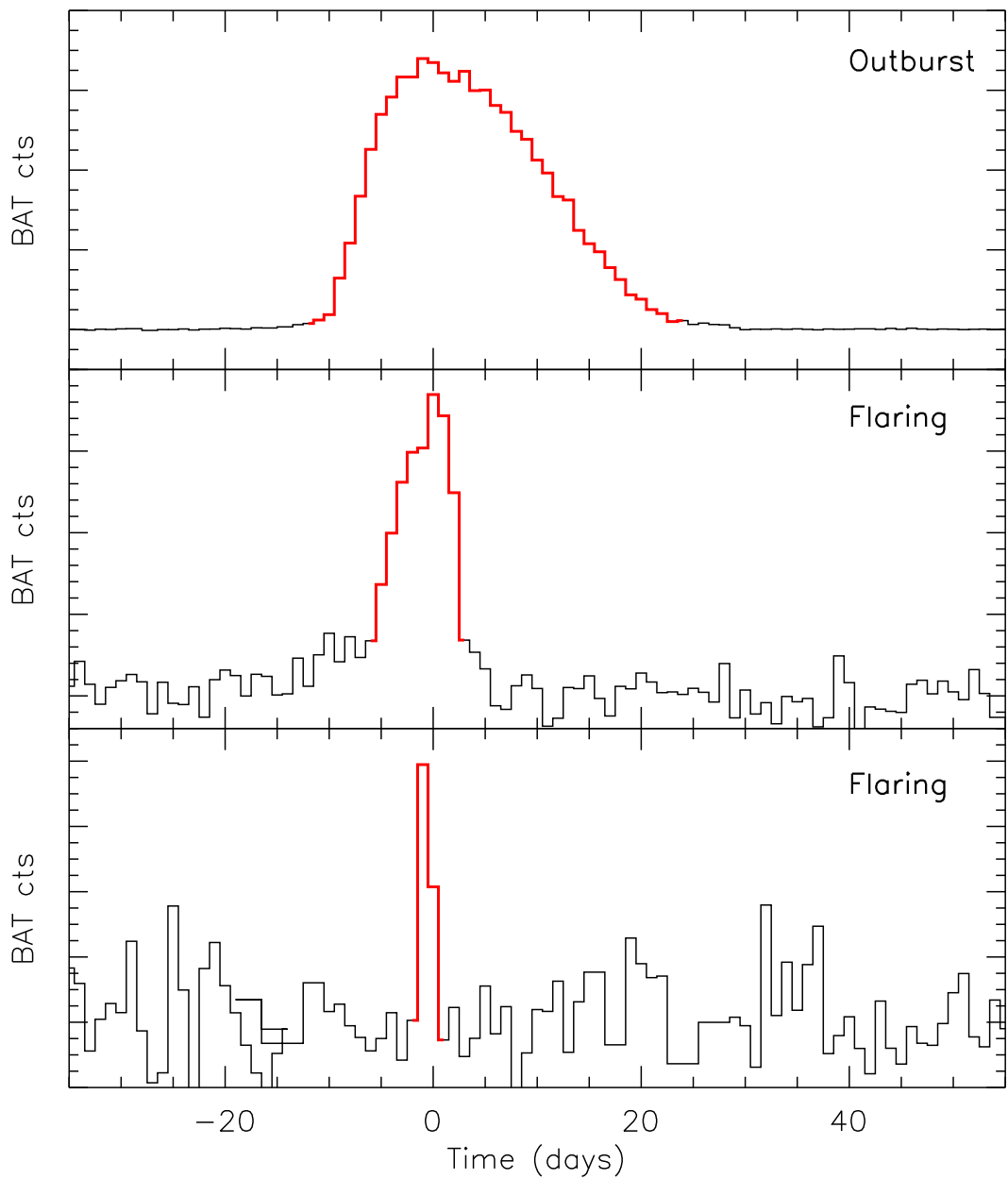}  
 \caption{Schematic light curves for one outburst source (top) and two flaring sources (middle and bottom).  The red areas indicate the duration of the outburst or flare.  This outburst lasts for $\approx$ 40 days (although much longer times are possible), while a typical flare has a duration of $\lesssim 10$\ days (middle) or $\approx 2$\ days (bottom). The light curves, which have been rescaled in the vertical axis and shifted in time, are based on events from the light curves of 1A~0535+262 (2011 outburst; top), XTE J1856+053 (2007 flare; middle), and IGR J17391$-$3021 (2006 flare; bottom)} \label{fig-schematic}
\end{figure}

\noindent Outburst class sources (green points in Figure~\ref{fig-var}) typically have larger variability than flare sources and larger excess variance than variable sources.  In terms of their light curves, these are sources that are not detectable most of the time, but that show significant episodes of high flux lasting typically 10 days or longer.   A typical outburst is shown in the top panel of Figure~\ref{fig-schematic}. This class includes 13 black holes (BH) or black hole candidates of which nine have had a single outburst during the {\em Swift} era. These include the BAT discoveries Swift J1842.5$-$1124 (Section~\ref{sw1842}), Swift J1910.2$-$0546 (Section~\ref{sw1842}),  and Swift J1745.1$-$2624 (Section~\ref{sw1745}), XTE J1752$-$223 \citep{shap10}, where only BAT was able to obtain observations throughout the peak of the 2009-2010 outburst, and two sources for which the peaks in the BAT emission preceded that of the softer X rays by $\sim 10$\ days: GRO J1655$-$40 \citep{broc06} and MAXI J1659$-$152 \citep{kenn11}.
The other four BH sources have had multiple outbursts:  GX 339$-$4 with major outbursts in 2006-07  \citep[e.g.][]{dels08} and 2010 \citep{debn10} and many smaller ones, IGR J17091$-$3624 with outbursts in 2007 \citep{capi09} and 2011 \citep{rodr11},  IGR J17464$-$3213 (H1743$-$322) with roughly yearly outbursts \citep[e.g.][]{prat09,blum10,mott10,mill12}, and 4U 1630$-$472, which had a fairly weak outburst in 2009-2010 \citep{atel2363} and a series of much stronger outbursts in 2011-2012 \citep[e.g.][]{atel3830,atel4077}.

Most of the remaining outburst sources are neutron star binaries, with 14 in HMXBs and 16 in low-mass X-ray binaries (LMXB)s.    The most significant BAT detections (cases in which a BAT outburst detection and announcement led to an observing campaign) among neutron stars are as follows.  1A~0535$+$262 had giant outbursts in 2005 \citep{atel504,coe06} and 2009 \citep{reyn10}, a somewhat less intense outburst in 2011 \citep{atel3166}, and multiple smaller outbursts.  The peak of the 2009 outburst at 5345~mCrab makes 1A~0535$+$262 the brightest object ever seen in the BAT monitor and even the minor outbursts can be brighter than 500 mCrab.  Aquila X-1 has had at least seven BAT-detected outbursts of $> 50$\ mCrab \citep[e.g.][]{atel1557,mill10}. Two Galactic HMXBs became particularly active in the {\em Swift} era.  GX 304$-$1 began a series of outbursts in 2010 April \citep{atel2538,deva11} spaced by the 132.5-day orbital period of the source \citep{prie83}. GRO J1008-57 had a large outburst in 2007 November \citep{atel1298,naik11} and continued to be detected in the BAT with minor outbursts at its 247.8-day orbital period \citep{coe07}.  This activity culminated in an unusually long and bright set of outbursts in 2012 \citep[e.g.][]{atel4319,atel4573}.   There was also a large outburst of the Be X-ray binary system GRO~J1750$-$27 (AX~J1749.1$-$2639) starting in 2008 February \citep{atel1376,shaw09}.

Three SFXTs fall into the outburst class by virtue of fairly low ($F_{\rm var} < $\ 5) excess variances, IGR J16479$-$4514 \citep{roma11a},  IGR J18483$-$0311 \citep{roma10}, and IGR J18450$-$0435  \citep[a.k.a. AX J1845.0$-$0433;][]{sgue07,atel2102,roma13}.  
The remaining outburst sources include the cataclysmic variable star GK Per, which had outbursts in 2006-07 \citep{evan09b} and 2010 \citep{atel2466}, the blazar Markarian 421, which in its low state is not detectable in the BAT on a daily basis, but which has had at least five major outbursts during the {\em Swift} era \citep[see][]{hora09,abdo11}, and the transient Swift J1922.7$-$1716 \citep{fala06} discovered in the BAT hard X-ray survey \citep{tuel10}, which had an outburst in 2011 August \citep{atel3567} and has recently been identified as a  neutron star low-mass X-ray binary by \citet{dege12}.

\subsubsection{Flaring sources}\label{flaring}

\noindent Flaring sources are characterized by low values of variability coupled with large values of excess variance and populate the upper left part of the  $F_{\rm var} - V$\ plot, within the dashed lines  (orange points in Figure~\ref{fig-var}). 
We note that the flare class will include both sources with intrinsically short high emission episodes such as supergiant fast X-ray transients (SFXTs) and soft gamma repeaters (SGRs) as well as faint sources that may have intrinsically longer episodes of increased emission, but for which only the peak of the outburst is detectable in the BAT.   Two example light curves for a fairly long and a short flare are shown in the middle and bottom panels of Figure~\ref{fig-schematic}, respectively. Most flare sources are neutron star (NS) systems (see Table~\ref{tab-class} for breakdown) including five high-mass X-ray binary (HMXB)/SFXTs, all of which triggered BAT on board (many several times) and for which the transient monitor light curves serve as important constraints on quiescent behavior.   
These SFXTs are IGR J17391$-$3021 \citep[a.k.a. XTE J1739$-$302;][]{sido09a,roma11a,fari12}, 
IGR J17544$-$2619 \citep{sido09a,roma11a,fari12}, 
IGR J08408$-$4503 \citep{roma09,sido09b}, 
IGR J18410$-$0535 \citep[a.k.a. AX J1841.0$-$0536;][]{roma09,roma11b} and
IGR J16328$-$4726 \citep{atel2588,roma13}.
Another HMXB/NS in the Large Magellanic Cloud, XMMU J054134.7$-$682550, had its first recorded outburst in 2007 August \citep{atel1169,inam09}.

There are seven black hole candidate (BHC) systems in this category. One is XTE J1818$-$245, which was discovered in 2005 August 
with an outburst that persisted for $\approx 50$\ days at low X-ray energies \citep{cado09}, but only $\approx 10$\ days at BAT and {\em INTEGRAL}/ISGRI energies.    The second BHC is XTE J1856$+$053, which was seen by {\em Swift}/BAT during both parts of  its 2007 outburst \citep{atel1093}, but BAT only detected the tops of the peaks observed in the {\em RXTE}/ASM \citep{sala08}.   The third BH is SAX J1819.3$-$2525 (V4641 Sgr) whose BAT flare in 2005 June was found in the archival light curve but was not reported at the time.  The other four are all sources with a single moderately long but weak ($M \lesssim 40$\ mCrab) outburst.  These are Swift J1539.2$-$6227 \citep[Section~\ref{sw1539};][]{krim11} and  Swift J1347.2.2$-$0933 \citep[Section~\ref{sw1357};][]{arma13},  BAT monitor discoveries with 2008-09 and 2011 outbursts, respectively,   XTE J1652$-$435 \citep{han11}, which had a $\sim 40$\ day outburst in 2009 and $F_{\rm var} = 6.6$, close to threshold for the flaring identification, and  MAXI J1543$-$564 \citep{stie12}, which underwent a weak $\sim 30$-day outburst in 2011.

Thirteen {\em Swift}/BAT discoveries are in the flaring class.  These include the transient X-ray pulsar 
Swift J1626.6$-$5156 \citep{reig08}, discovered in a flaring state in 2005 December \citep{atel678} before the start of the monitor.  The other {\em Swift}/BAT discovered flaring sources are the millisecond pulsar Swift J1756.9$-$2508 (Section~\ref{sw1756}), the pulsars Swift J1816.7$-$1613 (Section~\ref{sw1816}), Swift J0513.4$-$6547 (Section~\ref{sw0513}), Swift J1729.9$-$3437 (Section~\ref{sw1729}) and Swift J1843.5$-$0343 (Section~\ref{sw1843}), the black hole candidates Swift J1539.2$-$6227 and Swift J1357.2$-$0933 (Section~\ref{sw1357}), and the unidentified transients Swift J1713.4$-$4219 (Section~\ref{sw1713}), Swift J1836.6$+$0341 (Section~\ref{sw1836}), and Swift J1943.4$+$0228 (Section~\ref{sw1943}).   Also in the class are two tidal disruption flare events, Swift J164449.3$+$573451 \citep[Swift J1644;][]{burro11,levan11,zaud11} and Swift J2058.4$+$0516 \citep{cenk11}. Among the {\em Swift} flaring sources all have short and/or faint outbursts save for Swift J1626.6$-$5156 and Swift J1539.2$-$6227, which have $> 10$\ day outbursts, indicating that the classification method does not perfectly discriminate.

\subsection{New discoveries}\label{results-new}

\noindent Since short-term transients (e.g. gamma-ray bursts, soft gamma repeaters, supergiant fast X-ray transients) will trigger {\em Swift} onboard, discoveries in the transient monitor are longer term transients, particularly Galactic binaries, with typical outburst durations of weeks to months.  Since  2007 June, 17 previously unknown sources
have been discovered in the BAT transient monitor and confirmed by XRT or {\em RXTE}/PCA observations.  Most of these sources have also had further observations with other space- and ground-based telescopes.  Here we provide a summary of these discoveries and their interpretations.   The sources are summarized in Table~\ref{tab-loc} and discussed in detail below,  presented in the text in order of their discovery.

New discoveries are found through the daily mosaics (Section~\ref{methodology-mosaics}).  Automated software scans the mosaics using {\ttfamily batcelldetect} on each of five time scales (1-day, 2-day, 4-day, 8-day and 16-day) searching for excess flux at any location not corresponding to a BAT monitor catalog source.  Any excess at the $5\sigma$\ level or above on any time scaled is flagged as a possible transient and reported to the BAT monitor team via email, and a light curve with one-day cadence is automatically generated for the position of the possible new source, going back 30 days before the day of discovery.   A provisional {\em Swift} name is assigned and the source is added to monitor catalog as a provisional source.  The nominal criterion for announcement to the astronomical community is that the source is seen at $\geq 6\sigma$\ for 2 or more days in the 1-day mosaics or at $\geq 6\sigma$\ in a multi-day mosaic.   If the new detection is $\geq 8\sigma$, the coordinates are automatically distributed via the Gamma-ray Coordinates Network (GCN)\footnote{http://gcn.gsfc.nasa.gov.}.   In other cases, the BAT monitor team makes the decision whether or not to announce the new source, based on careful examination of the light curve and comparison of the source location with existing astronomical catalogs, images and databases.  In most cases, a first announcement is made as an Astronomer's Telegram\footnote{http://www.astronomerstelegram.org.} and a request is submitted for a {\em Swift} target of opportunity observation to confirm the source in the XRT, derive an improved position, and collect an initial spectrum.  If the BAT data alone are inconclusive, then an announcement is made only after confirmation with the XRT.

Each of the seventeen BAT monitor discoveries is discussed here in detail.  Sixteen of the sources (all save for Swift J1713.4$-$4219) were observed by the {\em Swift} XRT and UVOT.  All 16 of these were detected in the XRT and eight were detected in the UVOT.  All of the sources discovered prior to 2011 June were detected in the {\em RXTE}/PCA. 
The light curve and spectral fitting of the XRT data for all sources reported here were carried out using data and analysis based on \citet{evans09}. The enhanced positions for Swift J1729.9$-$3437 and Swift J1843.5$-$0343 used the method of \citet{goad07}.  For the four pulsars analyzed herein (Swift J1816.7$-$1613, Swift J0513.4$-$6547, Swift J1729.9$-$3437 and Swift J1843.5$-$0343) analysis is of data from the {\em RXTE}/PCA.  For the light curves of these sources all of the layers and operational proportional counter units (PCUs) are used to maximize the SNR.  The pulse periods are found by taking a weighted average of the 
measurements from each observation, and the pulse period of each 
observation is generally found using the harmonics of the signal if they 
are present. All of the PCA light curves, and hence pulse period detections, 
are barycenter-corrected.

\subsubsection{Swift J1756.9$-$2508}\label{sw1756}   
\noindent{\bf LMXB/NS; Discovered in 2007 June.} The first new discovery with the transient monitor, Swift J1756.9$-$2508 is an ultracompact binary containing an accretion-powered millisecond pulsar \citep{krim07,lina08} with what is most likely a He-dominated degenerate companion.   The $\approx 13$-day outburst was followed by {\em Swift} and {\em RXTE}.  The spin frequency is 182~Hz (5.5~ms) and the orbital period is 54.7~minutes.  Using models of white dwarf (WD) - neutron star ultracompact binaries, it was concluded that the donor star in Swift J1756.9$-$2508 has a mass between 0.0067 and 0.030 $M_{\odot}$ and that its thermal cooling has been slowed by irradiating flux generated by the accretion.   This source is described in detail in~\citet{krim07}.

In 2009 July, Swift J1756.9$-$2508 underwent a second outburst of roughly the same duration and peak rate ($\approx 50$\ mCrab) as the 2007 outburst.  \citet{patr10} studied this outburst in detail and reanalyzed the 2007 outburst using data from {\em Swift} and {\em RXTE}.  From this work they were able to set a constraint on the neutron star magnetic field of $0.4 \times 10^8 \lesssim B \lesssim 9 \times 10^8$, which the authors state is within the range of typical accreting millisecond pulsars.  They also constrain the spin frequency derivative to $|\dot{\nu}| \lesssim 3 \times 10^{-13}$\ Hz s$^{-1}$\ and derive an improved estimate of the mass accretion rate.   Although the close temporal proximity of the two outbursts of Swift J1756.9$-$2508 suggests a 2.1~y recurrence cycle, there were no comparable outbursts either in 2005 May or in 2011 August, suggesting that the accretion behavior of the source is more complex than originally thought.  

\subsubsection{Swift J1816.7$-$1613} \label{sw1816}  
\noindent{\bf XRB/NS; Discovered in 2008 March.} Although Swift J1816.7$-$1613 was first reported by the BAT monitor team \citep{atel1456}, it was found in archival data to have been detected with {\em Chandra} in 2007 \citep{atel1457}, {\em XMM} in 2003  \citep{atel1457}, where it was identified as  2XMM J181642.7$-$161320, and {\em BeppoSAX} in 1998 \citep{atel1462}.    Analysis of the {\em Chandra} results \citep{atel1457} show that Swift J1816.7$-$1613 is a pulsar with a barycentered period of $142.9 \pm 0.2$\ s.   Analysis of the two observations with the {\em RXTE} PCA on 2008 March 29 and April 7 (MJD 54554 and 54563) yields a weighted average pulse period of $143.2 \pm 0.1$\ s, consistent with the {\em Chandra} value.   Using this pulse period, we derive a 95\% confidence level upper bound for the accretion-driven luminosity \citep{joss84} of $L_x \leq 5.4\ \times 10^{38} \ {\rm erg\ s^{-1}}$.
There is weak evidence for a ``spin-up'' trend with these two observations of $\dot{P} = -5.93\ \times 10^{-7}\ {\rm s \cdot s^{-1}}\ (\chi^2 = 3.841/ 1$\ degrees of freedom (d.o.f.)).  The {\em RXTE} analysis used Standard2f data with 16-s timing resolution.

\begin{figure}
 \plotone{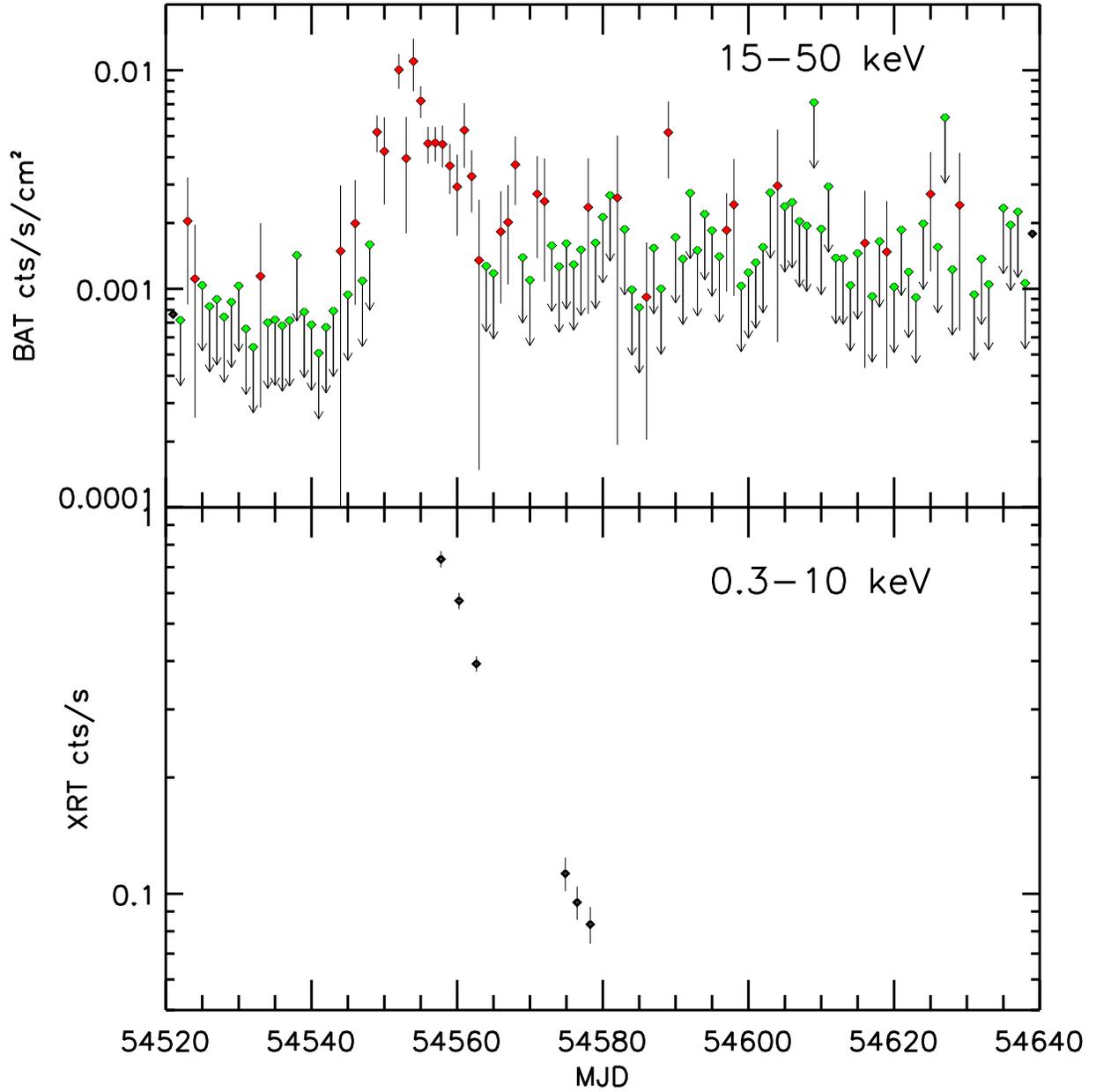} 
 \caption{Light curves for  Swift J1816.7$-$1613 from BAT (upper panel), XRT (bottom panel). In this, and in all subsequent X-ray light curve plots, red points are detections and green points with downward arrows are $1\sigma$\ upper limits. For the BAT, $0.01\ {\rm ct\ cm^{-2}\ s^{-1}} \approx 45$\ mCrab. }\label{sw1816_fig}
\end{figure}

The peak in the BAT (on March 29) was $0.008 \pm 0.002\ {\rm ct\ cm^{-2}\ s^{-1}}$ (35 mcrab; 15-50 keV).  A series of six {\em Swift} pointed observations were taken between 2008 April~1 and April~22.  Results for BAT and XRT are shown in Figure~\ref{sw1816_fig}.
The nature of the companion remains unknown.   UVOT observations on 2008 April 1 in multiple bands show no counterpart to the following $3\sigma$\ magnitude limits:  uvw2  $> 21.4$, u     $> 21.2$, uvm2  $> 21.2$, uvm2  $> 20.9$, u  $> 20.9$, uvm2  $> 21.3$.  An archival search reveals no optical or near IR counterpart at the location of the {\em Chandra} source.  This source has also had at least two other, less-intense, outbursts seen in the BAT, one from approximately 2009 July 21 (MJD 55033)  to 2009 August 10 (MJD 55053)  and the other from approximately 2011 June 18  (MJD 55730) to 2011 July  8 (MJD 455750). Both of these outbursts  peaked at $\approx 0.007\ {\rm ct\ cm^{-2}\ s^{-1}}$.  The second of these outbursts was also detected in the {\em RXTE} All-Sky Monitor.

\subsubsection{Swift J1842.5$-$1124}  \label{sw1842}
\begin{figure}
 \plotone{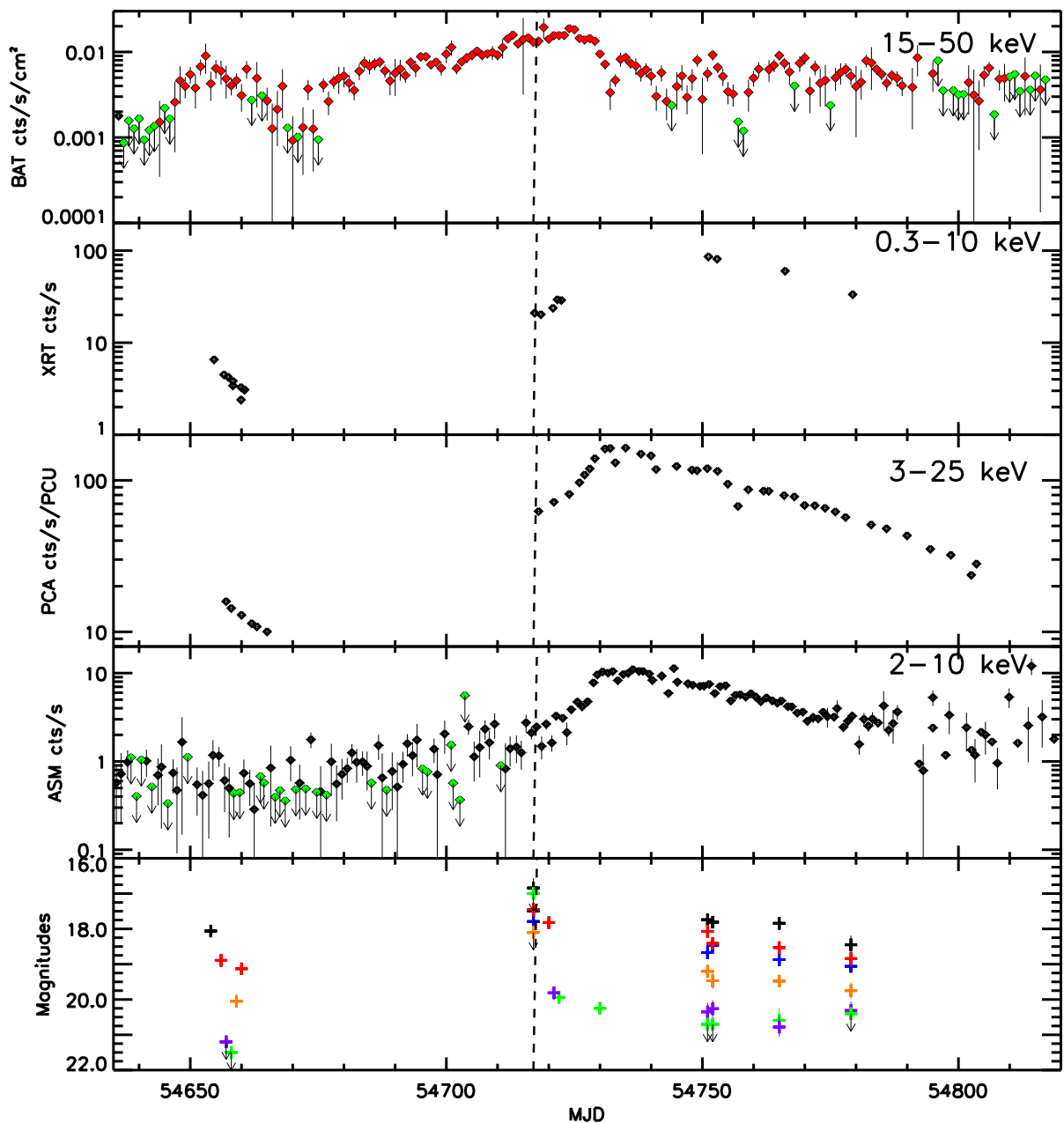}  
 \caption{Light curves for  Swift J1842.5$-$1124.  Light curves from top to bottom are for BAT, XRT, PCA, ASM and 
UVOT.  In the bottom plot crosses are magnitudes or $3\sigma$\ limits (downward pointing arrows) from Swift/UVOT filters: black = v, blue = b, red = u, orange = uvw1, green = uvm2, purple = uvw2.  The dashed line indicates the day (2008 September 8) of the three BAT triggers on this source, which is close to the peak of the BAT light curve.}\label{sw1842_fig}
\end{figure}

\noindent{\bf XRB/BHC; Discovered in 2008 July.} This source is a possible black hole with three on-board BAT triggers.   The source was first detected with the transient monitor on 2008 July 2 (MJD 54649) and reported by  \citet{atel1610}.  A series of {\em Swift} and {\em RXTE} observations were carried out (see Figure~\ref{sw1842_fig} for the light curve) and continued until the source became undetectable in the BAT around MJD 54662.  However, after a few days Swift J1842.5$-$1124 began to brighten again and on 2008 September 8 triggered the BAT on board three times \citep{gcn8199,gcn8200,gcn8201,atel1706}.  There were immediate automated observations with {\em Swift} and a renewed set of {\em RXTE} observations nearly to the end of the outburst.  Analysis of the {\em Swift} and {\em RXTE} data from 2008 September 9 was performed by \citet{atel1716} who suggested that the combined X-ray spectral and timing behaviors of the source are characteristic of a black hole in the hard spectral state, e.g. a combined black body and power law model (black body kT = 0.9 keV; photon index = 1.5), where the black body component contributes about 6\% of the total 2-40 keV flux.  At this time there is a strong quasi-periodic oscillation (QPO) present near a frequency of 0.8 Hz.   Preliminary investigations (T. Belloni, private communication) showed a weak QPO at 8 Hz on 2009 October 15 and a hardness-intensity diagram suggestive of a source transitioning through a hard spectral state toward a softer thermal state.  Note also that the peak of the hard X-ray (BAT) light curve (a few days after the time of the on-board triggers: dashed line in Figure~\ref{sw1842_fig}) precedes the peak of the softer X-ray (PCA and ASM) light curves by $\approx 10$\ days.  This type of lag has been seen in black hole sources:
e.g. Swift J1539.2$-$6227 \citep{krim11} and GRO J1655$-40$ \citet{broc06}.  
Therefore we tentatively suggest that Swift J1842.5$-$1124 is a candidate black-hole binary that reached a hard spectral state around MJD 54717.  Complete spectral and timing analysis of Swift J1842.5$-$1124 will be presented in a later paper.

The source was also detected in the {\em Swift} UVOT with a peak magnitude of v = $16.84 \pm 0.28$.  The source was still detectable in UVOT as late as 2008 November 9 (MJD 54779).  \citet{atel1720} report the detection of a near-infrared counterpart with the PANIC camera mounted on the 6.5m Baade telescope at Las Campanas Observatory.   The position of the counterpart is consistent with the UVOT position (see Table~\ref{tab-loc}) with a magnitude of ${\rm K_s} = 14.90 \pm 0.05$.  \citet{atel1720} also report a marginal detection at the same location in 2005 in the UKDSS (${\rm K_s} \approx 17.4$), which confirms brightening of the K-band source and its identification as the optical counterpart of Swift J1842.5$-$1124.  Although there is no detection of the source in either the BAT or ASM before the discovery, there was a second weaker outburst of the source in 2010 February, which is seen in the BAT and ASM monitors.  No pointed observations of the source were performed at that time.

\subsubsection{Swift J1539.2$-$6227} \label{sw1539}   
\noindent{\bf LMXB/BHC; Discovered in 2008 November.} This source is a low-mass X-ray binary black hole candidate that has been reported in detail in \citet{krim11}.    Observations with {\em Swift} and {\em RXTE} began immediately after discovery and continued for more than seven  months through the duration of the outburst.  Swift J1539.2$-$6227 was first found in a hard spectral state and showed a progression of spectral states typical of BH binaries, including a rise in the disk component during a thermal state and signatures in the power density spectra of transitions between hard and soft intermediate states.   This suite of behaviors, combined with the lack of observed pulsations led to the black hole identification; the faintness of  the quiescent source and lack of emission line features supported the LMXB model.  

\subsubsection{Swift J0513.4$-$6547} \label{sw0513}   
\begin{figure}
 \plotone{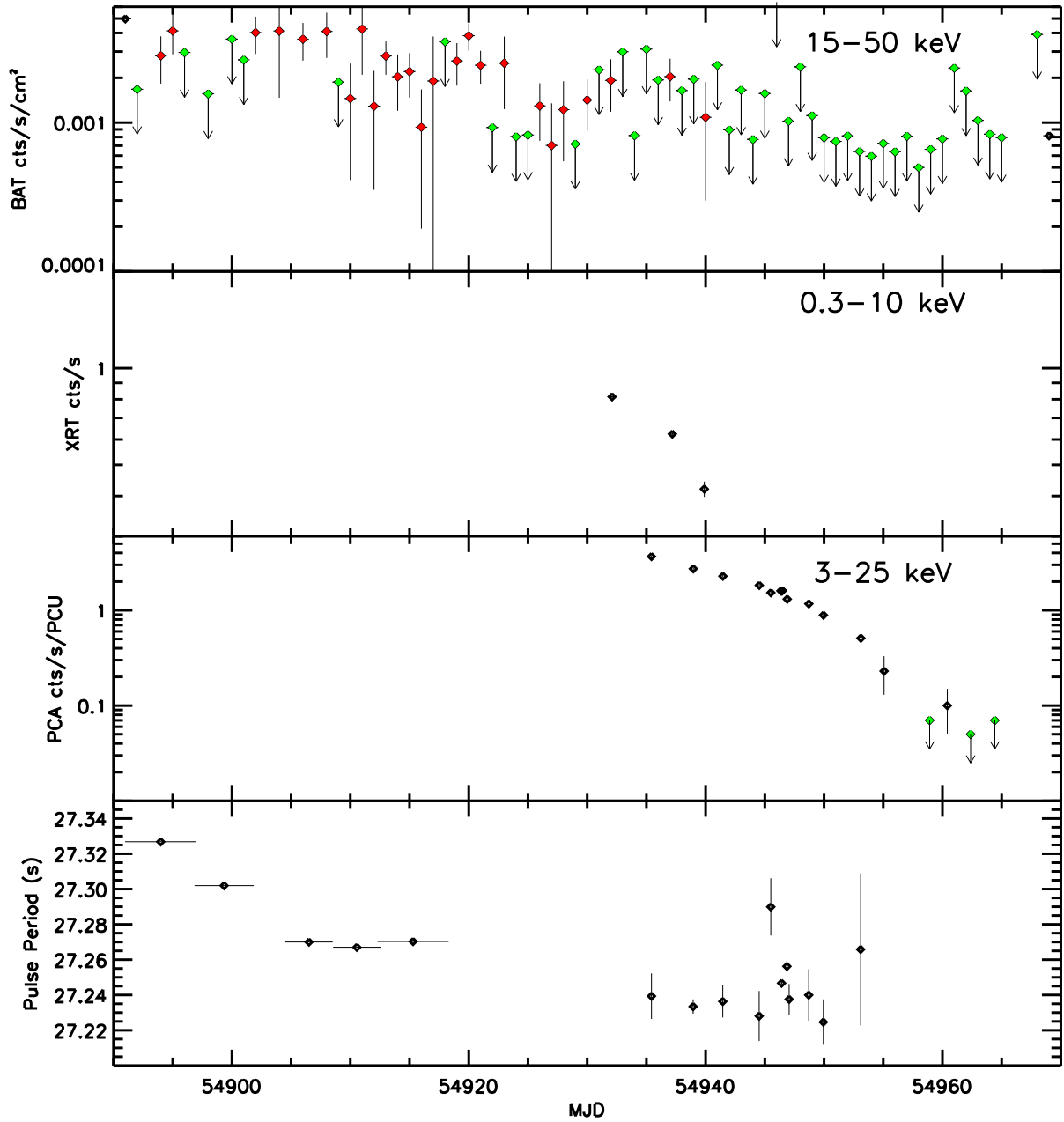}  
 \caption{Light curves and pulse period plot for  Swift J0513.4$-$6547.  Light curves from BAT (upper panel), XRT (second panel) and PCA (third panel). The bottom panel shows the variations in pulse period.  Points in the bottom panel before MJD 54920 are from the {\em Fermi}/GBM \citep{atel2023} and those after are from the {\em RXTE}/PCA (this work). }\label{sw0513_fig}
\end{figure}

\noindent{\bf HMXB/NS; Discovered in 2009 April.} Swift J0513.4$-$6547 is a 27.28-s period pulsar and likely Be star binary in the Large Magellanic Cloud.   There is strong evidence for pulsations in the power spectra derived from {\em RXTE} PCA observations from 2009 April  14 \citep[MJD 54935;][]{atel2011} to 2009 May 2 (MJD 54953).  The weighted average pulse period from the PCA observations is $P = 27.246 \pm 0.001$~s.   The pulse profile is double peaked, and the amplitude,  which is defined as (maximum - minimum) / (max + min), is large at about 85\%.   Over the course of the outburst, the gradient measured in the PCA (Bottom panel of Figure~\ref{sw0513_fig}, points after MJD 54935) shows a weak spin-down trend with period derivative \.{P}\ =\ ($1.81 \pm 0.55) \times 10^{-8}\ {\rm s\ s^{-1}}.$  The second harmonic of the pulse period was detected for observations through MJD 54949, but the third harmonic was only detected on MJD 54938.   As seen in Figure~\ref{sw0513_fig}, the outburst appears to have ended (or fallen below the PCA sensitivity) after MJD 54958 and no pulsations are detected after MJD 54955.  The PCA analysis used GoodXenon barycentered data with 0.5~s timing resolution for the pulse period detections.

A search for pulsations in the {\em Fermi}/GBM was carried out and the pulsar was detected over the period from MJD 54891$-$54918 \citep{atel2023}.  The GBM team found that the pulsation period varied by $\approx 0.2\%$\ over this period, showing a spin-up trend for the first 13 days followed by an increasing period after 54908.5, which \citet{atel2023} attribute to doppler shifts from the (unknown) binary orbit.  The calculated frequency rate, $(1.02 \pm 0.5)\ \times 10^{-10}\ {\rm Hz\ s^{-1}}$\ is reported as consistent with a pulsar accreting from a disk near the Eddington rate \citep{atel2023}.

The source is detected at wavelengths from the u to K bands.   The optical counterpart is identified as a source in the 2MASS catalog, 2MASS 05132826$-$6547187, with reported magnitudes B=15.3, R=15.5 (USNO-B1.0), J=15.2, H=15.1, K=14.8 (2MASS; Vega; exposures from 1998 December 10) and I = $15.09 \pm 0.04$ and J  = $15.20 \pm 0.14$\ (DENIS; Vega; exposures from 1996 December 22).  The optical brightness suggests that Swift J0513.4$-$6547 is a HMXB.  The 
optical magnitudes reported in \citet{atel2013} show that the source had brightened by $\approx 0.5$\ magnitudes since these archival observations.   The interpretation of the brightening and the colors by \citet{atel2013} is that the disk brightened considerably, which in turn suggests that increased mass loss from the donor can account for both the optical brightening and X-ray outburst.   No spectroscopic measurements of the expected emission lines have been reported.    The UVOT detections showed no variability.  On 2009 April 11, the source had a magnitude of u = $13.69 \pm 0.01$  \citep{atel2011} and on April 16, uvw2 = $12.84 \pm 0.01$.   One set of observations in all UVOT filters was carried out on 2009 April 19 with the following magnitudes:   v  = $15.08 \pm 0.02$, b  = $14.99 \pm 0.01$, u  = $13.67 \pm 0.01$, uvw1 = $13.16 \pm 0.01$,  uvm2  =  $12.89 \pm 0.01$,  uvw2  = $12.83 \pm 0.01$, showing no significant change in the u and uvw2 magnitudes.

The X-ray light curves are shown in the top three panels of Figure~\ref{sw0513_fig}.  Since the source was always relatively weak in the BAT, we required a fairly long integration for the flux to rise above the threshold for a new source discovery.  In fact, by the time the source was identified and follow-up observations were made, the source was already below the BAT threshold.  Both the XRT and PCA light curves show a fairly steady, featureless decline. The source spectrum in the first XRT observation can be fitted to a power law with photon index $0.999 \pm 0.099$, and N$_{\rm H} {\rm (intrinsic)}\ =\ 7.9 \pm 3.6 \times 10^{20}\ {\rm cm^{-2}}$. (N${\rm_H} {\rm (Galactic)}\ =\ 9.8 \times 10^{20}\ {\rm cm^{-2}}$). The unabsorbed flux (0.3-10 keV) is $4.8 \times 10^{-11}\ {\rm erg\ cm^{-2}\ s^{-1}}$. At the distance of the LMC (50 kpc), this corresponds to a luminosity of $1.5 \times 10^{37}\ {\rm erg\ s^{-1}}$. 

The presence of this HMXB source in the Large Magellanic Cloud is somewhat unusual given that the LMC hosts relatively few HMXB X-ray pulsars, compared to the SMC. \citet{liu05} note that there are 92 HMXBs in the SMC and 36 in the LMC.

\subsubsection{Swift J1713.4$-$4219} \label{sw1713}   
\noindent{\bf Unknown transient (likely Galactic); Discovered in 2009 November.} When Swift J1713.4$-$4219 was discovered \citep{atel2300}, it was too close in the sky to the Sun for observations with the {\em Swift} XRT and UVOT and was only visible to {\em RXTE} for 3 days starting on 2009 November 16 before {\em RXTE} too was in Sun-constraint.    The PCA power spectrum showed strong aperiodic variation, but no significant periodicities.  The energy spectrum was reported to be consistent with a black hole in the low (thermal) state  \citep{atel2300}.
The PCA light curve is flat at $\approx 15\ {\rm ct s^{-1} PCU^{-1}}$\ (3-25 keV) and the BAT light curve was flat at $\approx 0.005\ {\rm ct\ cm^{-2}\ s^{-1}}$\ for about six days starting on 2009 November 13 (MJD 55148).  Between MJD 55154 and 55179 when Swift J1713.4$-$4219 was very near the Sun, the BAT light curve had such large statistical errors that the light curve could not be followed.  By MJD 55180, the source was undetectable in the BAT.  No XRT observations were made.

\subsubsection{Swift J1729.9$-$3437}\label{sw1729} 
\begin{figure}
 \plotone{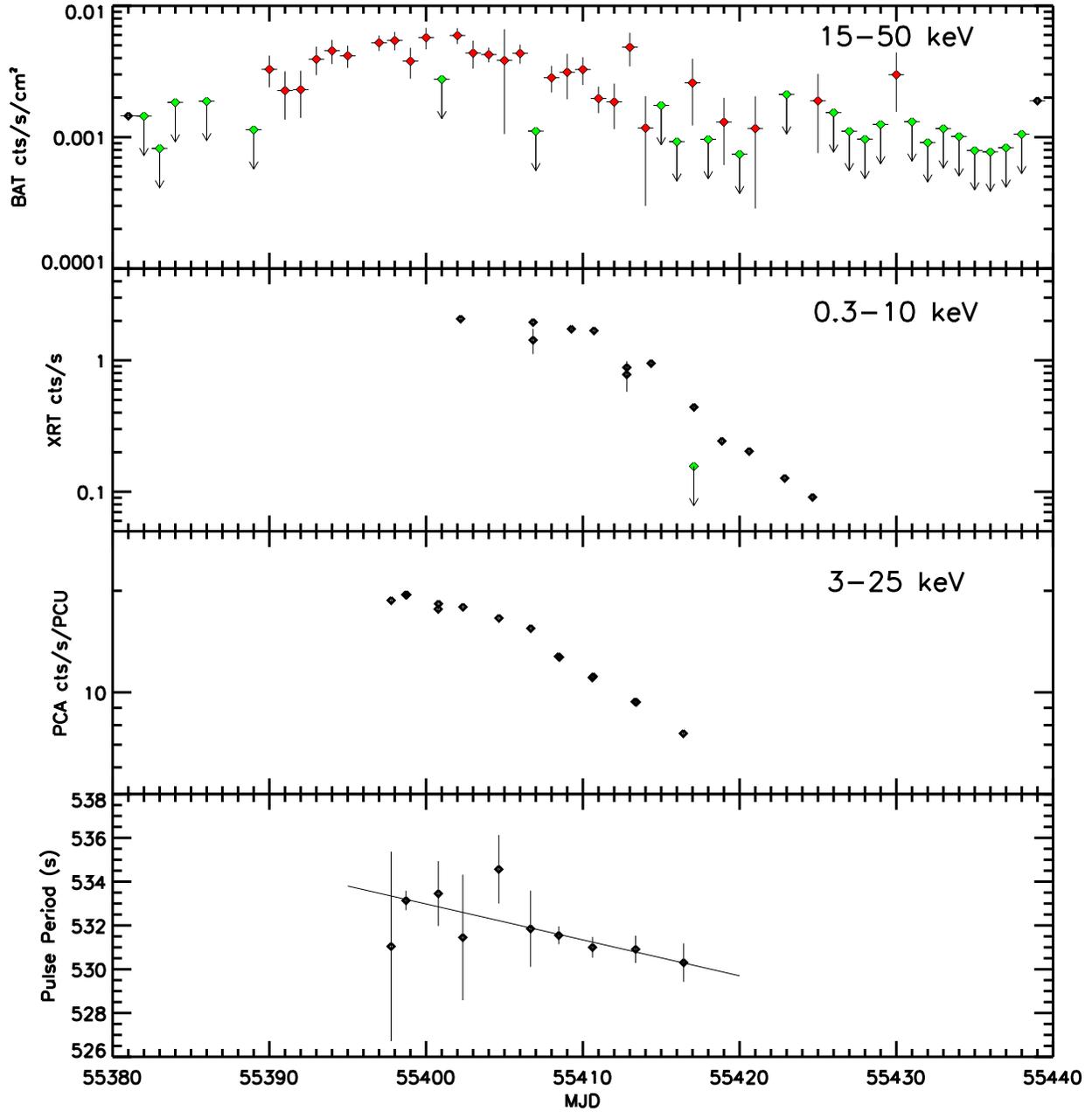}  
 \caption{Light curves and pulse period plot for  Swift J1729.9$-$3437.  Light curves from BAT (upper panel), XRT (second panel) and PCA (third panel).   The bottom panel shows the variations in the pulse period from the PCA observations.}\label{sw1729_fig}
\end{figure}

\noindent{\bf XRB/NS; Discovered in 2010 July.} This source was discovered independently by the BAT monitor and the {\em RXTE}  PCA in its Galactic center monitoring \citep{atel2747}.   {\em RXTE} PCA observations show a weighted average period of $531.8 \pm 0.2$\ s, which clearly identify it as an X-ray pulsar in a binary system.  The PCA observations used Standard2f data with 16-s timing resolution.  Second and third harmonics were also detected in all observations.   The pulse period plot (Figure~\ref{sw1729_fig} bottom panel) shows a clear indication of a spin-up trend with \.{P} = $(-1.93 \pm 0.45)\ \times 10^{-6}\ {\rm s\ s^{-1}}$. The value of $\chi^2_{\rm reduced}$\ is 0.40, which indicates that the errors might have been over estimated  somewhat.  It is unclear whether \.{P} is due to actual spin-up of the neutron star or Doppler modulation.  If we take the spin-up as entirely torque-driven, we calculate using the method of \citet{joss84}, a luminosity of $(9.97 \pm 2.70)\ \times 10^{37}\ {\rm erg\ s^{-1}}$.

The BAT light curve (Figure~\ref{sw1729_fig} top panel) shows a broad peak lasting for $\approx 20$\ days and then a slow fall-off.  The XRT and PCA light curves have a more rapid decline and the source was still detectable when observations were completed.    Examining the archives of the PCA Galactic bulge scans shows that Swift J1729.9$-$3437 was also active in mid-2001, with the PCA rate peaking at $14\ \pm\ 1.6\ {\rm ct\ s^{-1}}$\ (3-25 keV) on MJD 52090 (2001 June 30) and detectable until MJD 52101.

We searched the archival catalogs for an optical counterpart to Swift J1729.9$-$3437.  No source was found within the $1''.7$\ radius error circle (Table~\ref{tab-loc}).  There is a fairly bright source $4''.6$\ away, 2MASS 17300946$-$3436433, but with this distance it is unlikely that this is the counterpart.  The 2MASS star is clearly detected in the UVOT (b = $15.89 \pm 0.02$, u = $16.33 \pm 0.03$, uvm2 = $20.50 \pm 0.42$) with no sign of variation over the $\approx 20$\ days of the observations.  There is nothing detected in the UVOT at the X-ray position.  Contamination by the 2MASS source makes setting magnitude limits difficult.  Limits ($3\sigma$) for a nearby blank location are b $> 22.3$, u $> 21.9$, uvm2 $> 21.0$.

\subsubsection{Swift J1843.5$-$0343}\label{sw1843}   
\begin{figure}
 \plotone{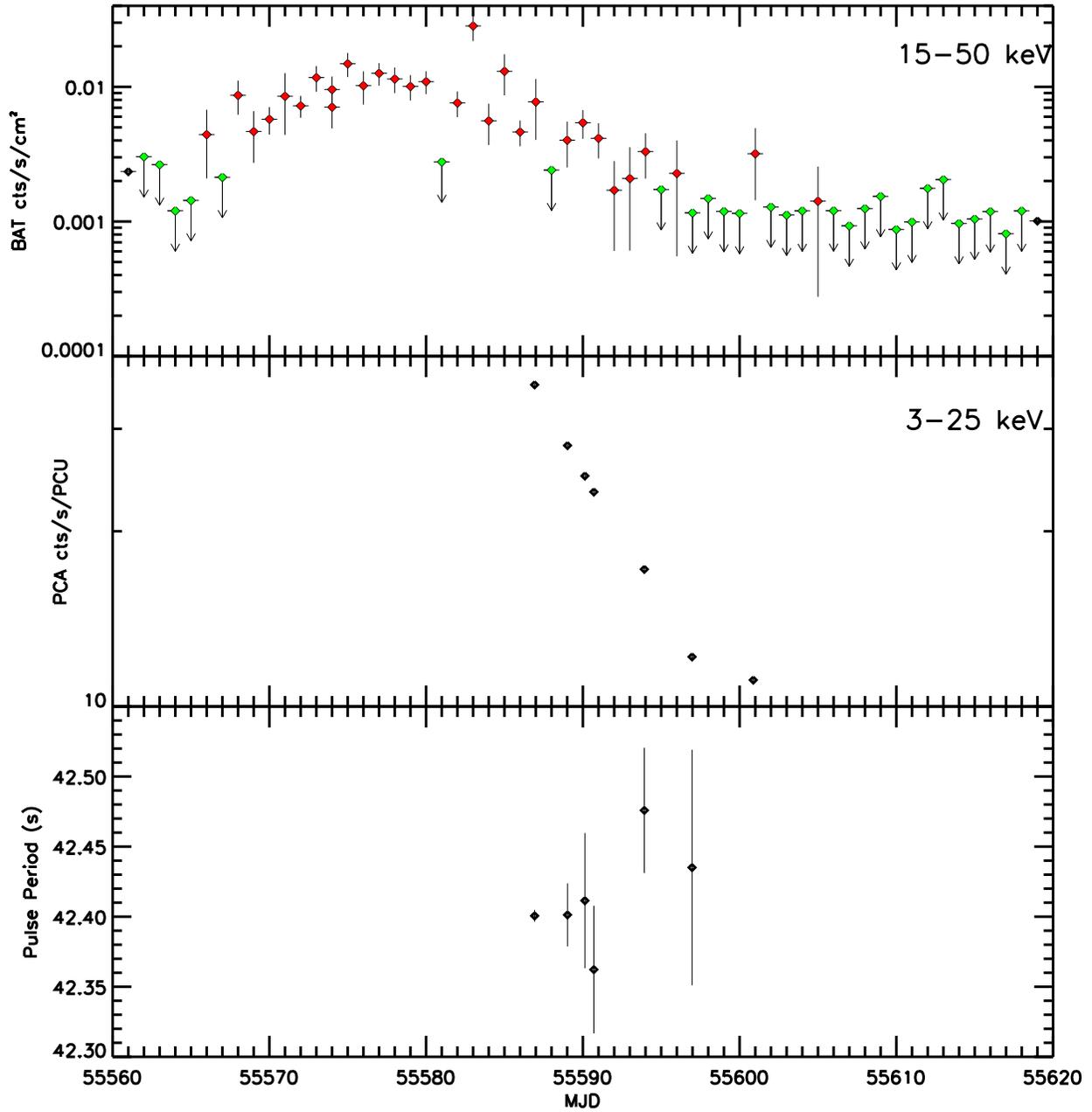}  
 \caption{Light curves and pulse period plot for  Swift J1843.5$-$0343.  Light curves from BAT (upper panel), PCA (center panel).  The bottom panel shows the variations in the pulse period from the PCA observations.}\label{sw1843_fig}
\end{figure}

\noindent{\bf XRB/NS; Discovered in 2011 January.}  Swift J1843.5$-$0343 is an X-ray pulsar that was discovered by the BAT transient monitor and reported on 2011 January  9 \citep{atel3109}.  At the time of discovery, the source was too near the Sun for follow-up observations with either the {\em Swift} XRT or {\em RXTE}, but it was confirmed by MAXI \citep{atel3114}. A later observation with {\em Swift} XRT on 2011 February 15 \citep{atel3169} measured the source at a count rate of $0.038 \pm 0.024\ {\rm ct\ s^{-1}}$\ (0.3-10 keV) and determined the position (see Table~\ref{tab-loc}).
Using a power spectral analysis of {\em RXTE} PCA observations starting on 2011 January 26, \citet{atel3121} found a strong pulsation with a period of 42.5 s (not barycentered).   A series of {\em RXTE} observations were carried out (see Figure~\ref{sw1843_fig}) and the pulse period was refined to $42.401 \pm 0.004$~s.  The analysis used Good Xenon data with a 0.5~s time resolution for all layers.

We note that the position of Swift J1843.5$-$0343 is consistent with an HII region and Galactic star-forming region IRAS 18408-0348.  None of the sources in this region (either previously detected X-ray sources or objects in the 2MASS or Digitized Sky Survey catalog) are coincident with Swift J1843.5$-$0343.

\subsubsection{Swift J1357.2$-$0933}\label{sw1357}   
\begin{figure}
 \plotone{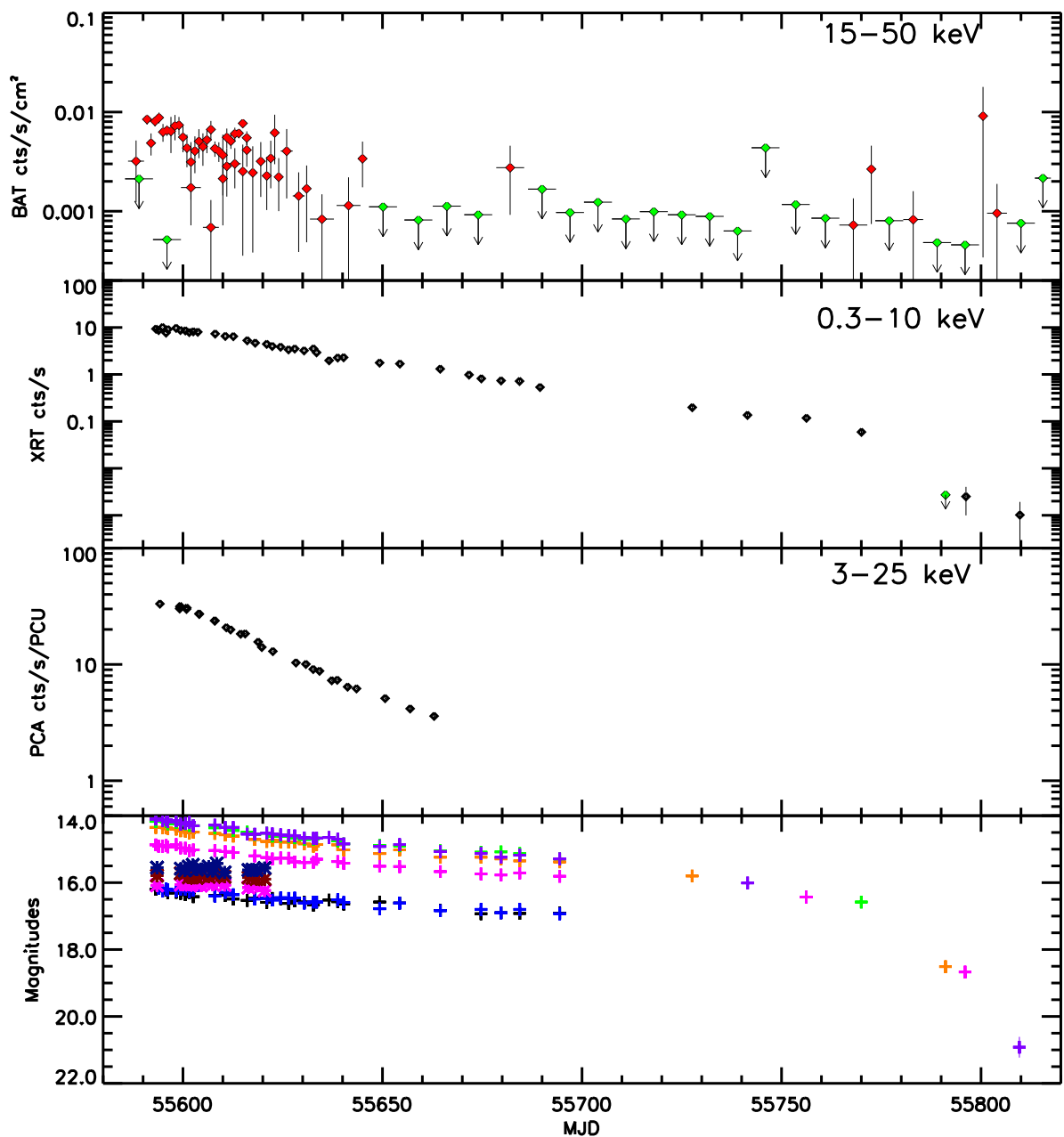}  
 \caption{Light curves for  Swift J1357.2$-$0933 in three X-ray bands and in optical magnitudes.   In the bottom plot crosses represent Swift/UVOT filters: black = v, blue = b, magenta = u, orange = uvw1, green = uvm2, purple = uvw2. The stars are PAIRITEL  \citep[except for earliest night, which is GROND,][converted to Vega magnitudes] {atel3140}: blue  = Ks, magenta = J, brick red  = H.  After MJD 55700, observations were made in the UVOT ``filter of the day," which is limited to one of the UV fiters.  The final observation is approaching the pre-outburst SDSS magnitude of the source \citep[u $=\ 22.83 \pm 0.63$,][]{atel3140}
 }\label{sw1357_fig}
\end{figure}

\noindent{\bf LMXB/BHC; Discovered in 2011 January.} Swift J1357.2$-$0933 is a Galactic binary transient source with a long series of multi-wavelength observations, as a possible identification as a BH transient.   The source was discovered in the BAT monitor \citep{atel3138}, confirmed with XRT observations and localized by the UVOT \citep{atel3142}. The source was also detected in the g$^{\prime}$r$^{\prime}$i$^{\prime}$z$^{\prime}$JHK bands by GROND \citep{atel3140} and by the PAIRITEL near infra-red telescope (this work).  Light curves for these instruments plus the {\em RXTE} PCA are shown in Figure~\ref{sw1357_fig}.   In the early part of the outburst, the BAT shows a steep rise, over $\approx 2$\ days and then a fairly flat peak is seen in all three X-ray instruments until roughly MJD 55602, when they, along with the optical and UV light curves begin a steady decline, all with an e-folding time of $\approx 30$\ days.  The count rate in the BAT fell below detectability by MJD 55650 and PCA observations were discontinued. In the XRT and UVOT there is a break in the decay at around MJD 55780 although the source is still detectable in the UVOT in the last observation.

The position is coincident with a source in the Sloan Digital Sky Survey (SDSS), and the 2006 SDSS magnitudes \citep{atel3140} are $\approx 6$\ magnitudes fainter than the initial GROND measurements.    Such a large increase in optical flux would be highly unusual for an AGN or blazar flare \citep[see e.g.][]{ulri97}, but is consistent with a Galactic X-ray binary.  Spectroscopic measurements \citep[e.g.][]{corr13}
give no indication of a cosmological red shift. For these reasons we believe that Swift J1357.2$-$0933 is Galactic, despite its high Galactic latitude (b = $+50^{\circ}.0$).  Since its apparent magnitude and $i-z$\ color are consistent with an M4 star at a distance of 1.5 kpc, we identify Swift J1357.2$-$0933 as a low-mass X-ray binary.  The high Galactic latitude is also an argument against a HMXB origin since HMXBs tend to have lower Galactic scale heights \citep[see][]{grim02}.  

Determination of the nature of the compact object has proven to be difficult.  One possible interpretation is that Swift J1357.2$-$0933 is an atoll neutron star binary.  Comparing the radio flux, $245 \pm  54\ \mu{\rm Jy}$ \citep{atel3147} to the peak X-ray flux, $4.1 \times 10^{-10}\ {\rm erg\ cm^{-2}\ s^{-1}} $\ \citep{arma13}, which at a source distance of 1.5 kpc translates to a peak luminosity of $L_X\ =\ 1.1 \times 10^{35}$\ erg s$^{-1}$, implying that the source is underluminous in the radio band by a factor of $\gtrsim 10$\ compared to typical black hole X-ray binaries.  Although there was only one reported radio observation, source spectroscopy is inconsistent with the alternative model that the low radio flux implies a transition between radiative inefficient and radiative efficient accretion flows.  Joint spectral fits to the XRT and PCA data show no clear sign of spectral evolution and for the first 60 days of the outburst, the spectrum is consistent with a power-law (PL) dominated spectrum with an average PL index of $1.7$.  An argument against the neutron star model is the absence of any pulsations in the timing data from {\em RXTE} PCA GoodXenon data in the 2.1-33 keV band.

A much more likely interpretation then is that Swift J1357.2$-$0933 is a LMXB black-hole binary.  The slow  evolution in a hard spectral state is more consistent with black holes than neutron stars \citep[see examples in][]{broc04}, as is the large optical outburst amplitude.  There is weak evidence for QPOs at frequencies ranging from $\approx 1$\ to $9$\ Hz and continuum power between $\approx 10$\ and $25$\%,  again suggestive of a black hole accretor.   

\citet{corr13} also argue that Swift J1357.2$-$0933 is a BH system, based on optical spectroscopy.  In studying the H$\alpha$\ emission line profile, these authors estimate a radial velocity semi-amplitude, $K_c \gtrsim 690\ {\rm km\ s^{-1}}$\ and note that the radial velocities of the H$\alpha$\  wings are modulated with a 2.8-hour period, which they interpret as the orbital period.  From $K_c$\ and the orbital period, \citet{corr13} calculate a lower limit to the mass of the compact object of $3.0 M_{\odot}$, which is greater than the maximum possible neutron star mass, confirming the black hole nature of the compact object.
Such large velocities have been seen in other black hole transients at high Galactic latitudes, both in outburst and quiescence, including Swift J1753.5$-$0127 \citep{atel566}, XTE J1118$+$480 \citep{torr02}  and GRO J0422$+$32 \citep{case95}.  
The detection of optical dips up to 0.8 magnitudes in the light curve presented by \citet{corr13} suggests that Swift J1357.2$-$0933 is viewed at a large inclination $i \gtrsim 70^{\circ}$.

\citet{arma13} performed a full spectral analysis of the XRT data and photometric analysis of the UVOT data.  These authors reached the conclusion that Swift J1357.2$-$0933 is a BHC LMXB and that its low peak X-ray luminosity ($\sim 10^{35}$\ (D/1.5 kpc)$^2$\ erg s$^{-1}$) classifies it as a very faint X-ray transient.  \citet{arma13} found that Swift J1357.2$-$0933 remained in a hard state throughout the outburst, but that as it returned to quiescence its X-ray spectrum softened, behavior that is consistent with numerous other BH systems studied \citep[see][for references]{arma13}.   These authors also show that there was a clear correlation between the X-ray flux both in the 0.5 - 10 keV and 2 - 10 keV bands and the UVOT magnitudes in all six bands. The other arguments that \citet{arma13} give for a BH nature of the compact object in Swift J1357.2$-$0933 are the low luminosity inferred from the final XRT non-detection upper limit, and the X-ray/optical correlation in the v band.

\subsubsection{Swift J2058.4$+$0516}  \label{sw2058} 
\noindent {\bf TDF; Discovered in 2011 May.} Swift J2058.4$+$0516 is an extragalactic  source \citep[red shift $z = 1.1853$;][]{cenk11} and believed to be a tidal disruption flare (TDF) event like Swift J164449.3$+$573451 \citep[Swift J1644;][]{burro11,levan11,zaud11}, which was discovered shortly before.  Since Swift J2058.4$+$0516 is at a higher red shift, it was not bright enough in the BAT to trigger on-board, but was instead found in the multi-day images in the BAT monitor.  The source continued to be detectable in the BAT for $\approx 16$\ days, although the statistics were too poor to see significant flaring.  In the XRT, there was a shallow decay ($t^{-2.2}$) with significant superimposed flares.  As reported in \citet{cenk11}, the source was also detected with the {\em Swift} UVOT, with the 7-channel near-infrared imager GROND \citep{grei08}, in the radio with the Expanded Very Large Array and with the {\em Chandra} X-ray observatory.  The consensus interpretation of these observations and numerous morphological similarities to Swift J1644 support the model that the outburst of Swift J2058.4$+$0516 is powered by tidal disruption of a non degenerate star on a black hole of mass, $M_{BH} \lesssim 10^8 M_{\odot}$, rather than by gas accretion onto an active galactic nucleus. \citet{cenk11} use several different calculations to estimate lower limits on $M_{BH}$, all of them consistent with a tidal disruption outside the event horizon. The strongest arguments for this scenario are the super-Eddington 0.3-10 keV X-ray luminosity $L_{X,iso} \approx 3 \times 10^{47}\ {\rm erg s^{-1}}$\ with a relatively faint magnitude (M $\approx 21$) optical absolute magnitude and a spectral energy distribution incompatible with the blazar sequence \citep{fosa98}.

\subsubsection{Swift J1112.2$-$8238} \label{sw1112}  
\noindent{\bf Possible TDF; Discovered in 2011 June.}  The source is at a high Galactic latitude and except for an early bright peak on MJD 55728, was seen only at a fairly faint level in the BAT monitor \citep{atel3463}.  The best position was derived from observations with GMOS on the Gemini-South 8-m telescope \citep[Table~\ref{tab-loc},][]{atel3469}. The source is described as faint and possibly extended.  No counterpart was found in {\em Swift} UVOT observations to a limiting magnitude of b $>\ 22.0\ (3\sigma)$.  The average photon-counting mode spectrum is well-described by an absorbed power-law with photon  index $\Gamma = 1.45^{+0.053}_{-0.089}$\ and N${\rm_H} = 1.60^{+0.29}_{-0.17} \times 10^{21}\ {\rm cm^{-2}}$ (C-stat = 593.9 for 567 d.o.f.).

\begin{figure}
 \plotone{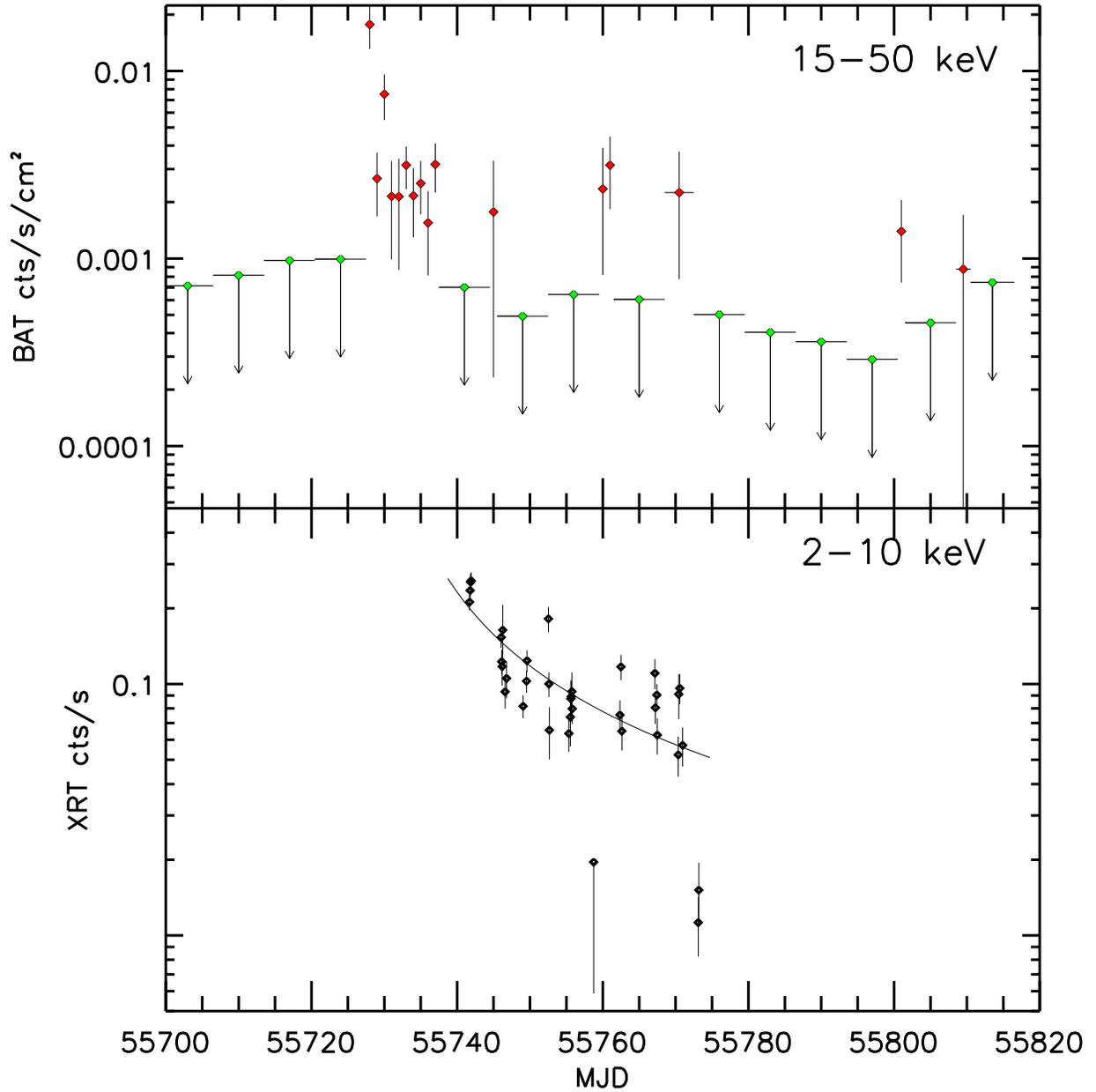}  
 \caption{Light curves for  Swift J1112.2$-$8238 from BAT (upper panel) and XRT (lower panel). The curve on the lower plot shows a fit to a $t^{-1.1}$\ decay from the time at which the BAT light curve peaks.  }\label{sw1112_fig}
\end{figure}

The current results do not allow us to clearly determine the nature of the source. The spectrum is consistent with a low-mass X-ray binary in a hard state. A high-mass X-ray binary interpretation is much less likely given the faintness of the optical counterpart and the high Galactic latitude.  Another possibility is that Swift J1112.2$-$8238 could be a similar TDF event to Swift J1644 or Swift J2058.4$+$0516.  The BAT and XRT light curves (Figure~\ref{sw1112_fig}) show features qualitatively similar to these earlier TDF.  In the BAT, there is a rapid rise to a possible flare at the start of the outburst and the XRT  light curve shows significant short timescale variations superimposed on a shallow decay. Comparing the XRT count rate for the three sources at a common time of 20 days after the onset of the outburst gives $\approx 2$\ for Swift J1644, $\approx 0.6$\ for Swift J2058.4$+$0516 and $\approx 0.06$\ for Swift J1112.2$-$8238, suggesting that it is at a considerably higher red shift.  The absence of a UVOT counterpart with relatively low absorption in the direction toward the source is also consistent with the TDF interpretation.  However, in the absence of a measured red shift for Swift J1112.2$-$8238, it is not possible to confirm this scenario.

\subsubsection{Swift J1836.6$+$0341} \label{sw1836}  
\begin{figure}
 \plotone{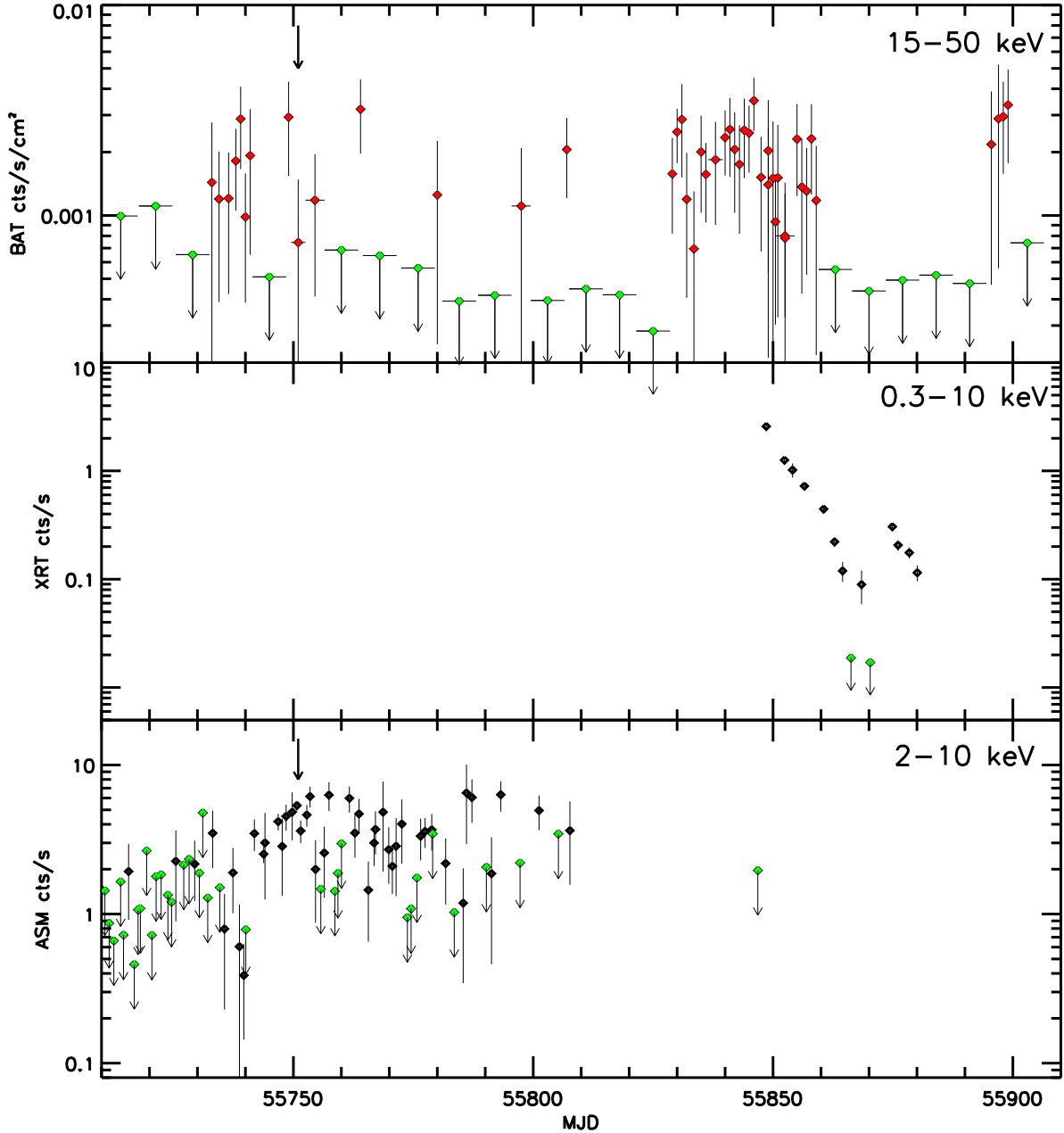}  
 \caption{Light curves for  Swift J1836.6$+$0341 from BAT (upper panel), XRT (middle panel) and ASM (lower panel).  There was only one ASM data point available after MJD 55807, so we are unable to track the main outburst in the ASM. The arrows on the BAT and ASM plots indicate the time of the first optical detection of Swift J1836.6$+$0341 with Pan-STARRS 1 \citep{atel1708}.}\label{sw1836_fig}
\end{figure}

\noindent{\bf XRB (likely Galactic); Discovered in 2011 October.} Since the source never reached a level above $\approx 16$\ mCrab, it was first detected by the BAT in a 16-day integration covering the days 2011 September 25 through 2011 October  10 (MJD 55829 - 55844).  Therefore no pointed observations were possible  before 2011 October 14 (see Figure~\ref{sw1836_fig}), when the source rate was already starting to decline.   The position of the source is consistent with that of XTE J1837+037, an unidentified source listed in the {\em RXTE}/ASM catalog. A literature search has revealed no previous reports on this object, save for a passing mention in \citet{remi06}.  Since the {\em RXTE} source was not studied in detail, this transient is given the additional name, Swift J1836.6$+$0341.

A series of observations with {\em Swift}/XRT was carried out.  Spectral analysis from the first observation \citep{atel3684} shows a good fit to an absorbed power-law model with the parameters: N${\rm_H} = (3.5 \pm 0.5) \times 10^{21} {\rm cm^{-2}}$, $\Gamma = 1.80 \pm 0.25$, Flux (0.3-10 keV) = $(1.2 \pm 0.16) \times 10^{-10} {\rm erg\ cm^{-2}\ s^{-1}}$\ with no evidence of any lines or other deviations from a smooth spectrum.  According to \citet{atel3687}, the value of N${\rm_H}$ suggests that Swift J1836.6$+$0341 lies near the end or behind the total Galactic column, which corresponds to a distance of at least 1 kpc.

An examination of the Vizier\footnote{http://vizier.u-strasbg.fr}  catalogs showed that there is no catalog source within the error radius. There is also no detection in the UVOT U band.
A pair of observations were carried out with the GROND telescope at the La Silla Observatory (Chile) on 2011 October 15 and 17 \citep{atel3687}.  An optical counterpart was found consistent with the XRT position (see Table~\ref{tab-loc}) and the source remained at constant brightness within photometric errors between the two observations. Comparison with the red DSS2-limit implies a brightening of a $>2$\ magnitudes.

\citet{atel3708} reported that Swift J1836.6$+$0341 underwent an optical outburst as measured during the Pan-STARRS 1 (PS1) 3Pi sky survey (${\rm r} = 18.88 \pm 0.02$), starting on 2011 July 9, more than two months before the hard X-ray outburst.  This magnitude is also significantly brighter than that seen on 2011 October 15 \citep[${\rm r}^{\prime} = 20.5 \pm 0.1$;][]{atel3687}.  It is also noted that both BAT and the ASM show a small but significant increase in flux around the time of the optical outburst, and examination of the ASM light curve suggests that this earlier outburst began around MJD 55740. By comparing with the r-band detection limit of the PS1 data before the outburst, \citet{atel3708} estimate that the quiescent optical magnitude of  Swift J1836.6$+$0341 is $\gtrsim 23$.

At a low Galactic latitude, b = +4$^{\circ}.96$, the source is most likely Galactic, and \citet{atel3687} speculate that the source could be a cataclysmic variable, based on the low luminosity $\approx 10^{34} ({\rm d}/1\ {\rm kpc})^2\ {\rm erg\ s^{-1}}$.  They also note that the extinction-corrected g$^{\prime}$-K (GROND) spectral energy distribution is very blue ($\approx \lambda^{-0.6}$), which is consistent with an accretion disk spectrum.  An extragalactic origin is also possible, although the $f_x/f_{opt}$\ ratio is not typical of AGN.  In short, the nature of Swift J1836.6$+$0341 remains unknown. 

\subsubsection{Swift J1943.4$+$0228}\label{sw1943} 
\noindent{\bf XRB (likely Galactic); Discovered in 2012 April.} This is a likely Galactic source, which was discovered in a 16-day integration covering the days 2012 March 19 through 2012 April 13 (MJD 56015 - 56020).  The source was detected in the {\em Swift} XRT and UVOT with magnitude b = $18.17 \pm 0.04$\ \citep{atel4049} and a position was found (see Table~\ref{tab-loc}).  An examination of the Vizier catalogs and Digitized Sky Survey images shows that there is no catalog source within the error radius.    The spectrum of the first XRT observation is well fitted by an absorbed power-law model with the following parameters: N${\rm_H} = 1.9 \pm 0.4 \times 10^{21} {\rm cm^{-2}}$, $\Gamma = 1.17 \pm 0.12$\ and observed flux (0.3-10 keV) of $8.3 \pm 0.64 \times 10^{-11} {\rm erg\ cm^{-2}\ s^{-1}}$.  There is no evidence of any lines or other deviations from a smooth power law in the X-ray spectrum.  

The current results do not allow us to determine the nature of the source. The position of Swift J1943.4$+$0228 is near the Galactic plane (Table~\ref{tab-loc}), suggesting that the source is Galactic. The faintness of the optical counterpart detected by GROND \citep[$10^{34} ({\rm d/1 kpc})^2\ {\rm erg\ s^{-1}}$;][]{atel4054} suggests that the source is a low-mass X-ray binary and possibly a cataclysmic variable.  However an extragalactic origin can not be ruled out.  

\subsubsection{Swift J1910.2$-$0546}\label{sw1910} 
\begin{figure}
 \plotone{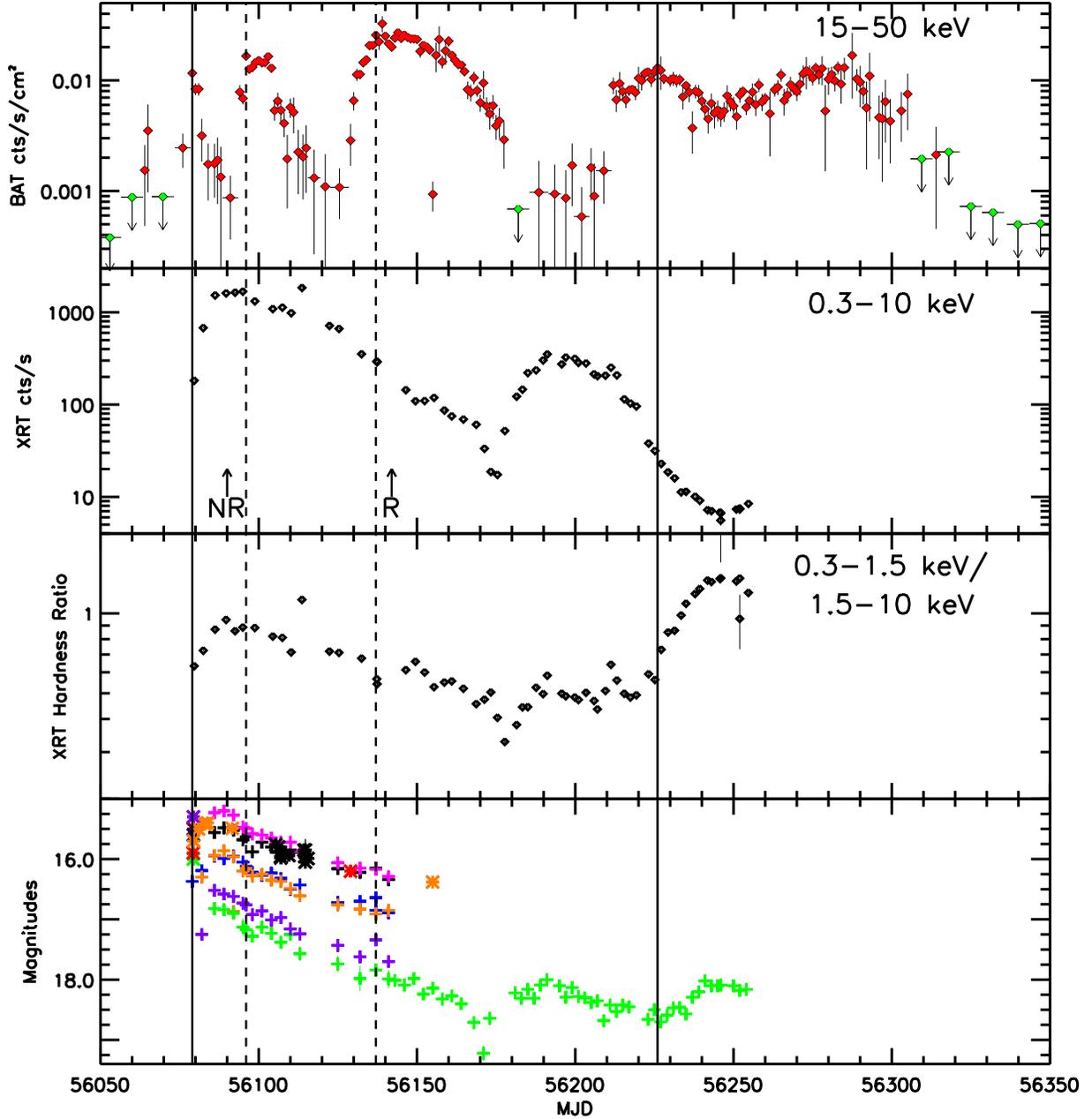}  
 \caption{\footnotesize Light curves for  Swift J1910.2$-$0546 from BAT (upper panel), XRT (second panel) and optical observations (lower panel) and XRT hardness ratio (third panel).  The vertical lines indicate the approximate dates of the peaks in the BAT light curve.  The dashed lines represent the two BAT on-board triggers on this source.   {\em Swift} was unable to observe Swift J1910.2$-$0546 with the NFIs after MJD 56254 due to the Sun observing constraint.  In the bottom plot crosses represent Swift/UVOT filters: black = v,	blue = b, magenta = u, orange = uvw1, green = uvm2, purple = uvw2. The stars indicate the following filters: black = V, red = R, orange = r$^{\prime}$, green = g$^{\prime}$, purple = z$^{\prime}$, blue  = K, brick red  = H.  The red stars are from \citet[][MJD 56079]{atel4146} and \citet[][MJD 56129]{atel4347}.  The black stars are from \citet{atel4246} and the orange stars are from \citet[][MJD 56079]{atel4144}, \citet[][MJD 56080 - 56083]{atel4195}, \citet[][MJD 56092]{atel4246}, and \citet[][MJD 56155]{atel4347}.  All other stars are from GROND \citep{atel4144}.  The arrows on the second panel indicate the times of a radio non-detection \citep[NR,][]{atel4171} and radio detection \citep[R,][]{atel4295}.}\label{sw1910_fig}
\end{figure}

\noindent{\bf LMXB/BHC; Discovered in 2012 May.} The source, a possible Galactic black hole candidate,  was first detected by the BAT in a 2-day integration covering the days 2012 May 30-31 \citep[MJD 56077$-$ 56078;][]{atel4139}.  It was simultaneously detected by MAXI \citep{atel4140}, who gave it the alternate name MAXI J1910$-$057.    This source also triggered the BAT onboard on two separate occasions (indicated by dashed vertical lines on Figure~\ref{sw1910_fig}), 2012 June 18 \citep{gcn13369} and 2012  July 29 \citep{gcn13527,gcn13538}.

Numerous optical observations were made with various telescopes over the first three months of the outburst (see Figure~\ref{sw1910_fig}, bottom panel, for magnitudes and references).  The most extensive optical measurements were made by UVOT, which observed in all six filters through MJD 56141 and in the uvm2 filter through MJD 56254.  The UVOT light curve is discussed in detail below.
An optical/near-infrared (IR) counterpart was detected and localized by GROND \citep{atel4144} on June 1, who reported magnitudes ranging from K = $15.6 \pm 0.1$\ to g$^{\prime} = 16.0 \pm 0.1$ and calculated an extinction-corrected \citep[E(B-V)=0.6;][]{schl98} g$^{\prime}-$K spectral energy distribution of $F_{\lambda} \approx \lambda^{-3}$, which is very blue, consistent with an accretion disk spectrum. \citet{atel4146} also detected the optical source on June 1 with the Palomar 48-inch Oschin Schmidt telescope, part of the Palomar Transient Factory, and reported a magnitude of $R = 15.9$.      Some authors reported flickering and also possibly periodic variation in the optical light curve, with a possible period of $\approx 2.2$\ hr \citep{atel4246} or $\approx 4$\ hr \citep{atel4347}.  The periodicity could be attributed to orbital variations, although \citet{atel4347} note that if measured H${\alpha}$\ variations are due to binary motion, the orbital period must be $> 6.2$~hr.  No group saw evidence of pulsations or QPOs  in the optical data.   

Spectroscopy was reported from three epochs. On 2012 June 18 (MJD 56097), during the soft (thermal) X-ray state (see below), \citet{atel4210} observed a spectrum typical of a LMXB, dominated by an almost featureless continuum, with very weak $\lesssim 0.2$\ \AA\ E.W.), broad He II 4686 emission and somewhat stronger (1.5 \AA\ E.W.) H${\alpha}$\ emission.  After Swift J1910.2$-$0546 entered the hard state, spectra were obtained by \citet{atel4347} on 2012 July 28 and 2012 August 16 (MJD 56136 and 56155, respectively).  These observations show significantly different spectra from the soft state with a weak ($\approx 2$\ \AA\ EW) \ion{He}{2} $\lambda$4686 emission line and broad (FWHM $\approx$2000-3000 km s$^{-1})\ {\rm H}{\beta}$\ and H${\gamma}$\ absorption features.  H${\alpha}$\ is seen as a wide absorption trough with a narrow (FWHM = $550 \pm 20$\ km s$^{-1}$) emission component exhibiting clear velocity motions.  \citet{atel4347} suggest that the H${\alpha}$\ is likely to arise from a turbulent region in the accretion disc, such as the hot spot. 

Extensive observations with all three {\em Swift} instruments show a complex light curve for Swift J1910.2$-$0546 (Figure~\ref{sw1910_fig}).
Although spectral analysis is beyond the scope of this paper, and will be reported in a later paper, we discuss the changes seen in the source light curves.  There are at least four separate peaks in the BAT light curve, indicated by the solid and dashed lines in Figure~\ref{sw1910_fig}, with the count rate falling below 0.002 ct cm$^{-2}$\ s$^{-1}$\ between each of them.  There is a possible weak early outburst at MJD 56065, but we count the peak at MJD 56079 as the first BAT peak.   At the time of the first BAT peak, the XRT rate and hardness ratio and the optical flux are just starting to rise, suggesting that the source is in an early hard state.  Over the next few days the BAT rate drops while the XRT rate peaks at approximately MJD 56092.  At this time, \citet{atel4198} suggest that Swift J1910.2$-$0546 is in a soft thermal state, based on MAXI spectroscopy.  But almost immediately the BAT rate rises to its second peak (and the first trigger), while the XRT rate remains roughly flat.  This can be attributed to the transition to an intermediate state around MJD 56095 \citep[also suggested by][]{atel4198}.  After this point the XRT rate drops steadily until $\approx$ MJD 56170, apart from a sharp and, as yet unexplained, rise on MJD 56113.  However, the BAT rate undergoes another rise from MJD 56129 - 56137, when there is a second on-board trigger.  At this point there are several indications that the source has re-entered a hard state.  \citet{atel4273} report on a spectral change in MAXI around MJD 56132,  \citet{atel4328} report that Swift J1910.2$-$0546 is detected up to $\approx 200$\ keV in INTEGRAL/ISGRI, and \citet{atel4295} report a detection at 2.5mJy with the Karl G. Jansky Very Large Array (JVLA) at 6 GHz.  The next significant light curve feature is a very sharp dip in all measured light curves, starting with the UVOT uvm2 filter from MJD 56168$-$56173, then in the XRT 0.3-10 keV from MJD 56171-56175, and finally in the BAT 15-50 keV from MJD 56179-56184.  Since this feature is not coincident in the three bands, it is unlikely to be due to an eclipse.  After the sharp drop, the XRT and UVOT rates recover while the BAT rate remains low.  Then once again around MJD 56211, the source ``pivots'' again, with the BAT rate rising and the XRT rate falling.  Also at this time the XRT hardness ratio begins a steady rise.  Near the end of the NFI observations at MJD 56254, the rates in all three instruments rise with the UVOT uvm2 leading the way at $\approx$ MJD 56239.  After $\approx$ MJD 56280 the BAT rate begins what appears to be a final decline, with the source becoming undetectable after MJD 56327.  A final XRT/UVOT observation was made on 2013 March 9 (MJD 56360; not shown in Figure~\ref{sw1910_fig})) and the source was still barely detected at a very low rate of $0.009 \pm 0.003$\ ct s$^{-1}$\ in the XRT and with a u magnitude of $18.39  \pm 0.08$.

The 2012 outburst of Swift J1910.2$-$0546 shows many of the signs of a black hole transient, most particularly the progression of state transitions and the mirroring of rises in the 0.3-10 keV band with falls in the 15-50 keV band (and vice versa).  However, without more extensive timing and spectral analysis, it cannot be stated with any certainty whether the compact object in Swift J1910.2$-$0546 is a BH or neutron star.

\subsubsection{Swift J1745.1$-$2624 (Swift J174510.8$-$262411)}\label{sw1745} 
\noindent{\bf LMXB/BHC; Discovered in 2012 September.}  This source was discovered when it triggered the BAT telescope onboard three times on  2012 September 16 and 17 \citep[MJD 56186$-$56187;][]{gcn13774,gcn13775,atel4383}.  Since it was initially localized by the XRT, it was given a name following the XRT convention, Swift J174510.8$-$262411.  However, for consistency with other Swift discoveries in this paper, we will refer to it with the truncated name Swift J1745.1$-$2624.  Note that it is also known in the literature as Swift J1745$-$26.  In addition to {\em Swift}, the X-ray source was also detected by {\em INTEGRAL} \citep{atel4381,atel4401,atel4450,atel4804}.  {\em INTEGRAL} also carried out serendipitous observations of the source fields in the days before the outburst \citep{atel4401}, with a reported $3\sigma$\ upper flux estimated to be 0.75 mCrab in the 20$-$60 keV band.

A near-infrared counterpart was detected by GROND \citep{atel4380} at magnitude J $\sim 16.5 \pm 0.5$.  It was recognized as the likely counterpart because the same star was found in an archival image in the same band \citep{atel4380} at a magnitude $\sim 3$\ times fainter.  An archival search for the quiescent counterpart was carried out by \citet{atel4417}, who report an upper limit of r$^{\prime} > 23.1 \pm 0.5$\ and estimate a quiescent color of r$^{\prime} - $\ J $> 3.6 \pm 0.7$, consistent with reddening maps of the Galactic bulge \citep{gonz12}.  The faintness and color of the quiescent counterpart is consistent with Swift J1745.1$-$2624 being a LMXB.   
\cite{muno13} carried out a 30-day optical monitoring campaign of this source. This group saw an optical peak $\approx 3$\ days later than the hard X-ray (15-50 keV) peak and an outburst magnitude $> 4.3$, which they determine is consistent with an orbital period $\lesssim 21$\ h and a companion star with a spectral type later than $\approx $\ A0.  Their interpretation of the broad H${\alpha}$\ line found in the optical spectroscopy is that Swift J1745.1$-$2624 is a black hole candidate, which was in the hard state the time of their observations.
There was no UVOT detection in the v filter in early observations \citep{atel4383} and further UVOT observations were made in the (ultraviolet) ``filter of the day,'' where, due to the large reddening, no source is detected. 

A strong radio source was detected in the VLA on 2012 September 17-18 (MJD 56187.99) by \citet{atel4394}, who report measured flux densities of $6.8 \pm 0.1$\ and $6.2 \pm 0.1$\ mJy at 5.0 and 7.45 GHz, respectively, and a spectral index of $-0.22 \pm 0.09$\ (defining spectral index $\alpha$\ via $S\ =\ k \nu^{+\alpha}$).  \citet{atel4410} made observations on 2012 September 19 with the Australia Telescope Compact Array yielding preliminary flux densities of $13.2 \pm 0.20$\ mJy at 5.5 GHz and $13.5 \pm 0.20$\ mJy at 9.0 GHz, giving a spectral index of $+0.05 \pm 0.04$. Both measurements are consistent with hard state emission from a compact jet.  A further radio observation was undertaken on 2013 January 11 (MJD 56303.08) with the Australia Telescope Compact Array \citep{atel4760}, finding preliminary flux densities of $0.68 \pm 0.07$\ mJy at 5.5 GHz and $0.70 \pm 0.05$\ mJy at 9 GHz (spectral index $\alpha = 0.06 \pm 0.25$), again suggestive of optically thick synchrotron emission from a compact jet from the source in a hard X-ray state.

\begin{figure}
 \plotone{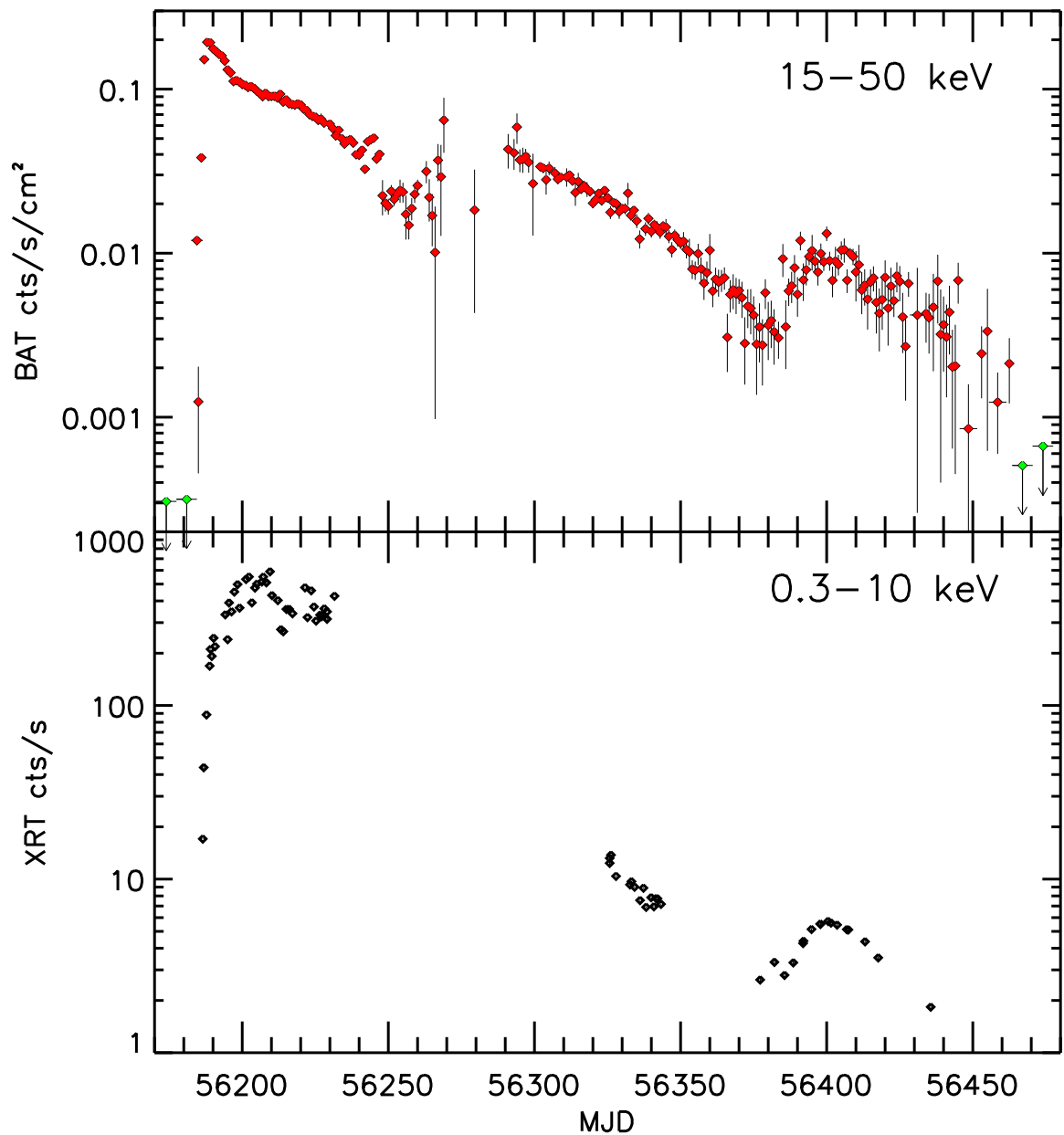}  
 \caption{Light curves for  Swift J1745.1$-$2624 from BAT (upper panel) and XRT (lower panel).  {\em Swift} was unable to observe Swift J1745.1$-$2624 with the NFIs between MJD 56231 and 56325 due to the Sun observing constraint.  Even the BAT light curve is affected (see Section~\ref{products}), with no observations between MJD 56269 and 56289 and larger error bars near the observation gap. }\label{sw1745_fig}
\end{figure}

Although full X-ray spectroscopic analysis will be carried out in another paper (Sbarufatti et al, in preparation), we can put together relevant reported outburst properties.  This combination of properties leads to the tentative conclusion that  Swift J1745.1$-$2624 is a black hole candidate.  First, both the BAT and XRT light curves (Figure~\ref{sw1745_fig}) have a very rapid rise, with the BAT light curve showing an increase of $\geq 3$\ orders of magnitude in the 15-50 keV band over five days.  Second, the early {\em INTEGRAL} spectra \citep{atel4381} can be fitted to a power-law with a high-energy exponential cutoff, $E_{\rm cut} = (122 \pm 10)$\ keV.  Third, a QPO is found in the XRT data, with a frequency of $0.250 \pm 0.003$\ Hz and width of $0.022 \pm 0.014$\ Hz at MJD 56188.8 \citep{atel4393} increasing to 2.4 Hz by MJD 56202.3 \citep{atel4450}. Fourth, in the same XRT data, \citet{atel4393} fit a power-law spectral index ($\Gamma = 1.53 \pm 0.02$) indicative of a BH transient in the hard state.   Fifth, combining the early radio measurements with near-simultaneous X-ray fluxes \citep{atel4394} and assuming that the source is at the distance of the Galactic center, yields a radio/X-ray flux ratio consistent with a black hole candidate, since a neutron star system would be expected to be much fainter in the radio.    Together these five measurements and inferences all support a black hole nature for the compact object in Swift J1745.1$-$2624.

The BAT light curve shows a very rapid rise early in the outburst, as described above, with a peak at MJD 56188 followed by a fairly steady decline for about 70 days.  At around MJD 56265, the BAT flux appears to start to rise again, but at this time the source was too close to the Sun to be observed, so it is unclear what occurred before MJD 56289, when the BAT flux is slightly lower than it was before the gap and declining.  After this, the decline is again steady until $\sim$\ MJD 56383, when the count rate began to rise again in both the BAT and XRT before leveling off at around MJD 56394.  The source was still detected in the BAT as of 2013 April 30.  The XRT light curve has an initial fast rise, then a roll-over to a more gradual rise. However, the XRT flux continues to increase until 56209, more than 20 days after the BAT peak.  The subsequent decline is also more shallow than in the BAT.  The XRT light curve is interrupted by a large gap due to a more stringent Sun-avoidance constraint than in the BAT.  After the gap the light curve resumes a decay with roughly the same slope before showing a rise near MJD 56380 roughly coincident with the rise in the BAT rate.  During the period up through MJD 56345, before and after the gap, there is considerable variation in the flux on time scales of $\sim$ 1 day.  The origin of these variations is unclear at this point.   

The nature of the X-ray spectral states of the Swift J1745.1$-$2624 outburst is also unclear.  Early XRT analysis \citep[MJD 56188;][]{atel4393} shows evidence of a hard state.  Analysis of {\em INTEGRAL} data shortly thereafter \citep{atel4401} suggests rapid spectral softening commencing by MJD 56189. \citet{atel4436} confirm continued spectral softening in the XRT at 56197, but see no sign of a thermal component in the spectral fit.  \citet{atel4450} infer, based on trends in the power-law spectral index, high-energy cutoff energy and QPO frequency, that by MJD 56201 the source was in a hard intermediate state and predicted that relativistic jet ejections might soon after occur. \citet{atel4456} also made such a prediction based on evolution in the R-i$^{\prime}$\ color.   No reports of such jet ejections have as yet been published.  After exiting from an observing constraint on MJD 56325, radio \citep{atel4760} and X-ray \citep{atel4782,atel4804} observations showed that Swift J1745.1$-$2624 had returned to a hard state.  

\subsubsection{Swift J1753.7$-$2544}\label{sw1757} 
\begin{figure}
 \plotone{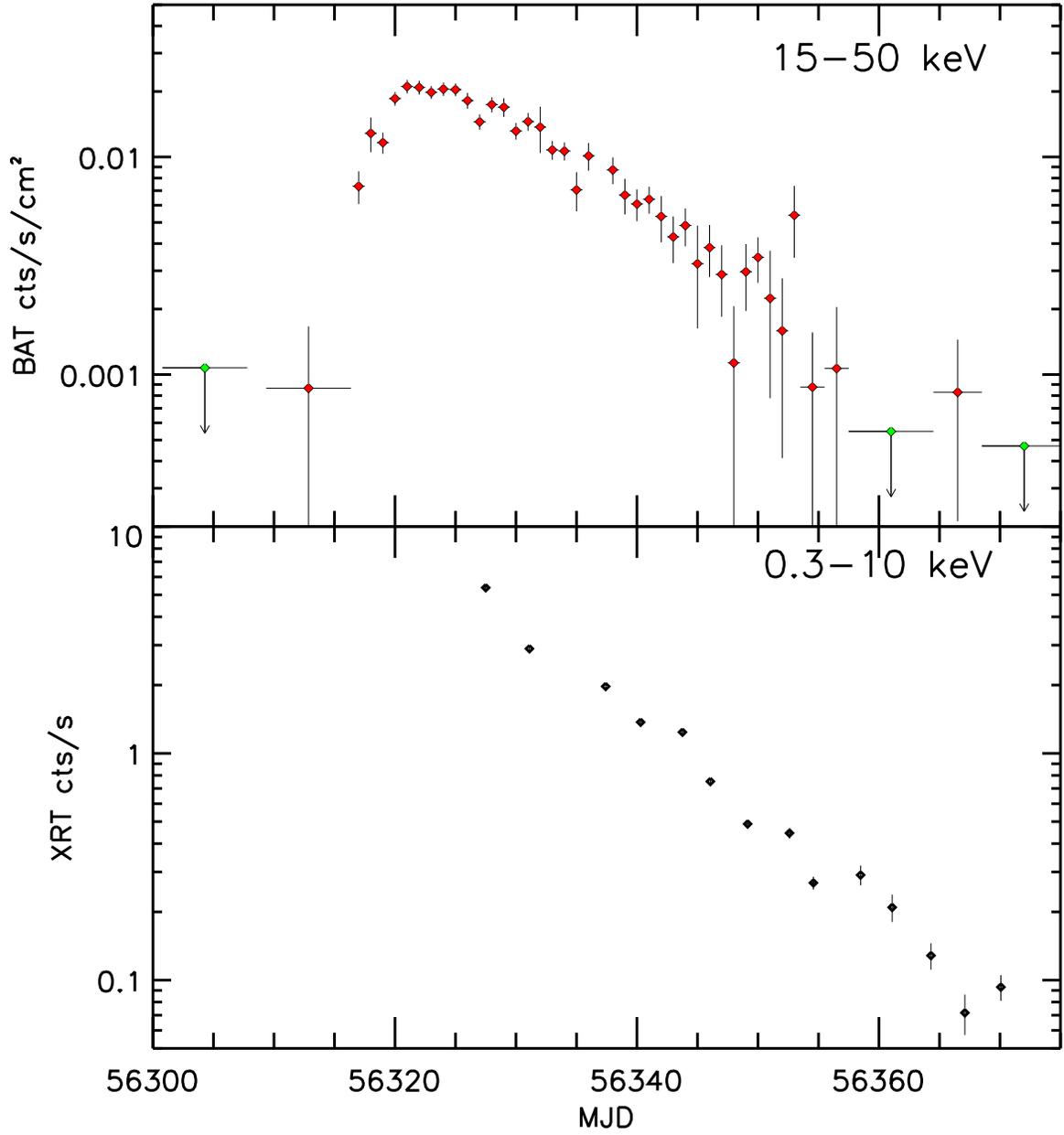}  
 \caption{Light curves for  Swift J1753.7$-$2544 from BAT (upper panel) and XRT (lower panel). }\label{sw1753_fig}
\end{figure}

\noindent{\bf XRB (likely Galactic); Discovered in 2013 January.} This is a likely Galactic source, which was first detected when it triggered {\em Swift}/BAT onboard on 2013 January 28 \citep[MJD 56320;][]{gcn14151,atel4769}.   Examination of the transient monitor light curves showed that it was first detected on 2013 January 24.  The count rate rose rapidly to a broad peak covering MJD 56321 - 56326, after which it exhibited a continual, mostly featureless, slow decline (Figure~\ref{sw1753_fig}).  Due to a {\em Swift} observing constraint, the first XRT and UVOT observations were not carried out until 2013 February 4 (MJD 56327).   A bright counterpart was clearly detected in the XRT, and spectrum of the first XRT observation is well fitted by an absorbed power-law model with the following parameters: N$_{\rm H} = 4.9 \pm 0.9 \times 10^{22} {\rm cm^{-2}}$, $\Gamma = 1.3 \pm 0.3$\ and observed flux (0.3-10 keV) of $5.7 \pm 0.5 \times 10^{-10} {\rm erg\  cm^{-2}\ s^{-1}}$.   Like the BAT, the XRT light curve (Figure~\ref{sw1753_fig}) shows a continual smooth decline.  The current results do not allow us to determine the nature of the source.

The source was localized by Chandra \citep{atel4899}. The absorbed (unabsorbed) 0.2-10 keV flux in the Chandra observation is reported as $1.9 \times 10^{-12}\ (3.4 \times 10^{-12})$ erg cm$^2$ s$^{-1}$ (assuming the X-ray spectrum measured by Swift/XRT for the early observations). This corresponds to an X-ray luminosity of $2.5 \times 10^{34}$ erg s$^{-1}$ for a distance of 8 kpc.   The position of the source on the sky and the high X-ray absorption suggest that it is located near the Galactic center.   \citet{atel4904} reported an optical counterpart in the K band consistent in position to the Chandra source, which was found to have decayed from $K \approx 16.5$\ on 2013 January 28 to $K \approx 17.5$\ on 2013 February 17.  The source was not detected in any other GROND band, nor was it detected in the UVOT.   Because the source lies in the direction of the Galactic Center, the actual reddening (or extinction) is unknown and probably large, which is consistent with the GROND and UVOT non-detections.

\subsubsection{Swift J1741.5$-$6548}\label{sw1741} 
\noindent{\bf XRB (likely Galactic neutron star); Discovered in 2013 March.} This is most likely a Galactic source, which was first detected in a 16-day mosaic covering 2013 February 26 - March 13 \citep[MJD 56349 - 56364;][]{atel4902} at an average count rate of  $0.0025 \pm 0.0003\ {\rm count\ cm^{-2}\ s}^{-1}$\ (11 mCrab).  A counterpart was found in the XRT (see below) and in the UVOT with b magnitude measurements \citep{atel4902} ranging from $18.56 \pm 0.1$\ to $19.20 \pm 0.16$\ over the first set of observations.  Detection of the optical source was confirmed by GROND \citep{atel4906} in all bands from g$^{\prime}$\ to K at comparable magnitudes.   The optical source does not match any catalog source in the Vizier database, but \citet{atel4906} report that it is coincident with an uncatalogued source in a Digitized Sky Survey\footnote{http://archive.eso.org/dss/dss} 2 blue image and that it had brightened by 1.5 - 2 magnitudes in g$^{\prime}$-r$^{\prime}$, suggesting that it is the optical counterpart to Swift J1741.5$-$6548.

The relative faintness of the source in the BAT and its location well away from the galactic plane (b = -17.823) gave rise to an initial speculation that it might be a distant tidal disruption flare, similar to Swift J2058.4$+$0516.  However, spectroscopy shows more similarities to a low-mass galactic X-ray binary.  Optical spectroscopy with the Gemini Multi-Object Spectrograph mounted on the 8-m Gemini South telescope on 2013 Mar 24 \citep{atel4919} shows a strong blue continuum with several emission lines, all consistent with zero red shift, and hence a galactic origin.  The initial XRT spectrum \citep{atel4902}, which is well fitted by an absorbed power-law model (Cstat = 483.2 for 552 d.o.f.) with N${\rm_H} = 1.51 \pm  0.31 \times 10^{21} {\rm cm}^{-2}$, $\Gamma = 1.54 \pm 0.093$\ and unabsorbed flux (0.3-10 keV)
of $1.78 \pm 0.079 \times 10^{-10} {\rm erg\ cm^{-2}\ s}^{-1}$, is consistent with a low-mass X-ray binary in a hard state.  After correcting for reddening, \citet{atel4906} find the optical source to be very blue, with a spectral slope ($F_{\nu} \sim \nu^{-\beta}$), $\beta = 0.69 \pm 0.17$, suggestive of an accretion disk spectrum.   

\begin{figure}
 \plotone{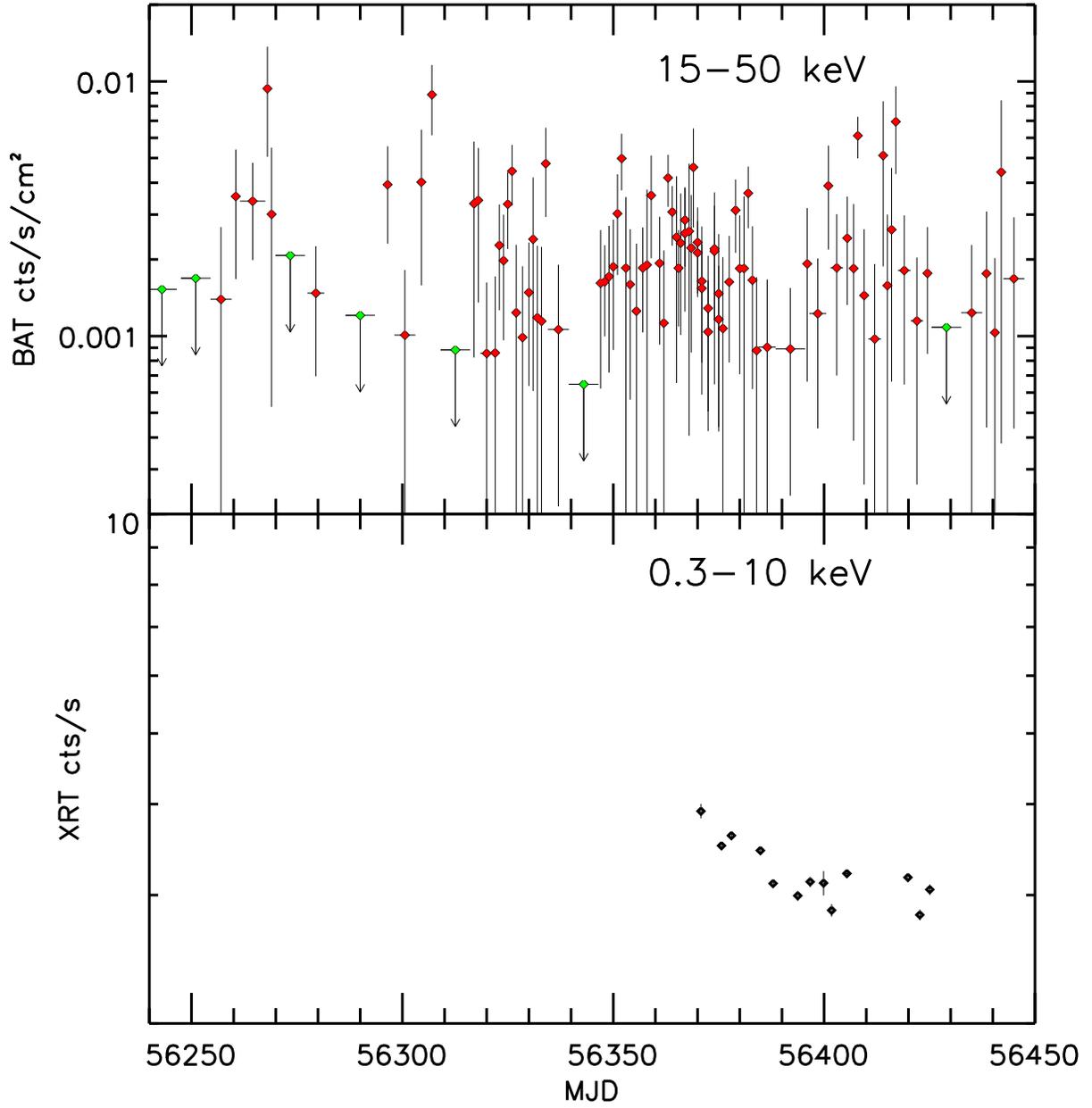}  
 \caption{Light  curves for  Swift J1741.5$-$6548 from BAT (upper panel) and XRT (lower panel). }\label{sw1741_fig}
\end{figure}

A further piece of evidence supporting the galactic transient interpretation for Swift J1741.5$-$6548 is a report of a possible previous outburst by \citet{atel4911}.    This outburst, on 2012 December 25, was initially reported by MAXI as GRB~121225A \citep{gcn14100}, with emission lasting at least 33 seconds.  However, based on a positional coincidence of GRB~121225A with Swift J1741.5$-$6548 and re-analysis of the data, \citet{atel4911} suggest that the 2012 December 25 event is instead a possible X-ray burst from a neutron star in the Swift J1741.5$-$6548 system.   \citet{atel4911} report that MAXI continued to detect this source at least until 2013 March 24, when the 2-20 keV flux was approximately 10 mCrab, consistent with the contemporaneous BAT flux.   The BAT light curve (Figure~\ref{sw1741_fig}, top panel) shows that there was activity around the time of the MAXI detection.  In fact, the highest rate seen in the BAT is on 2012 December 7 (MJD 56248) at  $0.009 \pm 0.004\ {\rm count\ cm^{-2}\ s}^{-1}$\ (40 mCrab).    Because Swift J1741.5$-$6548 was located near the Sun at the time, there were no {\em Swift} observations of the field coincident with the MAXI outburst.   After the 2012 December 7 peak, the BAT light curve drops for about 30 days before rising again to another peak on 2013 January 15 (MJD 56307), and then falling again.  Except for a brief period in 2013 February, Swift J1741.5$-$6548 remained detectable in the BAT through the end of 2013 April.  Examination of archival data reveals no previous outbursts from 2005 February to 2012 December.  The XRT light curve (Figure~\ref{sw1741_fig}, bottom panel) begins with the first {\em Swift} pointed observation on 2013 March 19 at a rate (0.3 - 10 keV) of $3.912 \pm   0.089\ {\rm count\ cm^{-2}\ s}^{-1}$\ followed by a steady decline with some variation until a possible upturn in the last observation on  2013 April 23.

\section{Conclusions}\label{discussion}

\noindent The {\em Swift}/BAT hard X-ray transient monitor\footnote{http://swift.gsfc.nasa.gov/docs/swift/results/transients/} has provided a continuous historical record of the variations in 15-50~keV X-ray flux of several hundred astrophysical hard X-ray sources from 2005 February 12 to the present time (2013 April 30), and its function is expected to continue as long as the {\em Swift}/BAT telescope is operational.    In total, 245 X-ray sources are considered to be detected in the BAT monitor during this period.  A source is considered detected if it meets one of three criteria.  The first two are systematized: either the mean rate $M$\ is $\geq 3$\ mCrab or the peak rate $P_7$\ is $> 30$\ mCrab and $\geq 7\sigma$.  Simple application of these two criteria finds 223 sources.  An additional 22 sources were reported to be in outburst either by {\em Swift}/BAT or by another group and then subsequently confirmed to be detected in the BAT monitor.  The detected sources are divided according to their variability, $V$, excess variance, $F_{\rm var}$, and $M$ into four categories:  outburst, flaring, steady and variable.  A subset of the variable sources are further classified as periodic.  Table~\ref{tab-class} shows the detected sources broken down by this classification (columns) and by source identification (rows).

This record shows that  99 sources that are normally not detected at the daily level in the monitor have exhibited one or more outbursts or flares during this period: 82 of these reached a level of 30~mCrab and 17 others had weaker events.  These 99 sources can be divided into two groups: 55 show least one outburst of $\gtrsim 10$-day duration and 44 exhibit flares of shorter duration.  For the most part these two groups can be reliably distinguished by calculating, for each source light curve, the variability, $V$, and the excess variance $F_{\rm var}$\ (defined in Section~\ref{results-known}).  Both groups have moderate to large $F_{\rm var}$, but the flaring sources exhibit low $V < 3$\ with particularly large $F_{\rm var} \geq 5$.  While Table~\ref{tab-class} shows that most of the different source types are represented in both of these groups, some very broad conclusions can be drawn. Among X-ray binaries for which the donor star type is known, most outburst sources are LMXB systems, while most flaring sources are HMXBs, including SFXTs.  Most of the other types of X-ray sources such as cataclysmic variables, TDFs, SGRs and blazars fall into the flaring category.

The BAT monitor also detects 146 persistent sources,  of which 88 are classified as steady (see Section~\ref{results-known} for the classification scheme) and 58 as variable or periodic.   140 of the persistent sources have a mean rate $M > 3$~mCrab. Of the other six, one, the SFXT SAX J1818.6$-$1703 has $P_7 = 57.7$, (above the $P_7$\ threshold) and the five others are among the 22 sources added to Table~\ref{tab-catalog} by hand.   Most (74/88) of the steady sources are weak $M < 10$\ mCrab and it is quite likely that some of these sources actually show variability, but at a level below that which can be detected in the BAT monitor.    Nearly half (42) of the steady sources are extragalactic: either AGNs, blazers or clusters and most of the rest (22) are LMXB/NS systems.  Despite its low frequency variations, the Crab Nebula is in the steady category as well.  Almost all of the 58 persistent variable or periodic sources are X-ray binaries (24 HMXB, 27 LMXB, 4 unclassified XRBs) and the remainder are extragalactic.  Five of these 58 sources show periodicity using the simple criterion described in Section~\ref{results-known}.  All but one of the periodic sources (Her~X-1) is a HMXB/NS system.

In addition to providing a real-time and archival data set of hard X-ray source light curves, the BAT transient monitor has also proven to be a very productive discovery tool.  Between the inception of the monitor and 2013 April 30, seventeen new sources have been discovered by the monitor. Nearly all of these sources have been extensively observed by X-ray, optical, and near IR telescopes; summaries and, for most sources, light curves, are provided in Section~\ref{results-new}.  Eleven of the new sources have been identified, five as neutron star systems, five as black hole candidates, and one as a tidal disruption flare event.   Of the ten X-ray binaries, five are LMXB, one HMXB and four have an as-yet unidentified donor star.   As improvements to the monitor have been implemented, the rate of discovery has accelerated, with ten of the 17 discoveries occurring within the past 2.5 years.  

With the advent of the BAT hard X-ray transient monitor, we have developed a powerful tool for studying, in near-real time, the variations in X-ray output from hundreds of astrophysical sources, as well as a discovery tool that has already led to the uncovering of 17 previously unknown transient sources.  Outbursts reported by the BAT monitor team or found by other observers on the public web pages have led to numerous observing campaigns and publications.  The BAT monitor also provides one of the most complete archives of fluctuations in the hard X-ray output of both Galactic and extragalactic sources.  The monitor is expected to continue to run as long as the {\em Swift} satellite is operating.

\acknowledgments 
\noindent The {\em Swift}/BAT transient monitor is supported by NASA under {\em Swift}  
Guest Observer grants NNX09AU85G, NNX12AD32G, NNX12AE57G and NNX13AC75G. H. A. K. also acknowledges these NASA grants for partial support. P. R. acknowledges ASI-INAF grant 1/004/11/0. We gratefully acknowledge the {\em RXTE} and {\em Swift} principal investigators for approving, and mission planners for scheduling, the many observations discussed in  this work.   This research has made use of data obtained from the High Energy Astrophysics Science Archive Research Center (HEASARC), provided by NASA's Goddard Space Flight Center and from the UK Swift Science Data Centre at the University of Leicester. This research has also made use of the SIMBAD database,
operated at CDS, Strasbourg, France.  Finally, the authors acknowledge helpful comments from an anonymous referee.


\begin{deluxetable}{lccccc}
\tablewidth{0pt} 	      	
\tabletypesize{\scriptsize} 
\tablecaption{Comparison of wide-field hard X-ray monitors.\label{tab-comparison}}
\tablehead{\colhead{Mission/Instrument} & \colhead{Type} & \colhead{Energy range} & \colhead{Sky coverage} & \colhead{Source position} & \colhead{1-day sensitivity}\\
& & {(keV)} & & {resolution} & {(3$\sigma$; mCrab)}}
\startdata
{\em $\dagger$CGRO}/BATSE\tablenotemark{a} & Earth occultation & 20-1800 & 80-90\% over 52 d & $> 0^{\circ}.2$ & 75 \\
{\em Fermi}/GBM\tablenotemark{b} & Earth occultation & 8-500 & 100\% over 26 days & $\approx 0^{\circ}.5$ & 150 (@ 20 keV)\\
{\em MAXI}/GSC\tablenotemark{c} & Gas slit camera & 2-20 & 95\% / day & $1^{\circ}.5$ & 9 \\
{\em $\dagger$RXTE}/ASM\tablenotemark{d} & Scanning shadow camera & 2-12 & \tablenotemark{e}& $5^{\prime}$ &  $\approx 15$\\
{\em Swift}/BAT\tablenotemark{f} & Coded aperture & 15-50 &  80-90\% / day& $2^{\prime}.5 (1\sigma)$ & 16\\
\enddata
\newline A dagger ($\dagger$) indicates that an instrument is no longer operating.
\tablenotetext{a}{\citet{harm02}.}
\tablenotetext{b} {\citet{wils12}.}
\tablenotetext{c}{ \cite{hiro11,sugi11}.}
\tablenotetext{d}{\citet{levi96}.}
\tablenotetext{e} {\citet{levi96} do not quote a sky coverage percentage, but state that a random source is scanned typically 5-10 times per day.}
\tablenotetext{f} {This work.}
\end{deluxetable}

\begin{deluxetable}{lc}
\tablewidth{0pt} 	      	
\tabletypesize{\scriptsize} 
\tablecaption{Transient Monitor source detection criteria\label{tab-criteria}}
\tablehead{\colhead{Criteria} & \colhead{Number meeting criteria}}
\startdata
(A) $M\tablenotemark{a} \geq 3$\ mCrab &         178\\
(B) $P_7\tablenotemark{b} \geq 30$\ mCrab &    154\\
(A) OR (B)  &         223\\
(A) AND (B)  &      109\\
\enddata
\tablenotetext{a}{Mean count rate.}
\tablenotetext{b}{Peak count rate for days when the source was found at $\geq 7 \sigma$\ significance.}
\end{deluxetable}

\begin{deluxetable}{llrllrrrrrrrr}
\tablewidth{0pt} 	      	
\tabletypesize{\scriptsize} 
\tablecaption{Sources detected in the BAT transient monitor\label{tab-catalog}}
\tablehead{\colhead{Source Name} & \colhead{R.A.} & \colhead{Decl.} & \colhead{Type} & \colhead{Class} & \colhead{$M$\tablenotemark{a}} & \colhead{$P_7$\tablenotemark{a}}  & \colhead{$V$\tablenotemark{b}} &  \colhead{$F_{var}$\tablenotemark{c}} &  \colhead{Error\tablenotemark{d}}}
\startdata
V709 Cas  &   7.204 &  59.289 & CV & Steady &   4.1 &  20.9 &         1.00 &    0.903 &   0.09905\\
IGR J00370+6122  &   9.250 &  61.367 & HMXB/NS & Flaring &   0.9 &  39.6 &         1.73 &   11.806 &   0.40115\\
NGC 262  &  12.196 &  31.957 & Sy2 & Steady &   5.3 &  16.7 &         1.00 &    0.691 &   0.06283\\
CF Tuc  &  13.283 & -74.652 & CV & Flaring &   0.5 &  46.9 &         1.01 &   15.033 &   0.64795\\
Gam Cas  &  14.177 &  60.717 & Star & Steady &   5.5 &  62.7 &         0.97 &    1.135 &   0.04459\\
SMC X-1  &  19.275 & -73.433 & HMXB/NS & Periodic &  27.2 &  89.9 &        18.64 &    0.757 &   0.00668\\
3A 0114+650  &  19.511 &  65.292 & HMXB/NS & Outburst &   8.3 & 129.9 &         5.40 &    1.463 &   0.02545\\
4U 0115+634  &  19.630 &  63.740 & HMXB/NS & Outburst &   5.4 & 483.7 &        16.45 &    9.981 &   0.05095\\
QSO B0241+62  &  41.240 &  62.468 & Sy1 & Steady &   3.6 &   0.0 &         1.01 &    0.827 &   0.13650\\
NGC 1275  &  49.950 &  41.517 & Sy2 & Steady &   4.4 &   0.0 &         0.95 &    1.011 &   0.09849\\
UX Ari  &  51.648 &  28.715 & CV & Flaring &  -0.3 &  38.9 &         1.45 &   20.679 &   1.51799\\
GK Per  &  52.799 &  43.905 & CV & Outburst &   4.0 &  54.1 &         4.08 &    2.857 &   0.11806\\
V 0332+53  &  53.750 &  53.173 & LMXB/NS & Outburst &   3.9 & 244.1 &        12.25 &    7.709 &   0.08425\\
X Per  &  58.850 &  31.050 & HMXB/NS & Variable &  30.4 &  78.0 &         3.29 &    0.256 &   0.01292\\
PKS 0405$-$385  &  61.746 & -38.441 & Quasar & Flaring\tablenotemark{e} &  -0.2 &   0.0 &         1.09 &   20.127 &   1.61169\\
3C 111  &  64.600 &  38.033 & Sy1 & Steady &   4.3 &  14.3 &         0.86 &    0.553 &   0.26795\\
3C 120  &  68.300 &   5.350 & Sy1 & Steady &   3.9 &   0.0 &         0.79 &    1.151 &   0.10409\\
LSV+44 17  &  70.247 &  44.530 & Star & Outburst &   3.0 & 254.1 &         6.50 &    6.700 &   0.13192\\
4U 0517+17  &  77.690 &  16.499 & Sy1.5 & Steady &   4.2 &   0.0 &         0.74 &    2.163 &   0.09975\\
SWIFT J0513.4$-$6547  &  78.368 & -65.788 & HMXB/NS & Flaring\tablenotemark{e} &   0.5 &   0.0 &         1.08 &   22.766 &   0.67024\\
4U 0513$-$40  &  78.528 & -40.041 & LMXB/NS & Steady &   3.1 &   0.0 &         0.76 &    0.210 &   0.59740\\
TV Col  &  82.356 & -32.818 & CV & Steady &   3.7 &   0.0 &         0.80 &    1.017 &   0.12891\\
LMC X-4  &  83.200 & -66.367 & HMXB/NS & Periodic &  20.5 &  79.9 &        14.86 &    0.913 &   0.00969\\
Crab Nebula  &  83.636 &  22.015 & PSR/PWN & Steady & 999.7 & 1293.4 &         1.44 &    0.044 &   0.00037\\
1A 0535+262  &  84.725 &  26.317 & HMXB/NS & Outburst &  88.1 & 5265.6 &       220.66 &    6.223 &   0.00352\\
XMMU J054134.7$-$682550  &  85.395 & -68.431 & HMXB/NS & Flaring &   1.4 &  50.3 &         1.86 &    5.825 &   0.21296\\
NGC 2110  &  88.047 &  -7.456 & Sy2 & Steady &  11.6 &  37.4 &         1.47 &    0.327 &   0.03718\\
MCG +8$-$11$-$11  &  88.725 &  46.433 & Sy1.5 & Steady &   5.8 &  13.4 &         0.88 &    0.874 &   0.07628\\
Mrk 3  &  93.901 &  71.037 & Sy2 & Steady &   4.9 &   0.0 &         0.87 &    0.368 &   0.16613\\
4U 0614+09  &  94.280 &   9.137 & LMXB/NS & Variable &  22.5 &  51.7 &         2.23 &    0.359 &   0.01689\\
MXB 0656$-$072  & 104.612 &  -7.263 & HMXB/NS & Outburst &   4.6 & 182.2 &        10.28 &    5.806 &   0.07181\\
EXO 0748$-$676  & 117.139 & -67.750 & LMXB/NS & Outburst &   6.7 &  38.2 &         5.11 &    1.586 &   0.04485\\
Vela Pulsar  & 128.850 & -45.183 & Pulsar & Steady &   6.7 &  35.3 &         0.80 &   -0.130 &   0.32288\\
GS 0834$-$430  & 128.979 & -43.185 & HMXB/NS & Outburst &  24.3 & 266.6 &        80.00 &    3.555 &   0.02696\\
IGR J08408$-$4503  & 130.197 & -45.058 & HMXB/SFXT & Flaring &   0.1 &  69.2 &         1.35 &   76.376 &   3.26594\\
Vela X-1  & 135.529 & -40.555 & HMXB/NS & Variable & 246.6 & 1720.4 &       409.45 &    0.769 &   0.00098\\
2S 0918$-$549  & 140.154 & -55.232 & LMXB/NS & Steady &   4.5 &   0.0 &         0.91 &   -0.143 &   0.52131\\
MCG $-$5$-$23$-$16  & 146.925 & -30.950 & Sy2 & Steady &   9.6 &  36.5 &         0.94 &    0.166 &   0.11514\\
GRO J1008$-$57  & 152.442 & -58.293 & HMXB/NS & Outburst &  18.0 & 978.4 &        39.98 &    6.269 &   0.01434\\
NGC 3227  & 155.878 &  19.865 & Sy1.5 & Steady &   4.5 &   0.0 &         0.79 &    0.968 &   0.07015\\
NGC 3281  & 157.967 & -34.854 & Sy2 & Steady &   3.8 &   0.0 &         0.74 &    1.256 &   0.10689\\
RXTE J1037.5$-$5647  & 159.397 & -56.799 & HMXB/NS & Steady &   3.7 &   0.0 &         0.93 &   -0.359 &   0.43566\\
Mrk 421  & 166.114 &  38.209 & Blazar & Outburst &   5.9 & 118.1 &         7.00 &    1.862 &   0.03666\\
NGC 3516  & 166.698 &  72.569 & Sy1.5 & Steady &   3.9 &   0.0 &         0.88 &    0.574 &   0.10782\\
SWIFT J1112.2$-$8238  & 167.949 & -82.646 & Unknown & Steady\tablenotemark{e} &   0.3 &   0.0 &         0.94 &   -4.855 &   5.00942\\
1A 1118$-$61  & 170.238 & -61.917 & HMXB/NS & Outburst &   3.9 & 527.1 &        11.88 &    9.626 &   0.07245\\
Cen X-3  & 170.300 & -60.617 & HMXB/NS & Variable &  73.8 & 345.0 &        66.86 &    0.764 &   0.00300\\
NGC 3783  & 174.750 & -37.733 & Sy1 & Steady &   6.4 &   0.0 &         0.84 &    0.775 &   0.06540\\
1E 1145.1$-$6141  & 176.869 & -61.954 & HMXB/NS & Variable &  18.5 & 171.2 &         4.75 &    0.666 &   0.01350\\
NGC 4151  & 182.650 &  39.417 & Sy1.5 & Variable &  23.1 &  51.1 &         4.81 &    0.355 &   0.00723\\
NGC 4388  & 186.450 &  12.650 & Sy2 & Variable &   9.0 &  34.8 &         1.28 &    0.647 &   0.03498\\
GX 301$-$2  & 186.650 & -62.767 & HMXB/NS & Periodic & 189.6 & 1588.5 &       247.39 &    1.099 &   0.00119\\
3C 273  & 187.275 &   2.050 & Blazar & Steady &  13.4 &  40.4 &         1.50 &    0.356 &   0.02699\\
IGR J12349$-$6434  & 188.728 & -64.565 & CV & Steady &   4.8 &   0.0 &         0.84 &    0.970 &   0.08800\\
NGC 4507  & 188.900 & -39.917 & Sy2 & Steady &   6.5 &   0.0 &         0.86 &    0.876 &   0.07108\\
AM 1236$-$270  & 189.727 & -27.308 & Sy2 & Steady &   3.2 &  36.9 &         0.80 &    0.530 &   0.46484\\
NGC 4593  & 189.914 &  -5.344 & Sy1 & Steady &   3.0 &   0.0 &         0.82 &    1.356 &   0.20854\\
H 1254$-$690  & 194.400 & -69.283 & LMXB/NS & Steady &   3.7 &   0.0 &         0.81 &    0.421 &   0.23611\\
GX 304$-$1  & 195.325 & -61.600 & HMXB/NS & Outburst &  37.9 & 1836.0 &        68.48 &    4.509 &   0.00676\\
NGC 4945  & 196.359 & -49.471 & Sy2 & Steady &   8.5 &  23.2 &         1.09 &    0.727 &   0.05299\\
MAXI J1305$-$704  & 196.735 & -70.451 & XRB/BHC & Variable &   5.9 &  49.2 &         3.26 &    1.388 &   0.07099\\
Cen A  & 201.365 & -43.019 & Sy2 & Variable &  48.0 & 146.4 &        11.02 &    0.355 &   0.00518\\
4U 1323$-$619  & 201.650 & -62.136 & LMXB/NS & Steady &  10.9 &  21.4 &         1.58 &    0.462 &   0.03654\\
MCG $-$6$-$30$-$15  & 203.975 & -34.300 & Sy1 & Steady &   3.3 &   0.0 &         0.73 &    2.866 &   0.13113\\
NGC 5252  & 204.567 &   4.542 & Sy2 & Steady &   3.6 &   0.0 &         0.94 &   -0.159 &   0.96564\\
1A 1343$-$60  & 206.854 & -60.643 & Sy1.5 & Steady &   3.7 &   0.0 &         0.75 &    0.908 &   0.18653\\
IC 4329A  & 207.325 & -30.317 & Sy1 & Steady &  11.3 &  20.2 &         0.94 &   -0.272 &   0.07051\\
SWIFT J1357.2$-$0933  & 209.320 &  -9.544 & LMXB/BHC & Flaring &   0.6 &  41.7 &         1.39 &   17.072 &   0.81443\\
MAXI J1409$-$619  & 212.011 & -61.984 & HMXB/NS & Flaring &   0.7 &  88.6 &         2.71 &   17.814 &   0.50557\\
ESO 97$-$13  & 213.292 & -65.323 & Sy2 & Steady &  12.6 &  32.2 &         0.88 &   -0.350 &   0.03379\\
NGC 5506  & 213.300 &  -3.217 & Sy2 & Steady &  10.2 &  19.3 &         0.95 &    0.536 &   0.03718\\
NGC 5548  & 214.500 &  25.133 & Sy1.5 & Steady &   3.0 &   0.0 &         0.91 &    1.264 &   0.12246\\
H 1417$-$624  & 215.300 & -62.700 & HMXB/NS & Outburst &   5.3 & 298.3 &         7.89 &    6.417 &   0.05871\\
NGC 5728  & 220.600 & -17.253 & Sy2 & Steady &   3.2 &   0.0 &         0.84 &    1.774 &   0.14435\\
QSO J1512$-$0906  & 228.211 &  -9.100 & Blazar & Flaring\tablenotemark{e} &   1.8 &   0.0 &         0.99 &    6.245 &   0.22024\\
PSR B1509$-$58  & 228.475 & -59.133 & PSR/PWN & Steady &   9.1 &  18.6 &         0.85 &   -0.494 &   0.05895\\
Cir X-1  & 230.170 & -57.167 & LMXB/NS & Outburst &   5.9 & 125.6 &         5.08 &    2.549 &   0.05910\\
SWIFT J1539.2$-$6227  & 234.818 & -62.459 & LMXB/BHC & Flaring &   0.8 & 107.5 &         2.00 &   18.657 &   0.48703\\
H 1538$-$522  & 235.597 & -52.386 & HMXB/NS & Variable &  20.9 &  94.4 &         3.22 &    0.475 &   0.01599\\
MAXI J1543$-$564  & 235.823 & -56.414 & XRB/BHC & Flaring &   0.3 &  36.9 &         1.39 &   50.769 &   1.95891\\
XTE J1543$-$568  & 236.021 & -56.762 & HMXB/NS & Variable\tablenotemark{e} &  -0.4 &   0.0 &         1.63 &   -3.930 &   8.32021\\
4U 1543$-$62  & 236.976 & -62.570 & LMXB/NS & Steady &   4.5 &   0.0 &         0.83 &    1.921 &   0.07693\\
IGR J15479$-$4529  & 237.060 & -45.479 & CV & Steady &   4.8 &   0.0 &         0.79 &   -0.338 &   0.33439\\
1E 1547.0$-$5408  & 237.726 & -54.307 & AXP & Flaring &   0.6 & 331.5 &         1.52 &   22.381 &   0.67070\\
H 1553$-$542  & 239.455 & -54.414 & HMXB/NS & Flaring &   0.5 &  45.2 &         2.21 &   21.114 &   0.90890\\
H 1608$-$522  & 243.175 & -52.417 & LMXB/NS & Outburst &  28.6 & 336.4 &        32.24 &    1.319 &   0.01000\\
Sco X-1  & 244.979 & -15.640 & LMXB/NS & Variable & 1225.5 & 3330.1 &        39.02 &    0.240 &   0.00020\\
IGR J16207$-$5129  & 245.175 & -51.483 & HMXB/SFXT & Steady\tablenotemark{e} &   2.6 &   0.0 &         0.89 &    2.328 &   0.22854\\
SWIFT J1626.6$-$5156  & 246.632 & -51.945 & LMXB/NS & Flaring &   1.7 & 131.9 &         1.72 &    6.608 &   0.26716\\
4U 1624$-$490  & 247.012 & -49.199 & LMXB/NS & Steady &   6.1 &   0.0 &         0.86 &    1.795 &   0.05246\\
IGR J16318$-$4848  & 247.967 & -48.803 & HMXB/NS & Variable &  23.9 & 273.7 &        10.42 &    0.918 &   0.01089\\
AX J1631.9$-$4752  & 248.000 & -47.878 & HMXB/NS & Variable &  19.4 & 155.6 &         5.69 &    0.753 &   0.01389\\
4U 1626$-$67  & 248.075 & -67.467 & LMXB/NS & Variable &  37.6 &  75.1 &         7.57 &    0.408 &   0.00521\\
IGR J16328$-$4726  & 248.192 & -47.437 & HMXB/SFXT & Flaring\tablenotemark{e} &   1.8 &   0.0 &         0.98 &    5.161 &   0.24792\\
4U 1630$-$472  & 248.502 & -47.394 & LMXB/BHC & Outburst &  11.0 & 462.3 &        27.89 &    5.282 &   0.03080\\
SGR 1627$-$41  & 248.968 & -47.587 & SGR & Flaring &   0.2 &  60.6 &         1.28 &   50.866 &   2.65212\\
IGR J16393$-$4643  & 249.825 & -46.717 & HMXB/NS & Steady &   6.0 &   0.0 &         0.87 &   -0.893 &   0.08074\\
4U 1636$-$536  & 250.231 & -53.751 & LMXB/NS & Variable &  25.2 & 115.1 &        12.19 &    0.782 &   0.00862\\
IGR J16418$-$4532  & 250.450 & -45.533 & HMXB/SFXT & Steady &   4.9 &  20.7 &         0.97 &    1.592 &   0.07612\\
Swift J164449.3+573451  & 251.205 &  57.581 & TDF & Flaring &   0.5 &  32.8 &         1.28 &   15.416 &   0.78233\\
GX 340+0  & 251.449 & -45.611 & LMXB/NS & Variable &  49.4 & 116.8 &         6.19 &    0.428 &   0.02128\\
IGR J16479$-$4514  & 251.975 & -45.233 & HMXB/SFXT & Outburst &   3.9 &  62.5 &         2.03 &    2.402 &   0.08253\\
MAXI J1647$-$227  & 252.051 & -23.015 & XRB/NS & Outburst &   4.1 &  36.0 &         4.03 &    1.366 &   0.20529\\
XTE J1652$-$453  & 253.192 & -45.353 & XRB/BHC & Flaring &   1.1 &  39.2 &         1.94 &    6.559 &   0.40747\\
Mrk 501  & 253.475 &  39.767 & Blazar & Variable &   4.0 &  36.5 &         1.28 &    1.040 &   0.10526\\
GRO J1655$-$40  & 253.501 & -39.833 & LMXB/BH & Outburst &   4.6 & 771.4 &         9.89 &   10.096 &   0.08317\\
Her X-1  & 254.457 &  35.342 & LMXB/NS & Periodic &  62.6 & 376.0 &       111.55 &    1.329 &   0.00297\\
MAXI J1659$-$152  & 254.760 & -15.258 & LMXB/BHC & Outburst &   2.6 & 233.9 &         8.87 &    8.479 &   0.12177\\
EXO 1657$-$419  & 255.200 & -41.673 & HMXB/NS & Variable &  59.3 & 464.5 &        50.79 &    0.916 &   0.00415\\
XTE J1701$-$462  & 255.243 & -46.186 & LMXB/NS & Outburst &   6.8 &  88.3 &         9.62 &    2.650 &   0.04701\\
XTE J1701$-$407  & 255.380 & -40.780 & LMXB/NS & Variable &   4.4 &   0.0 &         1.33 &    1.577 &   0.08509\\
GX 339$-$4  & 255.700 & -48.783 & LMXB/BH & Outburst &  26.8 & 698.0 &        48.55 &    3.446 &   0.01079\\
4U 1700$-$377  & 255.980 & -37.844 & HMXB/NS & Variable & 170.3 & 1556.4 &       177.81 &    0.759 &   0.00144\\
GX 349+2  & 256.450 & -36.417 & LMXB/NS & Variable &  74.6 & 195.3 &         6.76 &    0.245 &   0.00338\\
4U 1702$-$429  & 256.563 & -43.036 & LMXB/NS & Variable &  21.1 &  80.4 &         9.91 &    0.780 &   0.01052\\
IGR J17062$-$6143  & 256.567 & -61.711 & LMXB/NS & Steady\tablenotemark{e} &   2.8 &   0.0 &         0.87 &    2.401 &   0.12413\\
H 1705$-$440  & 257.225 & -44.100 & LMXB/NS & Variable &  24.0 & 125.1 &         6.54 &    0.581 &   0.01128\\
IGR J17091$-$3624  & 257.282 & -36.407 & LMXB/BHC & Outburst &   5.9 & 118.3 &         4.17 &    2.666 &   0.06857\\
Oph cluster  & 258.108 & -23.376 & Galaxycluster & Steady &   6.6 &   0.0 &         0.80 &    0.684 &   0.05274\\
SAX J1712.6$-$3739  & 258.136 & -37.632 & LMXB/NS & Steady &   6.4 &  27.4 &         0.99 &    0.684 &   0.07015\\
V2400 Oph  & 258.152 & -24.270 & CV & Steady &   3.1 &  19.8 &         0.91 &    1.278 &   0.13476\\
SWIFT J1713.4$-$4219  & 258.361 & -42.327 & Unknown & Flaring\tablenotemark{e} &  -0.2 &   0.0 &         1.21 &   30.547 &   2.30233\\
NGC 6300  & 259.209 & -62.792 & Sy2 & Steady &   4.3 &   0.0 &         0.81 &    1.668 &   0.09113\\
IGR J17191$-$2821  & 259.813 & -28.299 & XRB/NS & Flaring &   0.7 &  54.0 &         1.54 &   11.690 &   0.54975\\
IGR J17252$-$3616  & 261.308 & -36.273 & HMXB/NS & Variable &   7.7 &  66.7 &         2.44 &    0.886 &   0.04081\\
GRS 1724$-$308  & 261.900 & -30.800 & LMXB/NS & Variable &  19.2 &  65.1 &         3.19 &    0.338 &   0.01869\\
SWIFT J1729.9$-$3437  & 262.537 & -34.612 & XRB/NS & Flaring\tablenotemark{e} &   0.4 &  29.7 &         1.39 &   10.662 &   2.42489\\
IGR J17303$-$0601  & 262.590 &  -5.993 & CV & Steady &   3.7 &   0.0 &         0.79 &    1.764 &   0.09403\\
GX 9+9  & 262.934 & -16.962 & LMXB/NS & Steady &  20.9 &  56.5 &         1.74 &    0.226 &   0.01672\\
GX 1+4  & 263.000 & -24.750 & HMXB/NS & Variable &  59.1 & 340.1 &        35.25 &    0.830 &   0.00372\\
GX 354$-$0  & 263.000 & -33.833 & LMXB/NS & Variable &  53.4 & 202.7 &        32.78 &    0.662 &   0.00448\\
Rapid Burster  & 263.350 & -33.388 & LMXB/NS & Outburst &   2.6 &  59.6 &         2.87 &    3.249 &   0.17150\\
IGR J17361$-$4441  & 264.073 & -44.735 & Unknown & Flaring\tablenotemark{e} &  -0.1 &   0.0 &         1.15 &  133.570 &   5.78876\\
GRS 1734$-$292  & 264.369 & -29.131 & Sy1 & Steady &   4.6 &   0.0 &         0.83 &   -0.499 &   0.17815\\
SLX 1735$-$269  & 264.567 & -27.004 & LMXB/NS & Steady &  10.2 &  42.2 &         1.11 &    0.210 &   0.08274\\
4U 1735$-$44  & 264.743 & -44.450 & LMXB/NS & Variable &  32.2 &  73.6 &         3.81 &    0.375 &   0.00947\\
IGR J17391$-$3021  & 264.796 & -30.344 & HMXB/SFXT & Flaring &   1.0 &  46.6 &         1.34 &   13.059 &   0.47879\\
XTE J1739$-$285  & 264.975 & -28.480 & LMXB/NS & Flaring\tablenotemark{e} &   1.5 &   0.0 &         1.56 &    5.453 &   0.28721\\
SLX 1737$-$282  & 265.238 & -28.310 & LMXB/NS & Steady &   3.1 &   0.0 &         0.71 &    1.859 &   0.19243\\
SWIFT J1741.5$-$6548  & 265.348 & -65.791 & XRB/NS & Steady &   8.4 &   0.0 &         1.00 &    0.454 &   0.14687\\
1E 1740.7$-$2942  & 265.984 & -29.735 & LMXB/BHC & Variable &  37.0 &  92.1 &        11.82 &    0.513 &   0.00751\\
AX J1744.8$-$2921  & 266.240 & -29.336 & LMXB/NS & Variable &   4.9 &  35.4 &         2.05 &    1.479 &   0.09848\\
Granat J1741.9$-$2853  & 266.260 & -28.914 & LMXB & Outburst &   1.5 &  55.6 &         1.26 &    3.732 &   0.55557\\
Swift J1745.1$-$2624  & 266.295 & -26.403 & LMXB/BHC & Outburst & 156.9 & 874.3 &       251.92 &    1.160 &   0.00609\\
Sgr Astar  & 266.417 & -29.008 & Galacticcenter & Steady &   6.8 &   0.0 &         0.79 &    1.052 &   0.08030\\
1A 1742$-$294  & 266.525 & -29.517 & LMXB/NS & Variable &  11.5 &  49.3 &         2.60 &    0.789 &   0.02595\\
IGR J17464$-$3213  & 266.565 & -32.234 & LMXB/BHC & Outburst &  11.1 & 209.1 &        24.89 &    3.524 &   0.02832\\
1E 1743.1$-$2843  & 266.587 & -28.752 & LMXB & Steady &   6.7 &  28.5 &         0.76 &    1.195 &   0.05913\\
SAX J1747.0$-$2853  & 266.761 & -28.883 & LMXB/NS & Outburst &   1.8 &  34.6 &         1.26 &    3.180 &   0.40718\\
IGR J17473$-$2721  & 266.837 & -27.358 & LMXB/NS & Outburst &  12.2 & 380.5 &        35.59 &    5.867 &   0.02608\\
SLX 1744$-$300  & 266.856 & -30.045 & LMXB/NS & Steady &   9.5 &  24.7 &         1.01 &    0.511 &   0.04489\\
GX 3+1  & 266.975 & -26.567 & LMXB/NS & Variable &  24.7 &  57.6 &         2.73 &    0.277 &   0.01279\\
EXO 1745$-$248  & 267.022 & -24.780 & LMXB/NS & Outburst &   2.8 & 121.3 &         5.58 &    5.860 &   0.12544\\
AX J1749.1$-$2639  & 267.300 & -26.647 & HMXB/NS & Outburst &   7.0 & 277.8 &        24.19 &    6.157 &   0.04471\\
IGR J17497$-$2821  & 267.409 & -28.355 & LMXB/BHC & Outburst &   0.4 &  96.3 &         4.52 &   27.116 &   1.19124\\
IGR J17498$-$2921  & 267.481 & -29.322 & LMXB/NS & Flaring &   0.4 &  42.4 &         1.69 &   15.580 &   1.37010\\
4U 1746$-$370  & 267.553 & -37.052 & LMXB/NS & Steady &   5.3 &  20.7 &         0.82 &    1.478 &   0.07221\\
SAX J1750.8$-$2900  & 267.560 & -29.038 & LMXB/NS & Outburst &   1.5 &  87.5 &         4.02 &    7.120 &   0.27892\\
IGR J17511$-$3057  & 267.788 & -30.961 & LMXB/NS & Flaring &  -0.5 &  50.9 &         1.99 &   17.599 &   0.88595\\
XTE J1752$-$223  & 268.044 & -22.325 & LMXB/BHC & Outburst &  16.6 & 793.8 &        31.83 &    7.225 &   0.01881\\
SWIFT J1753.5$-$0127  & 268.368 &  -1.453 & LMXB/BHC & Variable &  64.0 & 396.6 &        16.20 &    0.512 &   0.00292\\
SAX J1753.5$-$2349  & 268.370 & -23.820 & LMXB/NS & Flaring &  -0.2 &  34.7 &         1.71 &   29.256 &   1.79504\\
SWIFT J1753.7$-$2544  & 268.429 & -25.742 & XRB & Outburst &  15.7 &  95.3 &        23.41 &    1.879 &   0.06977\\
IGR J17544$-$2619  & 268.605 & -26.331 & HMXB/SFXT & Flaring\tablenotemark{e} &   0.7 &  28.4 &         1.63 &    8.111 &   0.91605\\
SWIFT J1756.9$-$2508  & 269.218 & -25.125 & LMXB/NS & Flaring &   0.2 &  47.3 &         1.64 &    9.509 &   5.45584\\
IGR J17586$-$2129  & 269.658 & -21.327 & HMXB & Outburst &   3.3 &  49.8 &         2.05 &    2.613 &   0.10314\\
GX 5$-$1  & 270.275 & -25.083 & LMXB/NS & Variable &  83.9 & 191.9 &        34.17 &    0.446 &   0.00268\\
GRS 1758$-$258  & 270.300 & -25.733 & LMXB/BHC & Variable &  42.1 & 116.3 &         6.33 &    0.340 &   0.00547\\
GX 9+1  & 270.375 & -20.533 & LMXB/NS & Steady &  37.0 &  64.0 &         1.79 &    0.167 &   0.00783\\
IGR J18027$-$2016  & 270.692 & -20.294 & HMXB/NS & Variable &   6.3 &  30.3 &         1.46 &    1.359 &   0.04239\\
SAX J1806.5$-$2215  & 271.634 & -22.238 & LMXB/NS & Outburst\tablenotemark{e} &   3.0 &  28.5 &         1.82 &    3.578 &   0.11497\\
SAX J1808.4$-$3658  & 272.115 & -36.979 & LMXB/NS & Outburst &   1.0 &  80.2 &         3.28 &   10.766 &   0.40786\\
XTE J1810$-$189  & 272.586 & -19.070 & LMXB/NS & Outburst &   4.5 & 105.2 &         6.14 &    3.900 &   0.07728\\
SAX J1810.8$-$2609  & 272.651 & -26.153 & LMXB/NS & Outburst &   0.6 &  74.5 &         4.59 &   13.050 &   0.78682\\
GX 13+1  & 273.630 & -17.157 & LMXB/NS & Variable &  20.4 &  57.3 &         3.12 &    0.415 &   0.01353\\
4U 1812$-$12  & 273.800 & -12.083 & LMXB/NS & Steady &  26.7 &  42.8 &         1.34 &    0.122 &   0.02297\\
GX 17+2  & 274.000 & -14.033 & LMXB/NS & Variable &  92.1 & 243.6 &        18.88 &    0.359 &   0.00265\\
AM Her  & 274.055 &  49.868 & CV & Steady &   3.3 &   0.0 &         0.97 &    1.014 &   0.12591\\
SWIFT J1816.7$-$1613  & 274.176 & -16.222 & XRB/NS & Flaring &   0.4 &  35.0 &         1.32 &   21.639 &   0.89720\\
XTE J1817$-$330  & 274.425 & -33.018 & LMXB/BHC & Outburst &   1.2 & 146.7 &         3.33 &    9.350 &   0.33954\\
XTE J1818$-$245  & 274.605 & -24.542 & LMXB/BHC & Flaring &   0.4 &  70.8 &         1.17 &   28.484 &   1.33545\\
SAX J1818.6$-$1703  & 274.663 & -17.052 & HMXB/SFXT & Variable &   1.1 &  57.7 &         1.49 &    0.612 &   3.42352\\
SAX J1819.3$-$2525  & 274.825 & -25.417 & LMXB/BH & Flaring &   0.5 & 153.3 &         1.42 &   23.183 &   0.96045\\
XMMSL1 J182155.0$-$134719  & 275.479 & -13.791 & XRB & Variable\tablenotemark{e} &   1.6 &  24.0 &         1.27 &   -3.867 &   1.00749\\
H 1820$-$303  & 275.925 & -30.367 & LMXB/NS & Variable &  56.8 & 162.9 &         8.64 &    0.311 &   0.00415\\
IGR J18245$-$2452  & 276.138 & -24.879 & LMXB/NS & Outburst &  16.2 &  70.2 &        13.82 &    1.378 &   0.06718\\
H 1822$-$000  & 276.350 &  -0.017 & LMXB/NS & Steady &   3.1 &  21.8 &         0.88 &    1.034 &   0.17194\\
4U 1822$-$371  & 276.445 & -37.105 & LMXB/NS & Steady &  34.2 &  62.8 &         1.56 &    0.250 &   0.00942\\
Ginga 1826$-$238  & 277.368 & -23.797 & LMXB/NS & Variable &  72.8 & 136.4 &         4.50 &    0.171 &   0.00352\\
SNR 021.5$-$00.9  & 278.383 & -10.560 & SNR & Steady &   3.1 &   0.0 &         0.86 &    0.569 &   0.32701\\
4C 32.55  & 278.764 &  32.696 & Sy1 & Steady &   3.5 &   0.0 &         0.82 &    1.069 &   0.13293\\
MAXI J1836$-$194  & 278.931 & -19.320 & XRB/BHC & Outburst &   2.1 &  73.1 &         5.33 &    5.467 &   0.16122\\
XB 1832$-$330  & 278.933 & -32.982 & LMXB/NS & Steady &   7.1 &   0.0 &         1.16 &    0.266 &   0.14414\\
SWIFT J1836.6+0341  & 279.164 &   3.683 & XRB & Flaring\tablenotemark{e} &   0.3 &   0.0 &         1.14 &   15.767 &   2.16627\\
ESO 103$-$035  & 279.585 & -65.428 & Sy2 & Steady &   5.3 &   0.0 &         0.76 &   -0.467 &   0.11810\\
Ser X-1  & 279.990 &   5.036 & LMXB/NS & Steady &  16.7 &  43.7 &         1.68 &    0.328 &   0.01995\\
IGR J18410$-$0535  & 280.252 &  -5.596 & HMXB/SFXT & Flaring &   0.8 &  37.0 &         1.56 &    6.707 &   0.52962\\
3C 390.3  & 280.550 &  79.767 & Sy1 & Steady &   4.2 &   0.0 &         0.83 &    1.116 &   0.05685\\
SWIFT J1842.5$-$1124  & 280.573 & -11.418 & XRB/BHC & Outburst &   2.3 &  93.3 &         4.59 &    6.177 &   0.18658\\
SWIFT J1843.5$-$0343  & 280.895 &  -3.716 & XRB/NS & Flaring &   0.4 &  64.8 &         1.32 &   17.957 &   1.14115\\
IGR J18450$-$0435  & 281.250 &  -4.583 & HMXB/SFXT & Outburst\tablenotemark{e} &   1.7 &   0.0 &         1.11 &    4.257 &   0.22459\\
Ginga 1843+00  & 281.412 &   0.891 & HMXB/NS & Outburst &   2.8 & 103.0 &         4.44 &    3.996 &   0.11557\\
XMMSL1 J184555.4$-$003941  & 281.449 &  -0.633 & XRB & Flaring\tablenotemark{e} &  -0.0 &   0.0 &         1.28 & 1112.850 &  192.20800\\
GS 1843$-$02  & 282.074 &  -2.420 & HMXB/NS & Outburst &   2.4 &  82.8 &         2.32 &    3.657 &   0.47958\\
IGR J18483$-$0311  & 282.075 &  -3.161 & HMXB/SFXT & Outburst &   4.5 &  57.2 &         2.23 &    2.492 &   0.07481\\
4U 1850$-$087  & 283.270 &  -8.706 & LMXB/NS & Steady &   6.1 &   0.0 &         0.96 &    0.830 &   0.05861\\
4U 1849$-$31  & 283.750 & -31.167 & CV & Steady &   8.0 &  19.7 &         0.94 &    1.195 &   0.04392\\
XTE J1855$-$026  & 283.880 &  -2.607 & HMXB/NS & Variable &  11.3 &  51.4 &         2.16 &    0.790 &   0.02343\\
XTE J1856+053  & 284.163 &   5.330 & LMXB/BHC & Flaring &   0.4 &  66.9 &         2.37 &   13.130 &   1.19348\\
XTE J1858+034  & 284.650 &   3.350 & HMXB/NS & Flaring &   0.4 &  65.4 &         1.59 &    5.532 &   1.97935\\
HETE 1900.1$-$2455  & 285.036 & -24.921 & LMXB/NS & Variable &  24.0 &  65.1 &         6.22 &    0.549 &   0.00939\\
H 1907+097  & 287.400 &   9.833 & HMXB/NS & Variable &  12.2 &  59.4 &         2.89 &    0.701 &   0.01954\\
SWIFT J1910.2$-$0546  & 287.595 &  -5.799 & LMXB/BHC & Outburst &  23.7 & 122.5 &        21.99 &    1.245 &   0.02176\\
4U 1909+07  & 287.699 &   7.598 & HMXB/NS & Steady &  13.7 &  72.7 &         1.78 &    0.390 &   0.02212\\
Aql X-1  & 287.825 &   0.583 & LMXB/NS & Outburst &   7.7 & 204.7 &        11.51 &    3.056 &   0.03606\\
SS 433  & 287.956 &   4.990 & HMXB/uQUASAR & Variable &   7.0 &  30.9 &         1.88 &    0.644 &   0.05984\\
IGR J19140+0951  & 288.508 &   9.888 & HMXB/NS & Variable &   8.1 &  77.7 &         2.84 &    1.097 &   0.03207\\
GRS 1915+105  & 288.800 &  10.940 & LMXB/BH & Variable & 293.8 & 681.4 &       407.55 &    0.457 &   0.00073\\
4U 1916$-$053  & 289.700 &  -5.236 & LMXB/NS & Steady &   9.8 &  20.7 &         0.96 &    0.424 &   0.03720\\
SWIFT J1922.7$-$1716  & 290.679 & -17.283 & LMXB/NS & Outburst\tablenotemark{e} &   2.1 &  26.7 &         2.04 &    3.338 &   0.17168\\
IGR J19294+1816  & 292.483 &  18.311 & HMXB/NS & Flaring &   1.8 &  53.2 &         1.43 &    9.072 &   0.19029\\
NGC 6814  & 295.675 & -10.317 & Sy1.5 & Steady &   3.2 &  56.2 &         0.88 &    1.239 &   0.22331\\
SWIFT J1943.4+0228  & 295.892 &   2.465 & XRB & Flaring\tablenotemark{e} &   0.1 &   0.0 &         1.17 &   48.682 &   8.61044\\
XTE J1946+274  & 296.414 &  27.365 & HMXB/NS & Outburst &   3.6 & 141.0 &         9.95 &    5.764 &   0.08992\\
4U 1954+31  & 298.929 &  32.100 & LMXB/NS & Outburst &  14.5 & 192.0 &         9.20 &    1.062 &   0.01665\\
Cyg X-1  & 299.591 &  35.202 & HMXB/BH & Variable & 613.6 & 1708.1 &      1356.89 &    0.478 &   0.00042\\
3C 405.0  & 299.868 &  40.734 & Sy2 & Steady &   4.8 &   0.0 &         0.87 &    0.395 &   0.11405\\
EXO 2030+375  & 308.064 &  37.637 & HMXB/NS & Periodic &  58.2 & 1188.2 &        90.20 &    2.549 &   0.00381\\
Cyg X-3  & 308.107 &  40.958 & HMXB & Variable & 149.8 & 288.2 &       126.43 &    0.438 &   0.00128\\
Mrk 509  & 311.050 & -10.717 & Sy1 & Steady &   4.5 &   0.0 &         0.77 &    1.232 &   0.12071\\
SWIFT J2058.4+0516  & 314.583 &   5.226 & TDF & Flaring\tablenotemark{e} &   0.5 &   0.0 &         1.02 &    9.934 &   1.61701\\
GRO J2058+42  & 314.698 &  41.777 & HMXB/NS & Flaring\tablenotemark{e} &   0.4 &   0.0 &         1.18 &   28.754 &   0.85790\\
SAX J2103.5+4545  & 315.899 &  45.757 & HMXB/NS & Outburst &   3.9 & 169.1 &         5.71 &    3.722 &   0.06734\\
IGR J21247+5058  & 321.175 &  50.967 & Blazar & Steady &   8.1 &  14.5 &         1.11 &    0.636 &   0.03163\\
XB 2127+119  & 322.493 &  12.167 & LMXB/NS & Steady &   4.0 &   0.0 &         0.79 &   -0.843 &   0.18942\\
Ginga 2138+56  & 324.878 &  56.986 & HMXB/NS & Flaring &   0.8 &  78.3 &         1.55 &   13.523 &   0.36793\\
Cyg X-2  & 326.172 &  38.322 & LMXB/NS & Variable &  39.0 &  79.2 &        10.12 &    0.349 &   0.00510\\
NGC 7172  & 330.507 & -31.872 & Sy2 & Steady &   5.7 &   0.0 &         0.89 &    1.100 &   0.10363\\
4U 2206+54  & 331.984 &  54.518 & HMXB/NS & Variable &   8.3 &  91.3 &         3.57 &    1.062 &   0.02970\\
3C 454.3  & 343.490 &  16.148 & Blazar & Steady &   3.4 &  37.3 &         1.16 &    0.817 &   0.19591\\
QSO B2251$-$179  & 343.525 & -17.582 & Sy1 & Steady &   4.4 &   0.0 &         0.75 &    0.596 &   0.17634\\
NGC 7469  & 345.816 &   8.874 & Sy1 & Steady &   3.1 &   0.0 &         0.75 &    1.456 &   0.16423\\
Mrk 926  & 346.181 &  -8.686 & Sy1.5 & Steady &   4.6 &   0.0 &         0.76 &    1.868 &   0.07458\\
NGC 7582  & 349.600 & -42.367 & Sy2 & Steady &   3.2 &   0.0 &         0.77 &   -0.806 &   0.31578\\
Cas A  & 350.800 &  58.817 & SNR & Steady &   4.4 &   0.0 &         0.91 &    0.661 &   0.10333\\
\enddata
\tablenotetext{a}{Flux in mCrab.} 
\tablenotetext{b}{Scaled variability index as defined in Equation~\ref{eq-1}.}
\tablenotetext{c}{Excess variance as defined in Equation~\ref{eq-2}.}
\tablenotetext{d}{Error on excess variance.}
\tablenotetext{e}{Number of outbursts using the criteria defined in Table~\ref{tab-vfvar} for one-month intervals.}
\end{deluxetable}

\begin{deluxetable}{lcccccc}
\tablewidth{0pt} 	      	
\tabletypesize{\scriptsize} 
\tablecaption{Classification of BAT Monitor detected sources\label{tab-class}}
\tablehead{\colhead{Classification\tablenotemark{a}} & \colhead{Overall} & \colhead{Steady} & \colhead{Variable (Persistent)} & \colhead{Periodic} & \colhead{Outburst} & \colhead{Flaring}}
\startdata
HMXB/NS (incl. SFXT) &           58 &            5 &           18 &            4 &           17 &           14\\
HMXB/BH &            2 &            0 &            2 &            0 &            0 &            0\\
LMXB/NS &           69 &           22 &           22 &            1 &           18 &            6\\
LMXB/BH/BHC &           20 &            0 &            4 &            0 &           11 &            5\\
XRB/NS &            6 &            1 &            0 &            0 &            1 &            4\\
XRB/BHC &            5 &            0 &            1 &            0 &            2 &            2\\
XRB (other) &            9 &            1 &            2 &            0 &            3 &            3\\
Pulsar/PWN/SGR/AXP &            5 &            3 &            0 &            0 &            0 &            2\\
Stars (incl. CV) &           13 &            9 &            0 &            0 &            2 &            2\\
AGN (Seyferts) &           42 &           39 &            3 &            0 &            0 &            0\\
Blazar/Quasar &            7 &            3 &            1 &            0 &            1 &            2\\
Other\tablenotemark{b} &            6 &            4 &            0 &            0 &            0 &            2\\
Unknown\tablenotemark{c} &            3 &            1 &            0 &            0 &            0 &            2\\
TOTAL &          245 &           88 &           53 &            5 &           55 &           44\\

\enddata
\tablenotetext{a}{Acronyms:  HXMB = high-mass X-ray binary, NS = neutron star, SFXT = supergiant fast X-ray transient, BH = black hole, LMXB = low-mass X-ray binary, BHC = black hole candidate, XRB = X-ray binary, PWN = pulsar wind nebula, SGR = soft gamma repeater, AXP = anomalous X-ray pulsar, CV = cataclysmic variable, AGN = active galactic  nucleus.  The XRB classification is for sources that have not yet been classified as either LMXB or HMXB.  The XRB (other) designation means that the nature of the compact object is not known.}
\tablenotetext{b}{Includes supernova remnants, galaxy clusters, tidal disruption flares and the Galactic center.} 
\tablenotetext{c}{Sources for which the nature is undefined.}
\end{deluxetable}

\begin{deluxetable}{lllll}
\tablewidth{0pt} 	      	
\tabletypesize{\scriptsize} 
\tablecaption{Criteria for classifying BAT monitor sources.\label{tab-vfvar}}
\tablehead{\colhead{Category} & \multicolumn{4}{c}{Criteria\tablenotemark{a}}}
\startdata
Steady &  $M < 10:$ & $V < 1.2\ \ {\rm AND}\ \ F_{var} < 3$  & $M \geq 10:$ & $V < 2\ \ {\rm AND}\ \ F_{var} < 3$ \\
Variable/Periodic &  $M < 10:$ & $(1.2 \leq V < 2\ \ {\rm AND}\ \ F_{var} < 3)$ & $M \geq 10:$ & $(2 \leq V < 4\ \ {\rm AND}\ \ F_{var} < 2)$ \\
& & OR $(2 \leq V < 4\ \ {\rm AND}\ \ F_{var} < 2)$  & & OR $( V \geq 4\ \ {\rm AND}\ \ F_{var} < 1)$\\
& & OR $( V \geq 4\ \ {\rm AND}\ \ F_{var} < 1)$  & & \\
Outburst & \multicolumn{2}{l}{$(V < 3\ \ {\rm AND}\ \ 3 \leq F_{var} < 5)$ } & & \\
&  \multicolumn{2}{l}{OR $(2 \leq V < 3\ \ {\rm AND}\ \ 2 \leq F_{var} < 3)$ } & & \\
&  \multicolumn{2}{l}{OR $(3 \leq V < 4\ \ {\rm AND}\ \  F_{var} \geq 2)$ } & & \\
&  \multicolumn{2}{l}{OR $(V \geq 4\ \ {\rm AND}\ \ F_{var} \geq 1)$ } & & \\
Flaring & \multicolumn{2}{l}{$V < 3\ \ {\rm AND}\ \ F_{var} \geq 5$ } &  & \\
\enddata
\tablenotetext{a}{$M$\ is the mean flux in mCrab, $V$\ is the scaled variability index, and $F_{var}$\ is the excess variance.  See text (Section~\ref{results-known}) for full definitions of $V$\ and $F_{var}$.}
\end{deluxetable}

\begin{deluxetable}{lllcllll}
\tablewidth{0pt} 	      	
\tabletypesize{\scriptsize} 
\tablecaption{Localizations of {\em Swift}/BAT discovered transients\label{tab-loc}}
\tablehead{\colhead{{\em Swift} Source} & \colhead{RA (J2000)}  & \colhead{Declination (J2000)}   & \colhead{Error\tablenotemark{a}}& \colhead{Gal. lon.}  & \colhead{Gal. lat.}  & \colhead{Instrument\tablenotemark{b}}}
\startdata
J0513.4$-$6547 & $78^{\circ}.36787$\ \ ($05^{\rm h}13^{\rm m}28^{\rm s}.29$) & $-65^{\circ}.78858$ ($-65^{\circ}47'18.9''$) & $0''.3$ & $275^{\circ}.98641$ &  $-34.55411$ &  GROND\tablenotemark{1}  \\ 
J1112.2$-$8238 & $167^{\circ}.94915$ ($11^{\rm h}11^{\rm m}47^{\rm s}.797$) & $-82^{\circ}.64575$  ($-82^{\circ}38'44.71''$) & $0''.1$ & $299^{\circ}.63384$ & $-20^{\circ}.42062  $ &  Gemini South\tablenotemark{2} \\ 
J1357.2$-$0933 & $209^{\circ}.32026$ ($13^{\rm h}57^{\rm m}16^{\rm s}.86$) &  $-9^{\circ}.54414$\ \ ($-09^{\circ}32'38.9''$) & $0''.42$ &  $328^{\circ}.70219$ & $+50^{\circ}.00418$ & UVOT\tablenotemark{3}   \\ 
J1539.2$-$6227 & $234^{\circ}.79985$  ($15^{\rm h}39^{\rm m}11^{\rm s}.963$) & $-62^{\circ}.46731$ ($-62^{\circ}28'02.30''$) & $0''.5$ & $321^{\circ}.018595$ &  $-5^{\circ}.642750$&  UVOT\tablenotemark{4} \\ 
J1713.4$-$4219 & $258^{\circ}.36$\ \ \ \ \ ($17^{\rm h}13^{\rm m}27^{\rm s}$) & $-42^{\circ}.32$\ \ \ \ \ ($-42^{\circ}19'37''$) & $3'.0$ & $345^{\circ}.24$ & $-1^{\circ}.96$  & BAT\tablenotemark{5}  \\ 
J1729.9$-$3437 &  $262^{\circ}.5379$\ \ ($17^{\rm h}30^{\rm m}09^{\rm s}.10$) &   $-34^{\circ}.6122$\ \ ($-34^{\circ}36'43.8''$) &  $1''.7$ &  $353^{\circ}.4476$ & $ -0^{\circ}.2651$ &   XRT\tablenotemark{6}\\ 
J1741.5$-$6548 & $265^{\circ}.35046$ ($17^{\rm h}41^{\rm m}24^{\rm s}.11$) & $-65^{\circ}.79094$\ \ ($-65^{\circ}47'27.4''$)  & $0''.43$ & $ 327^{\circ}.18604$ & $-17^{\circ}.82337$ & UVOT\tablenotemark{7} \\
J1745.1$-$2624 & $266^{\circ}.295204$ ($17^{\rm h}45^{\rm m}10^{\rm s}.849$) & $-26^{\circ}.403500$ ($-26^{\circ}24'12.60''$)  & $0''.01$\tablenotemark{c} & $2^{\circ}.110863$ & $+1^{\circ}.403220$ & VLA\tablenotemark{8} \\
J1753.7$-$2544 & $268^{\circ}.41604$ ($17^{\rm h}53^{\rm m}39^{\rm s}.85$) & $-25^{\circ}.7539$\ \ ($-25^{\circ}45'14.2''$)  & $0''.3$ & $3^{\circ}.64768$ & $+0^{\circ}.10351$ & GROND\tablenotemark{9} \\
J1756.9$-$2508 &  $269^{\circ}.239$\ \ \ \ ($17^{\rm h}56^{\rm m}57^{\rm s}.35$) & $+25^{\circ}.108$\ \ \ \ ($+25^{\circ}06'27.8''$)  & $3''.5$ & $50^{\circ}.605$ & $+22^{\circ}.536$ &  XRT\tablenotemark{10} \\ 
J1816.7$-$1613 & $274^{\circ}.17775$ ($18^{\rm h}16^{\rm m}42^{\rm s}.66$) & $-16^{\circ}.22317$ ($-16^{\circ}13'23.4''$)  &  $1''.0$\tablenotemark{d} & $14^{\circ}.58724$ &  $+0^{\circ}.09156$ &   Chandra\tablenotemark{11} \\ 
J1836.6$+$0341 & $279^{\circ}.16433$ ($18^{\rm h}36^{\rm m}39^{\rm s}.44$) &   $+3^{\circ}.68350$\ \ ($+03^{\circ}41'00.6''$) & $0''.3$ & $34^{\circ}.53387$ & $+4^{\circ}.96424$ &  GROND\tablenotemark{12} \\ 
J1842.5$-$1124 &  $280^{\circ}.57271$ ($18^{\rm h}42^{\rm m}17^{\rm s}.45$)  & $-11^{\circ}.41775$ ($-11^{\circ}25'03.9''$)   &   $0''.6$ & $21^{\circ}.72714$ &   $-3^{\circ}.17916$ &  UVOT\tablenotemark{13}  \\ 
J1843.5$-$0343 & $280^{\circ}.8948$\ \ ($18^{\rm h}43^{\rm m}34^{\rm s}.75$) &  $-3^{\circ}.7157$\ \ \  ($-03^{\circ}42'56.6''$) & $2''.7$ & $28^{\circ}.7297$ &	 $+0^{\circ}.0514$  &  XRT\tablenotemark{14} \\ 
J1910.2$-$0546 & $287^{\circ}.59500$ ($19^{\rm h}10^{\rm m}22^{\rm s}.80$) & $-5^{\circ}.79886$\ \ ($-05^{\circ}47'55.9''$)  & $0''.3$ & $29^{\circ}.90265$ & $-6^{\circ}.84416$ & GROND\tablenotemark{15}, PTF\tablenotemark{16} \\
J1943.4$+$0228 & $295^{\circ}.89221$ ($19^{\rm h}43^{\rm m}34^{\rm s}.13$)  & $+2^{\circ}.46528$\ \ ($+02^{\circ}27'55.0''$)  & $0''.42$ & $41^{\circ}.17541$ & $ -10^{\circ}.42217$ & UVOT\tablenotemark{17} \\
J2058.4$+$0516 & $314^{\circ}.582908$	 ($20^{\rm h}58^{\rm m}19^{\rm s}.898$) &  $+5^{\circ}.225625$ ($+05^{\circ}13'32.25''$) 	& $0''.05$ & $53^{\circ}.617079$ &	 $-25^{\circ}.118892$  & EVLA\tablenotemark{18}  \\ 
\enddata
\tablenotetext{a}{Radius, 90\% confidence level}
\tablenotetext{b}{Telescope that provided the best position measurement.}
\tablenotetext{c}{The larger dimension of the elliptical error region quoted by \citet{atel4394}.}
\tablenotetext{d}{Error radius for Swift J1816.7$-$1613, J. Halpern, private communication.}
\tablerefs{
(1) {\citet{atel2013}.} 
(2) {\citet{atel3469}.} 
(3) {\citet{atel3138}.} 
(4) {\citet{krim11}.} 
(5) {\citet{atel2300}.} 
(6) {This work.} 
(7) {\citet{atel4902}.} 
(8) {\citet{atel4394}.} 
(9) {\citet{atel4904}.} 
(10) {\citet{krim07}.} 
(11) {\citet{atel1457}.}
(12) {\citet{atel3687}.} 
(13) {\citet{atel1716}.} 
(14) {\citet{atel3169}.} 
(15) {\citet{atel4144}.} 
(16) {\citet{atel4146}.} 
(17) {\citet{atel4049}.} 
(18) {\citet{cenk11}.} 
}
\end{deluxetable}

\clearpage

\clearpage

\clearpage

\clearpage

\clearpage

\clearpage

\clearpage

\clearpage

\clearpage

\clearpage

\clearpage

\clearpage

\clearpage

\clearpage

\clearpage

\clearpage

\clearpage

\clearpage
 \begin{figure}
 \plotone{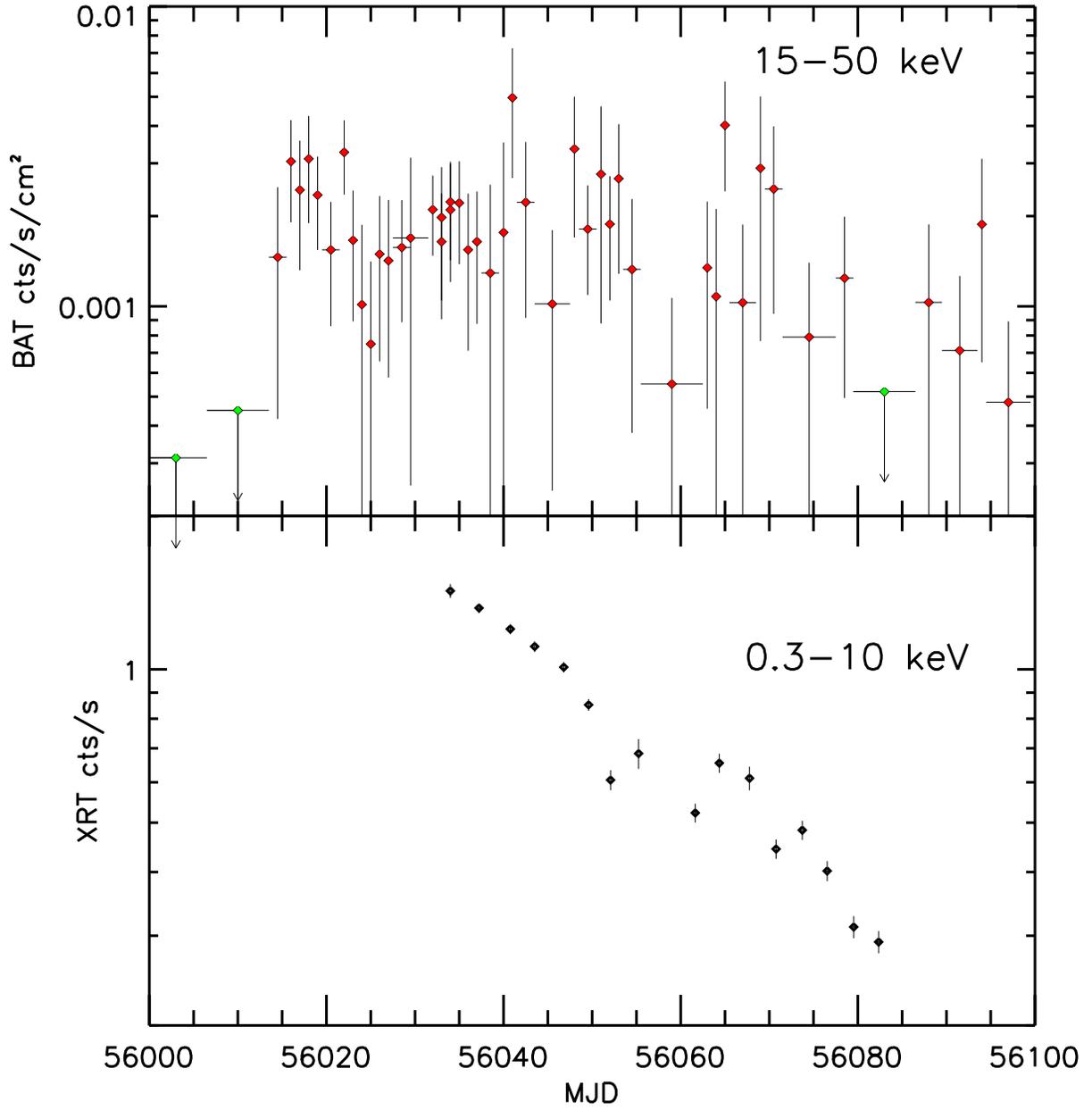}  
 \caption{Light curves for  Swift J1943.4$+$0228 from BAT (upper panel) and XRT (lower panel).} \label{sw1943_fig}
\end{figure}

\clearpage

\clearpage

\clearpage

\clearpage

\end{document}